\documentclass[manuscript,nonacm]{acmart}

\newif\ifDEBUG
\DEBUGtrue
\newif\ifEXTENDED
\EXTENDEDfalse

\usepackage{geometry} 
\usepackage{graphicx}
\usepackage{hyperref} 
\usepackage{cleveref}
\usepackage{caption}
\usepackage{pifont}
\usepackage{xcolor}
\usepackage{listings}
\usepackage{subcaption}
\usepackage{caption}

\usepackage{booktabs}
\usepackage{multirow} 
\usepackage{subcaption}
\usepackage{makecell}
\usepackage{tabularx}
\usepackage{ragged2e}
\usepackage{xspace}
\usepackage{soul}
\usepackage{array} 
\usepackage{float}
\usepackage{fancyhdr}

\usepackage{amsmath,amsfonts}
\usepackage{textcomp}
\usepackage{colortbl}
\usepackage{longtable}
\usepackage{array}
\usepackage{adjustbox}
\usepackage{algorithm}
\usepackage{threeparttable}
\usepackage{tablefootnote}
\makeatletter
\@ifundefined{r@tocindent4}{\expandafter\def\csname r@tocindent4\endcsname{10pt}}{}
\@ifundefined{r@tocindent3}{\expandafter\def\csname r@tocindent3\endcsname{10pt}}{}
\makeatother

\usepackage{xltabular}   % longtable + tabularx hybrid
\renewcommand{\arraystretch}{1.2}
\newcolumntype{L}{>{\raggedright\arraybackslash}X}

\newcolumntype{R}{>{\raggedleft\arraybackslash}X}

\usepackage{tikz}
\usetikzlibrary{arrows.meta,positioning,shapes.multipart}

\usepackage[noend]{algpseudocode}
\algrenewcommand\algorithmiccomment[1]{\hfill$\triangleright$~#1} % pretty right-margin comments
\usepackage[skins,breakable,listings]{tcolorbox}

\newcommand{\myprompt}[2]{%
  \begin{tcolorbox}[
    enhanced,
    colback=gray!10,
    colframe=black!20,
    boxrule=0.4pt,
    arc=1mm,
    left=2mm,right=2mm,top=1mm,bottom=1mm,
    fonttitle=\bfseries,
    coltitle=black,
    title={#1},
    before skip=3pt plus 2pt minus 1pt,
    after skip=3pt plus 2pt minus 1pt
  ]
  \small #2
  \end{tcolorbox}%
}

\usepackage{booktabs} % For formal tables
\usepackage{amsmath}
\usepackage{algorithm}
\usepackage{algpseudocode}
\usepackage[normalem]{ulem} %sout
\usepackage{xspace}
\usepackage{relsize}
\usepackage{multirow} % Fancy tables
\usepackage{balance} % Reference balancing
\usepackage{fancyvrb} % Verbose mode in footnotes
\usepackage{caption}
\usepackage{pifont}

\usepackage{enumitem}
\setlist[itemize]{leftmargin=*,noitemsep,topsep=0pt}
\setlist[enumerate]{leftmargin=*}

\usepackage{etoolbox}
\makeatletter
\patchcmd{\@makecaption}
	{\scshape}
	{}
	{}
	{}
\makeatletter
\patchcmd{\@makecaption}
	{\\}
	{.\ }
	{}
	{}
\makeatother

\newcommand{\Rplus}{\protect\hspace{-.1em}\protect\raisebox{.35ex}{\smaller{\smaller\textbf{+}}}}
\newcommand{\Cpp}{\mbox{C\Rplus\Rplus}\xspace}

\newcommand{\ie}{\textit{i.e.,}\xspace}
\newcommand{\eg}{\textit{e.g.,}\xspace}
\newcommand{\etal}{\textit{et al.}\xspace}

\usepackage{mdframed}
\mdfsetup{skipabove=0.5\topskip,skipbelow=0.5\topskip,align=center}

\usepackage{amsthm}
\newtheorem{thm}{Theorem}
\DeclareMathSymbol{\mlq}{\mathord}{operators}{``}
\DeclareMathSymbol{\mrq}{\mathord}{operators}{`'}

\newif\ifSAVESPACE
\SAVESPACEfalse

\ifSAVESPACE

\else

\fi

\usepackage{todonotes}
\usepackage{soul}
\usepackage{fontawesome}
\usepackage{ragged2e}

\ifDEBUG
    \newcommand{\AH}[1]{\todo[color=cyan,inline]{AH:#1}}
    \newcommand{\AM}[1]{\todo[color=red,inline]{Machiry:#1}}
    \newcommand{\JD}[1]{\todo[color=yellow,inline]{JD:#1}}
    \newcommand{\SA}[1]{\todo[color=green,inline]{SA:#1}}
    \newcommand{\PA}[1]{\todo[color=orange,inline]{PA:#1}}
    
    \newcommand{\KR}[1]{\todo[color=yellow,inline]{Kyle:#1}}
    \newcommand{\LS}[1]{\todo[color=green,inline]{LS:#1}}
    \newcommand{\HP}[1]{\todo[color=cyan,inline]{HP:#1}}
    \newcommand{\NJE}[1]{\todo[color=red,inline]{NJE: #1}}
    \newcommand{\GKT}[1]{\todo[color=red,inline]{GKT:#1}}
    \newcommand{\KL}[1]{\todo[color=teal,inline]{KL:#1}}
    \newcommand{\RH}[1]{\todo[color=red,inline]{RH:#1}}
    \newcommand{\WJ}[1]{\todo[color=SkyBlue,inline]{Wenxin:#1}} 
    \newcommand{\AG}[1]{\todo[color=orange,inline]{AG:#1}}
    \newcommand{\PJ}[1]{\todo[color=lime,inline]{PJ:#1}}
    \newcommand{\AZ}[1]{\todo[color=SkyBlue,inline]{AZ:#1}}
    \newcommand{\PT}[1]{\todo[color=pink,inline]{Parth:#1}}
\else
    \newcommand{\AH}[1]{}
    \newcommand{\AM}[1]{}
    \newcommand{\JD}[1]{}
    \newcommand{\SA}[1]{}
    \newcommand{\PA}[1]{}
    \newcommand{\KR}[1]{}
    \newcommand{\LS}[1]{}
    \newcommand{\HP}[1]{}
    \newcommand{\NJE}[1]{}
    \newcommand{\GKT}[1]{}
    \newcommand{\KL}[1]{}
    \newcommand{\RH}[1]{}
    \newcommand{\WJ}[1]{}
    \newcommand{\AG}[1]{}
    \newcommand{\PJ}[1]{}
    \newcommand{\PT}[1]{}
    \newcommand{\AZ}[1]{}
    
\fi

\newenvironment{RQList}{
   \setlength{\topsep}{0pt}
   \setlength{\partopsep}{0pt}
   \setlength{\parskip}{0pt}
   \begin{description}[style=unboxed]
   \setlength{\leftmargin}{1in}
   \setlength{\parsep}{0pt}
   \setlength{\parskip}{0pt}
   \setlength{\itemsep}{0pt}
   }
   {\end{description}}

\usepackage{hyperref}

\usepackage{cleveref}
\crefformat{section}{\S#2#1#3}
\crefname{figure}{Figure}{Figures}
\crefname{table}{Table}{Tables}
\crefname{theorem}{Theorem}{Theorems}
\crefname{thm}{Theorem}{Theorems}
\crefname{lemma}{Lemma}{Lemmata}
\crefname{equation}{Eqt.}{Eqts.}
\crefformat{Grammar}{Grammar #1}
\crefname{appendix}{Appendix}{Appendices}
\crefname{listing}{Listing}{Listings}
\crefname{algorithm}{Algorithm}{Algorithms}

\usepackage{url}

\usepackage{tcolorbox}

\makeatletter
\newcommand{\linebreakand}{%
  \end{@IEEEauthorhalign}
  \hfill\mbox{}\par
  \mbox{}\hfill\begin{@IEEEauthorhalign}
}
\makeatother

\usepackage{pifont}

 \usepackage{pifont}% http://ctan.org/pkg/pifont

\newcommand{\cmarkg}{\textcolor{green!50!black}{\ding{51}}}
\newcommand{\xmarkr}{\textcolor{red}{\ding{55}}}  
\definecolor{codegray}{gray}{0.95}

\newcolumntype{C}{>{\centering\arraybackslash}X}

\makeatletter
\@ifpackageloaded{endfloat}{\AtBeginDocument{\let\efloat@iwrite\relax}}{}
\makeatother

\def\tsc#1{\csdef{#1}{\textsc{\lowercase{#1}}\xspace}}
\tsc{WGM}
\tsc{QE}
\tsc{EP}
\tsc{PMS}
\tsc{BEC}
\tsc{DE}
\title{SysLLMatic: Large Language Models are Software System Optimizers}

\author{Huiyun Peng}
\affiliation{%
  \department{Department of Electrical and Computer Engineering}
  \institution{Purdue University}
  \city{West Lafayette}
  \state{Indiana}
  \postcode{47907}
  \country{USA}}
\email{hpeng@purdue.edu}

\author{Arjun Gupte}
\affiliation{%
  \department{Department of Electrical and Computer Engineering}
  \institution{Purdue University}
  \city{West Lafayette}
  \state{Indiana}
  \postcode{47907}
  \country{USA}}

\author{Ryan Hasler}
\affiliation{%
  \department{Department of Computer Science}
  \institution{Loyola University Chicago}
  \city{Chicago}
  \state{Illinois}
  \postcode{60660}
  \country{USA}}

\author{Nicholas John Eliopoulos}
\affiliation{%
  \department{Department of Electrical and Computer Engineering}
  \institution{Purdue University}
  \city{West Lafayette}
  \state{Indiana}
  \postcode{47907}
  \country{USA}}

\author{Chien-Chou Ho}
\affiliation{%
  \department{Department of Electrical and Computer Engineering}
  \institution{Purdue University}
  \city{West Lafayette}
  \state{Indiana}
  \postcode{47907}
  \country{USA}}

\author{Rishi Mantri}
\affiliation{%
  \department{Department of Electrical and Computer Engineering}
  \institution{Purdue University}
  \city{West Lafayette}
  \state{Indiana}
  \postcode{47907}
  \country{USA}}

\author{Leo Deng}
\affiliation{%
  \department{Department of Computer Science}
  \institution{Purdue University}
  \city{West Lafayette}
  \state{Indiana}
  \postcode{47907}
  \country{USA}}

\author{Konstantin Läufer}
\affiliation{%
  \department{Department of Computer Science}
  \institution{Loyola University Chicago}
  \city{Chicago}
  \state{Illinois}
  \postcode{60660}
  \country{USA}}

\author{George K. Thiruvathukal}
\affiliation{%
  \department{Department of Computer Science}
  \institution{Loyola University Chicago}
  \city{Chicago}
  \state{Illinois}
  \postcode{60660}
  \country{USA}}

\author{James C. Davis}
\affiliation{%
  \department{Department of Electrical and Computer Engineering}
  \institution{Purdue University}
  \city{West Lafayette}
  \state{Indiana}
  \postcode{47907}
  \country{USA}}

\begin{document}
% \let\WriteBookmarks\relax
% \def\floatpagepagefraction{1}
% \def\textpagefraction{.001}
% \shorttitle{SysLLMatic: Large Language Models are Software System Optimizers}
% \shortauthors{Huiyun Peng et~al.}

% \title [mode = title]{SysLLMatic: Large Language Models are Software System Optimizers}                       
% \maketitle

% \author[1]{Huiyun Peng}
% \author[1]{Arjun Gupte}
% \author[3]{Ryan Hasler}
% \author[1]{Nicholas John Eliopoulos}
% \author[1]{Chien-Chou Ho}
% \author[1]{Rishi Mantri}
% \author[2]{Leo Deng}
% \author[3]{Konstantin Läufer}
% \author[3]{George K. Thiruvathukal}
% \author[1]{James C. Davis}

% \credit{Conceptualization, Methodology, Software, Writing, Investigation}

% \affiliation[1]{organization={Department of Electrical and Computer Engineering, Purdue University},
%     addressline={West Lafayette}, 
%     city={Indiana},
%     postcode={47907}, 
%     country={USA}}

% \affiliation[2]{organization={Department of Computer Science, Purdue University},
%     addressline={West Lafayette}, 
%     city={Indiana},
%     postcode={47907}, 
%     country={USA}}
    
% \affiliation[3]{organization={Department of Computer Science, Loyola University Chicago},
%     addressline={Chicago}, 
%     city={Illinois},
%     postcode={60660},
%     country={USA}}

% \def\printorcid{}

% --- ABSTRACT ---
\begin{abstract}
Automatic software system optimization can improve software speed, reduce operating costs, and save energy.
Traditional approaches to optimization rely on manual tuning and compiler heuristics, limiting their ability to generalize across diverse codebases and system contexts.
Recent methods using Large Language Models (LLMs) introduce automation on simple programs, but they do not scale effectively to the complexity and size of real-world software systems. %, leaving a gap in practical applicability.

We present SysLLMatic, a system that integrates LLMs with performance diagnostics and a curated catalog of 43 optimization patterns to automatically optimize software systems.
By leveraging profiling to identify performance hotspots, our approach enables LLMs to optimize real-world software beyond isolated code snippets.
We evaluate it on three benchmark suites: HumanEval\_CPP (competitive programming in \Cpp),
  SciMark2 (scientific kernels in Java),
  and
  DaCapo (large-scale software systems in Java).
Results show that SysLLMatic can improve software system performance, including latency, throughput, energy efficiency, memory usage, and CPU utilization.
It consistently outperforms state-of-the-art LLM baselines on microbenchmarks.
On large-scale application codes, to which prior LLM approaches have not scaled, it surpasses compiler optimizations, achieving average relative improvements of 1.54$\times$ in latency (vs. 1.01$\times$ for the compiler) and 1.24$\times$ in energy (vs. 1.08$\times$ for the compiler).
Our findings demonstrate that LLMs, guided by performance knowledge through the optimization pattern catalog and appropriate performance diagnostics, can serve as viable software system optimizers.
We further identify limitations of our approach and the challenges involved in handling complex applications. 
This work provides a foundation for generating optimized code across various languages, benchmarks, and program sizes in a principled manner.
\end{abstract}



\maketitle
\thispagestyle{fancy}

\section{Introduction}
Software performance directly affects reliability~\cite{Jain1991}, user experience~\cite{Wilke2013}, energy use~\cite{Dayarathna2015, krasner2021cost, garraghan2014}, and sustainability~\cite{Ferme2017}; performance failures pose downtime and security risks~\cite{Liu2022Acquirer,mitre2024cwe1132,Deng2025AISecurity}.
Software performance optimization becomes increasingly essential as data centers consume 3--4\% of electricity in major regions~\cite{powering_intelligence2024, eu2024sustainability, Chen2025energy}.
Despite its importance, software performance optimization often loses out to organizational goals like feature development, time-to-market, and business objectives~\cite{paulsen2018criticality}.
This is partly because efficiency, while valuable, is costly to achieve --- it requires deep expertise, tooling, and time --- and is prioritized only when performance bottlenecks become urgent or business-critical~\cite{currie2024green}.

The research community has extensively studied software optimization, including via compiler techniques~\cite{Muchnick1998}, algorithmic improvements~\cite{Bader2002}, and hardware-aware tuning~\cite{Buchty2012}.
While effective, these approaches are limited by their need for deep domain expertise and significant manual effort~\cite{Balaprakash2018}.
% Large Language Models (LLMs) offer a promising alternative by automating software engineering tasks like debugging and code generation~\cite{Ozkaya2023}, overcoming many of the limitations inherent in traditional methods.
% More recently, researchers have begun to apply LLMs to software optimization, with studies showing promising improvements in metrics such as runtime, memory, and code quality~\cite{gong2025}.
Large Language Models (LLMs) have demonstrated strong capabilities across a range of software engineering tasks, including debugging, code generation, and software optimization~\cite{Ozkaya2023,gong2025}, overcoming many of the limitations inherent in traditional methods.
Recent studies show that LLM-based optimization approaches can improve metrics such as runtime, memory, and code quality~\cite{gong2025}.
However, most prior work has focused on isolated code snippets from competitive programming benchmarks, with limited attempts on larger software systems~\cite{gong2025}.
Existing approaches also neglected to systematically incorporate performance optimization knowledge (\cref{sec:GapAnalysis}).

To address these limitations, we present SysLLMatic, an LLM-based software optimization pipeline that integrates system performance knowledge and runtime feedback.
SysLLMatic refactors software to improve metrics like latency, resource utilization, and energy efficiency.
It consists of three components:
  a catalog of performance optimization patterns (\cref{sec:perf_optim_pattern}),
  a performance hotspot identification module (\cref{sec:perf_hotspot}),
  and
  a refinement module that applies the Diagnosis Cycle to iteratively optimize code (\cref{sec:formulation,sec:Instrumentation,sec:data-collection-refinement}).
SysLLMatic uses pre-trained LLMs and relies on prompting.
Information from the performance hotspot identification module and performance optimization catalog is injected into prompts, which are then fed to the LLMs to yield optimized code.
Thus, SysLLMatic readily generalizes across languages and benchmark suites.

We evaluate SysLLMatic on synthetic benchmarks (HumanEval in~\Cpp~\cite{chen2021evaluatinglargelanguagemodels} and SciMark2 in Java~\cite{scimark2}) and on real-world applications (DaCapo benchmarks in Java)~\cite{Blackburn2025}, ensuring coverage of both performance-oriented and widely used application domains.
We choose ~\Cpp because it is known for efficiency, and Java because it is widely used in industry-scale systems.
We compare it to prior LLM-based optimizations and traditional compiler-based optimizations (\cref{sec:EQ1}). 
Overall, SysLLMatic outperforms baselines across three benchmarks on multiple performance metrics.
It achieves the largest performance gains and optimizes the most programs on SciMark2.
On HumanEval, SysLLMatic increases the number of optimized programs by up to 8.9\% compared to the LLM optimization baseline while maintaining comparable performance improvements.
Meanwhile, the compiler baseline performs strongly on HumanEval, indicating the continued strength of traditional compiler optimizations.
On five DaCapo applications, SysLLMatic achieves substantial performance gains, including a 3.62$\times$ latency and 2.15$\times$ energy improvement in BioJava through parallelization.
% \JD{I think the next sentence needs to be reworded, it is not how such results are typically reported and there is no need for commentary about `hidden costs' --- hogwash.} Beyond raw gains, we model the net benefits of optimization by comparing SysLLMatic’s one-time cost against long-term improvements in energy efficiency and latency, making explicit the hidden costs of LLM-based optimization (\cref{sec:eq4_cost}) ~\cite{Coignion2025}. 
Our findings demonstrate that LLMs can serve as practical software system optimizers with tangible benefits for green computing.
Looking ahead, we further discuss the limitations of SysLLMatic and the challenges inherent to real-world software optimization, highlighting directions for future research to move closer to push-button optimization.

\ul{In summary, we contribute}:
\begin{itemize}
    \item \textbf{A Catalog of Performance Optimization Patterns (\cref{sec:perf_optim_pattern}):}
      We present a systematic catalog of 43 performance optimization patterns targeting various performance metrics. The catalog serves as a reusable knowledge base that guides LLMs in performing software optimizations.
    \item \textbf{A Novel Software System Performance Optimization Framework (\cref{sec:methodology}):}
    Our language-agnostic LLM framework combines domain-specific knowledge from the catalog with program analysis to apply performance optimization patterns. 
   The framework specifically targets large-scale applications with realistic workloads.
    \item \textbf{Multi-Metric Evaluation on Real-World Software (\cref{sec:experiment} and \cref{sec:result}):}
    We apply the proposed framework to synthetic microbenchmarks and to real-world applications from the DaCapo benchmark suite, which exemplifies large-scale systems and contrasts with the small, synthetic benchmarks commonly used in prior work (\cref{tab:dacapo-apps-merged,tab:line_diff_stats}). 
    We evaluate along multiple dimensions, including performance metrics (\eg latency, throughput, and energy) as well as maintainability metrics (\eg cyclomatic complexity).
    Our results show an average of 1.54$\times$ and 1.24$\times$ improvements in latency and energy, respectively, on DaCapo.
\end{itemize}

\vspace{0.1cm}
\noindent
\underline{\textbf{Significance:}}
Software optimization has been studied for over five decades, traditionally driven by compiler-based techniques and, more recently, by LLMs.
In this work, we explore LLMs as a tool for optimizing large-scale applications.
This approach moves closer to the goal of push-button optimization, where code can be significantly improved with minimal manual effort.
Such a capability allows researchers and software engineers to focus on integrating software components rather than spending extensive time debugging and tuning performance.

\section{Background and Related Works}
\label{sec:background}
% \JD{Are you citing the MSR-Challenge paper anywhere? Should.}

This section provides background on systems performance (\cref{sec:rel-work-systems-performance}), then summarizes the state-of-the-art approaches to traditional (\cref{sec:rel-work-traditional-optimization}) and LLM-based code optimization (\cref{sec:rel-work-llm-optimization}).

\subsection{System Performance: Motivation and Methodologies}
\label{sec:rel-work-systems-performance}
Performance is a critical concern in modern computing systems.
For server workloads, like distributed and cloud-based environments,
performance bottlenecks undermine responsiveness, throughput, and cost-efficiency~\cite{Gregg_2021}.
On the client side, performance affects energy consumption and perceived quality, especially on resource-constrained devices~\cite{Selakovic2016}.
In High Performance Computing (HPC), optimized code enables large-scale simulations and data analysis that are otherwise computationally infeasible~\cite{Gupta2016}.
Across these and other domains, effective performance engineering enables high-quality software.

Foundational methods in system performance analysis, such as scalability modeling~\cite{Amdahl1967}, design heuristics~\cite{Jain1991}, and capacity planning~\cite{Gunther1997}, underpin modern approaches.
Gregg summarizes best practices in \textit{Systems Performance} (2\textsuperscript{nd} ed.)~\cite{Gregg_2021}.
We draw on this body of work to inform our system design; the specific methodologies employed in this study are discussed below.

\begin{figure}[htbp]
    \centering
    \includegraphics[width=0.6\textwidth, trim={1cm 24cm 1cm 1cm}, clip]{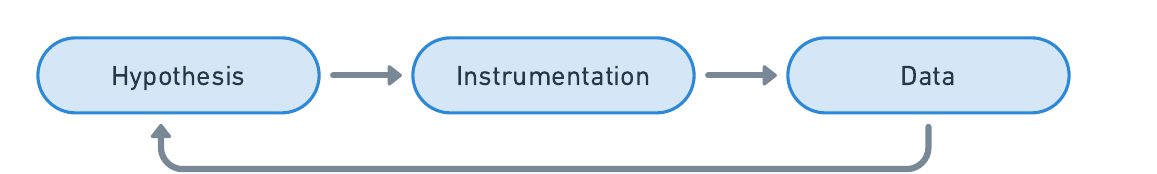}
    \caption{
    The Diagnosis Cycle~\cite{Gregg_2021} for performance optimization. 
    The process iteratively forms a hypothesis, instruments the program, and collects data to validate or refine it. 
    Because performance improvements can shift the system’s limiting factor, mitigating one bottleneck may reveal another, requiring repeated cycles of hypothesis generation and refinement. Our system design (\cref{fig:design-merged}) operationalizes this cycle for automated software optimization.
    % \JD{This caption should be updated based on the JD earlier in this section (along with new prose)}
    % \JD{Add a cref to Figure 4.}
    }
    \label{fig:cycle}
\end{figure}

\textit{Diagnosis Cycle}:
As summarized by Gregg, the Diagnosis Cycle methodology follows an iterative process (\cref{fig:cycle}).
In performance optimization, improvements often shift the system’s limiting factor, so mitigating one bottleneck may expose another.
New measurements therefore lead to additional rounds of hypothesis generation and validation.
Hypotheses are continually tested and refined through data collection, balancing principled reasoning with observed behavior.

% \JD{Performance diagnosis is typically iterative because improvements often shift the limiting factor and require re-measurement. Consider making that explicit in the surrounding prose (so long as it matches Gregg’s presentation), so it doesn’t read like a “there's one bottleneck --- hypothesize until you find the bottleneck and fix it” story. You might add notes here that also say `this is different from fixing a correctness defect, where the diagnosis cycle is a search for the right explanation...here we have multiple searches for distinct explanations over time. Enhancing this presentation will give us a much more direct map between Figure 1 and Figure 4.}

\textit{Profiling}:
Profiling is essential for analyzing and optimizing application performance, offering visibility into resource usage across CPU, GPU, and system components~\cite{mckenney1999differential, Graham1983, Graham1982}.
Tools like \texttt{perf} use timed sampling to capture user and kernel stack traces~\cite{shende1999profiling}.
These profiles are often analyzed with Flame Graphs~\cite{gregg2016flame}, where frame width visually indicates frequency to help identify performance hotspots.
Beyond CPU profiling, tools like Intel RAPL provide data on energy use and power consumption~\cite{Marcus2012,thorat2017energy,khan2018rapl}.
In the context of whole-system benchmarking, tools such as 
Babeltrace~\cite{babeltrace},
Omnitrace~\cite{amd_omnitrace},
NSight~\cite{iyer2016gpu}, and THAPI~\cite{bekele2025thapi} offer useful traces or visualizations.
Performance engineers employ such tools in the diagnosis cycle to identify performance bottlenecks and guide targeted optimizations~\cite{future_soft_perf_eng_2017,perf_eng_case_smith_1982}.

\subsection{Non-LLM Approaches for Code Optimization}
\label{sec:rel-work-traditional-optimization}

\textit{Code optimization} refactors programs to improve performance or efficiency without changing functionality, often prioritizing runtime metrics over readability or maintainability. 
Prior to the use of AI or LLMs, most work on optimization relied on established techniques such as compiler transformation and program analysis~\cite{bacon1994compiler}. 
At the compiler level, classic transformations include
global common-subexpression elimination~\cite{John1970}, loop-invariant code motion~\cite{Aho1986}, and register allocation~\cite{CHAITIN1981}.
Later work introduced advanced data-flow techniques such as sparse conditional constant propagation, enabling efficient constant folding across complex control flow with near-linear time~\cite{Wegman1991}.
Cytron \etal~\cite{Cytron1991} introduced static single-assignment (SSA) form for unified data-flow analyses.
Profile-guided optimization (PGO) followed, with path profiling~\cite{Ball1996} enabling code layout improvements based on runtime traces.
These and other advances such as SSA~\cite{Cytron1991} and PGO~\cite{Ball1996}
are integrated into compilers such as LLVM~\cite{Lattner2004}, a SSA-based infrastructure supporting compile-, link-, and run-time optimization.
Beyond traditional compilers, tools like OpenTuner~\cite{Ansel2014} automate the search for optimal compiler configurations.

Complementing compiler-level transformations, source-code-level optimizations directly modify source code to improve performance.
These source-level changes are often described as recurring \emph{performance patterns}, which capture common optimization strategies that developers apply to improve efficiency.
Common techniques include loop transformations (\eg unrolling~\cite{bacon1994compiler} and tiling~\cite{Lam1991}) to enhance instruction-level parallelism and cache locality.
Other key techniques include memoization~\cite{bacon1994compiler,ford2002packrat, Michie1968Memo}, dead code elimination~\cite{Cytron1991}, precision tuning~\cite{Khalifa2022}, and data structure selection~\cite{Couto2017} to improve memory access and algorithmic complexity.
These optimizations can be applied manually or through source-to-source transformation tools, but both approaches often require significant expertise and developer effort~\cite{Ansel2014, Selakovic2017}.
Genetic programming has been explored to automate such optimizations; however, a key challenge for such optimizations is the vast search space~\cite{white_evo_improvement_code_2011}.
This motivates the need for automated techniques that can explore this space more efficiently.

\subsection{LLM Approaches for Code Optimization}
\label{sec:rel-work-llm-optimization}
% Large language models have been increasingly applied across software engineering tasks. In this section, we review prior work on general software engineering applications and recent efforts in code optimization.

% % \subsubsection{General Software Engineering Tasks}
Large Language Models (LLMs), such as ChatGPT, Claude, and Gemini, have been widely adopted to automate software engineering tasks~\cite{Hou2024}. 
They have demonstrated effectiveness in tasks such as code generation and completion~\cite{Li2023, nijkamp2023codegen, austin2021programsynthesislargelanguage, Jiang2025}, test case generation~\cite{wang2024, yuan2024}, and defect detection and repair~\cite{wu2023, kang2023, chen2023, huang2023}.
More recently, researchers have applied LLMs to code optimization~\cite{gong2025}, spanning high-level optimizations~\cite{shypula2024learning, liu2024evaluatinglanguagemodelsefficient}, low-level compiler-style optimizations~\cite{cummins2023largelanguagemodelscompiler, cummins2024metalargelanguagemodel, taneja2024}, and domain-specific languages~\cite{wei2025improvingparallelprogramperformance}.
% These efforts show that LLMs can improve runtime performance~\cite{Chen2024, Gao2025, shypula2024learning}, reduce memory usage~\cite{Chen2024, Garg2022}, enhance energy efficiency~\cite{dearing2025leveragingllmsautomateenergyaware, peng2024largelanguagemodelsenergyefficient}, and raise overall code quality~\cite{li2025falconfeedbackdrivenadaptivelongshortterm}. 
% Importantly, LLMs are capable of handling complex code structures and supporting higher-level transformations such as loop restructuring and algorithm substitution.
In this section, we first review work on function-level code optimization with LLMs, then examine their limited applicability to real-world software systems, and finally discuss recent work on energy-efficient code optimization.

\subsubsection{Function-Level Code Optimization with LLMs}
Recent work has explored various approaches to LLM-based code optimization.
Shypula \etal established a foundation with the PIE dataset, consisting of paired slow-fast program variants from competitive programming, and demonstrated the baseline effectiveness of prompting and fine-tuning~\cite{shypula2024learning}.
Chen \etal developed SUPERSONIC, which uses a supervised seq2seq model to predict diff-based patches that apply small, targeted source-level changes while maintaining functional equivalence~\cite{Chen2024}. 
Beyond direct generation, recent approaches have incorporated structural guidance: Gao \etal formulate code optimization as a search problem guided by LLMs~\cite{Gao2025}, and Zhang \etal introduce a bidirectional tree–based progressive training paradigm for code acceleration that generates structured multi-level optimization data and enables LLMs to learn hierarchical optimizations~\cite{Zhang2025ACL}.

Across these approaches, feedback-based iterative optimization~\cite{madaan2023selfrefine} has emerged as a dominant strategy, providing a general mechanism for refining candidate optimizations through repeated evaluation and revision~\cite{gong2025}.
For example, EFFI-LEARNER~\cite{huang2024} of Huang \etal employs a self-optimization pipeline that leverages the overhead profile to improve code efficiency. Peng \etal introduced PerfCodeGen~\cite{peng2024perfcodegenimprovingperformancellm}, which leverages unit test runtime to guide performance improvements in generated code.
Static analysis has also been incorporated in the feedback, enabling refinements beyond functional correctness~\cite{blyth2025staticanalysisfeedbackloop}.
Beyond single-model usage, LLM-based agents have emerged that coordinate multiple reasoning steps and tools, enabling more autonomous and adaptive approaches to software optimization~\cite{dong2025surveycodegenerationllmbased}.
More recently, Rahman \etal propose MARCO~\cite{rahman2025marcomultiagentoptimizinghpc}, a multi-agent framework for optimizing HPC code.
These works show promise on small benchmarks (\eg single-function LeetCode-style tasks) but their scalability to multi-language, application-level codebases remains unclear.

% \JD{Tune the transition here --- make this topic sentence read more clearly if I haven't read the preceding paragraph.}
\subsubsection{Optimization on Real-World Software}
A recent survey by Gong \etal~\cite{gong2025} echoes concerns about the limited applicability of LLM-based optimization techniques to real-world software systems. It catalogs limitations of current LLM-based optimizers and notes that only 5 out of 55 (9\%) studies evaluate full real-world projects, underscoring the gap between proof-of-concept results and real-world potential.
Among the three works targeting source-level code optimization, Garg \etal propose RAPGen, a retrieval-augmented prompting framework for repairing performance bugs in C\# programs~\cite{garg2025rapgenapproachfixingcode}. 
However, it is confined to \.NET-specific APIs and evaluates fixes only via static metrics like Verbatim Match, without execution. 
Choi \etal~\cite{Choi2024} explore iterative LLM-based refactoring to reduce cyclomatic complexity in Java projects from the Defects4J benchmark. 
Though the method is nominally language-agnostic, its evaluation is limited to static maintainability metrics and does not assess runtime performance or execution-level improvements. 
DeepDev-PERF~\cite{Garg2022}, introduced by Garg \etal, fine-tunes BART on real-world C\# projects and assesses performance via pre-defined benchmark tests, but does not report the impact on end-to-end application behavior.

More recent benchmarks attempt to improve evaluation realism by grounding performance in executable outcomes. SWE-Perf~\cite{he2025sweperflanguagemodelsoptimize} and SWE-fficiency~\cite{ma2025swefficiencylanguagemodelsoptimize} require models to generate concrete patches that improve runtime efficiency under realistic project structures, build systems, and workloads. 
However, these benchmarks abstract away the performance diagnosis stage by assuming that optimization targets are already identified, which remains one of the core challenges in real-world performance engineering~\cite{Graham1982}. 
Complementing benchmark efforts, Yi \etal~\cite{yi2025experimentalstudyreallifellmproposed} mine human-authored performance pull requests as ground truth and compare them against LLM-generated optimizations, finding that expert-written patches consistently outperform LLM outputs. 
This finding reveals the need for frameworks that strengthen LLMs’ ability to identify, localize, and improve performance optimizations in complex, real-world systems.

In response to these limitations, our language- and framework-agnostic framework enables end-to-end optimization on real-world codebases, with performance evaluated at the application level.
Unlike prior work that assumes pre-identified optimization targets or relies on static proxies, our approach unifies hotspot discovery, transformation, and workload-driven evaluation within a single pipeline.
This integration advances LLM-based optimization beyond isolated code edits toward practical performance engineering in complex software systems.

% \GKT{This revision reads well. One possible improvement to the first sentence: In response to these limitations, our language- and framework-agnostic framework enables...}
% \JD{Now that we have added a lot more text, you need to write 1-2 more sentences here to clarify the contribution w.r.t. all dimensions you've brought up in the preceding.}

\subsubsection{Energy-Efficient Code Optimization with LLMs}
% \JD{In the title and prose, I don't like `sustainable', the word means too many things and thus it is meaningless. Should we consider the sustainability of the cars used to drive the engineers to work and the lightbulbs in their office?}
\label{sec:rel-work-sustainable-llm}

% \JD{The next sentences feel like a laundry list that are disconnected --- I cannot, as a reader, easily see how these papers fit together in way that helps me interpret them and contextualize your work. The way to fix this is to introduce some kind of inline framework for walking through each one, \eg ``As a first-order approximation, the relevant formula is INSERT, weighing the cost of the tool against the resulting benefit. Several approaches have been explored in this vein. To lower the cost of the model, Ashraf \etal have compared small and large... To lower the cost of the resulting code, LLMs can be trained (green-code) or prompted (peng) to generate more energy-efficient code. Coignion \etal observed that the relationship between speed and energy is complex, ...}
While LLM-based code optimization has largely focused on runtime and memory improvements, LLMs themselves entail substantial energy consumption~\cite{strubell-etal-2019-energy}. A growing line of work emphasizes the need to evaluate optimization outcomes in terms of energy efficiency.
To reduce the cost of using the model itself, Ashraf \etal~\cite{ashraf2025energyawarecodegenerationllms} examine tradeoffs between small and large LLMs, finding that smaller models can produce more energy-efficient code when correctness is preserved.
To reduce the cost of the generated code, Peng \etal~\cite{peng2024largelanguagemodelsenergyefficient} demonstrate that LLMs can be guided through tailored prompting to optimize code for energy efficiency; and 
Ilager \etal~\cite{greencode2025} propose
GREEN-CODE, a framework that further tunes models for energy-efficient code generation without sacrificing accuracy.
Complementing these efforts, Coignion \etal~\cite{coignion2025faster} emphasizes the need to consider optimization results against the sustainability costs of using LLMs, showing that achieving net savings may require hundreds to hundreds of thousands of executions.

Our work aligns with both efforts to lower the cost of the generated code and efforts to account for the sustainability costs of using LLMs.
Prior work has not evaluated these questions on large-scale software systems.
We address this gap by measuring the break-even point of energy savings by modeling inference costs across multiple deployment scenarios (\cref{sec:eq4_resource_consumption}).
This allows us to examine when the upfront overhead of LLM-guided optimization is amortized by downstream efficiency gains, offering new insights into the practical sustainability of LLM-based code optimization.

% \JD{Real-world large-scale are generally synonymous terms. Avoid saying both in one breath, especially since (1) we don't evaluate on any real-world, just DaCapo which is a benchmark; and (2) our `large-scale' is (to my eyes) more medium-scale (not 1MLoC+).}
\section{Problem Statement}

\begin{tcolorbox}[colback=yellow!5,colframe=black!75!black,title=Problem Statement]
This work aims to develop a system that leverages LLMs to automatically optimize real-world software applications in a principled manner.
The system targets improvements in key performance metrics such as latency, throughput, and energy consumption while preserving program correctness.
\end{tcolorbox}

\subsection{Gap Analysis}
\label{sec:GapAnalysis}
Our work addresses three key gaps in LLM-based software optimization, as shown in~\cref{tab:related-works-llm-opt-merged}.
\begin{enumerate}
    \item \textit{Scalability challenges in LLM-based optimization:} Existing LLM-based methods automate code transformation but typically operate on small, isolated code segments.
    We refer to these approaches as \textit{optimizing synthetic benchmarks}, which show potential but fail to scale to complex, multi-module software systems.
    \item \textit{Limited generalizability:} Current approaches are often fine-tuned for narrow tasks or rely on language-specific heuristics. This limits their applicability across different languages and workload types~\cite{amusuo2025falsecrashreducermitigatingfalsepositive}.
    \item \textit{Limited metric scope:} Most prior works focus on evaluating on a single metric --- typically runtime --- without considering multi-objective trade-offs (\eg energy, maintainability) essential for real-world software systems~\cite{gong2025}. 
    Additionally, in the context of large-scale application-level optimization, existing approaches~\cite{Choi2024, garg2025rapgenapproachfixingcode} often rely on static code metrics rather than runtime performance measurements.
\end{enumerate}

\begin{table}[h!]
    \centering
    \captionsetup{type=table}
    \small 
    \caption{
        Comparison of related LLM-based works across three dimensions: \textbf{Applications} indicates whether the method is applied to real-world software systems rather than isolated synthetic tasks; \textbf{Generalizable} reflects whether the approach is language- and task-agnostic; \textbf{Performance Metrics} indicates whether the method uses runtime performance metrics rather than only static code metrics (\eg cyclomatic complexity).
        Overall, few existing works perform large-scale software optimization, and those that do either rely solely on static code metrics or lack generalizability across domains.
    }
    \label{tab:related-works-llm-opt-merged}
    \footnotesize 
    
    \begin{tabular*}{\textwidth}{l @{\extracolsep{\fill}} ccc}
        \toprule
        \textbf{Related works} & \textbf{Applications} & \textbf{Generalizable} & \textbf{Uses Performance Metrics} \\
        \midrule
        \multicolumn{4}{l}{\textit{Synthetic Benchmark Optimization (Competitive programming benchmarks like LeetCode)}} \\ \eg~\cite{shypula2024learning,Chen2024,Gao2025,Zhang2025ACL,huang2024,peng2024perfcodegenimprovingperformancellm,rahman2025marcomultiagentoptimizinghpc} & \xmarkr & \cmarkg & \cmarkg \\
        \midrule
        \multicolumn{4}{l}{\textit{Large-Scale Optimization (Real-world applications)}} \\
        Choi \etal~\cite{Choi2024} & \cmarkg & \cmarkg & \xmarkr \\
        DeepDev-PERF~\cite{Garg2022} & \cmarkg & \xmarkr & \cmarkg \\
        RAPGen~\cite{garg2025rapgenapproachfixingcode} & \cmarkg & \xmarkr & \xmarkr \\
        \midrule
        \textbf{SysLLMatic (ours)} & \cmarkg & \cmarkg & \cmarkg \\
        \bottomrule
    \end{tabular*}

\end{table}

\subsection{Our Approach}
\label{sec:problem_Statement_our_approach}
In addressing these gaps, we follow the feedback-based iterative optimization approach standard in this literature.
Our main innovations are:
(1) grounding iterative software optimization with expert systems performance principles. 
Our ablation shows that incorporating the optimization pattern catalog leads to improvements in 4 out of 5 metrics in DaCapo benchmarks, with gains up to 49\% compared to running without the catalog (\cref{tab:ablation_study_metrics});
and
(2) developing an optimization framework that enables LLM-driven transformations across multiple performance metrics for real-world software systems.
We first describe a catalog of performance optimization patterns (\cref{sec:perf_optim_pattern}), and then a system design that leverages it (\cref{sec:methodology}).

% \JD{Note that cref chokes on the `label' for the paragraph-level cref}
Our design adopts a controlled LLM-based optimization approach rather than a fully autonomous agent model, as this allows us to systematically guide transformations and evaluate the impact of each component. We revisit this distinction in Discussion (\cref{sec:discussion_agent}).

\section{Performance Optimization Pattern Catalog}
\label{sec:perf_optim_pattern}

\begin{figure}[htbp]
     \centering
     \includegraphics[width=0.7\textwidth, trim={1cm 18.5cm 1cm 2.5cm}, clip]{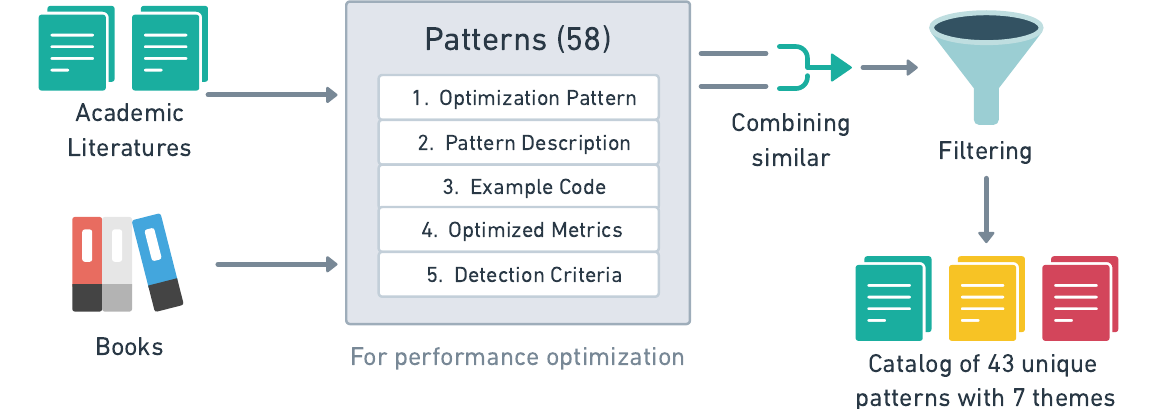}
       \caption{
       Construction of the Performance Optimization Pattern Catalog: We annotate, merge, and filter patterns from literature to derive 43 unique optimization patterns across 7 themes. 
       }
     \label{fig:patterncatalog-merged}
 \end{figure}

Our first goal was to construct an LLM-friendly catalog of software performance optimization patterns.
Our method and results are summarized in~\cref{fig:patterncatalog-merged}.

\subsection{Methodology}

To systematically incorporate performance optimization domain knowledge, we conducted a literature review.
We began with three authoritative performance engineering textbooks~\cite{gerber2006software,Gregg_2021,Kukunas2015}, from which we identified 28 optimization techniques.
We then reviewed 355 papers, initially gathered from a major survey paper on software non-functional properties~\cite{Blot2025}.
The survey, published in ACM Computing Surveys (Feb 2025), employs a rigorous selection process that aligns with our focus on performance optimization. 
It covers 425 papers published up to July 2024, with most appearing in top-tier venues such as ICSE and IEEE TSE.

To focus on actionable and broadly applicable patterns, we manually applied exclusion filters to remove papers outside our scope—specifically those that target low-level or domain-specific optimizations, as well as tool-centric works lacking generalizable insights. 
While this subset does not fully represent the entire field of performance optimization, it reflects our emphasis on extracting reusable source-level patterns.
After applying these filters, we reduced the set to 110 papers.
We then reviewed each remaining paper and extracted 30 optimization patterns from the papers, resulting in a total of 58 patterns.
To reduce subjectivity in pattern identification and extraction, two authors independently reviewed the papers and cross-checked the extracted patterns to ensure consistency.

\subsection{Results}
\begin{figure*}[htbp]
    \centering
    \includegraphics[width=\linewidth, trim={1.2cm 16.5cm 1.2cm 3.6cm}, clip]{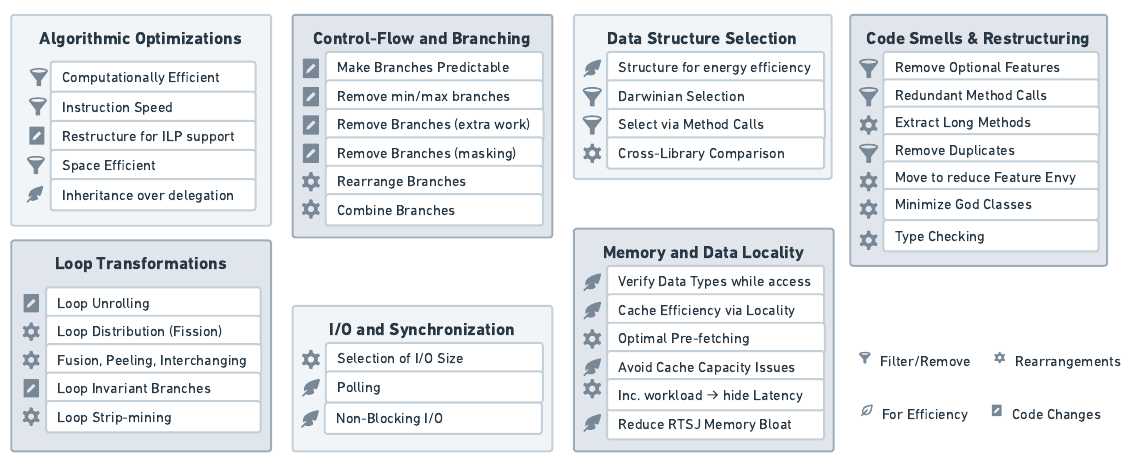}
    \caption{Catalog of optimization patterns grouped into seven themes.
    For readability, pattern names are shortened and in some cases merged to conserve space. 
    Symbols indicate four change types: 
    Filter/Remove code, Rearrange code, Energy-focused changes, and Rewrite/Modify code.}
    \label{tab:taxonomy-merged}
\end{figure*}

The result was a final catalog of 43 unique optimization patterns.
Imitating software patterns~\cite{gangof4}, each catalog entry includes attributes such as the pattern name, intent, applicability, the primary affected performance metric(s), and a worked example.
This catalog represents expert knowledge in performance optimization and can be consumed by an LLM as contextual knowledge for optimization tasks.
These patterns are organized into seven categories that reflect common optimization goals and strategies: 
\textit{Algorithm-Level Optimizations}, 
\textit{Control-Flow and Branching Optimizations}, 
\textit{Memory and Data Locality Optimizations}, 
\textit{Loop Transformations}, 
\textit{I/O}, 
\textit{Data Structure Selection and Adaptation}, and 
\textit{Code Smells and Structural Simplification}.
% The categories are illustrated in~\cref{tab:taxonomy-merged}, with two examples of patterns shown in~\cref{tab:pattern-merged}.
Four example patterns are shown in~\cref{tab:pattern-merged}, and the categories are summarized in~\cref{tab:taxonomy-merged}.
The full catalog is available in the artifact.

% \JD{The next paragraph reads a bit oddly/defensively. Can we just delete it? What value is it adding?}
% Our classification follows categories established in authoritative performance engineering texts and corroborated across multiple sources.
% Patterns are grouped by the nature of the optimization technique rather than by specific performance outcomes, as many optimizations affect multiple metrics simultaneously.
% While loop constructs relate to control flow, \textit{Loop Transformations} (\eg unrolling, interchange, fusion) are a well-recognized, standalone category in compiler and performance optimization literature.
% They are especially important in scientific computing, where many algorithms, such as Jacobi SOR (\cref{fig:jacobi-dual}), rely on iterative solving.

\begin{table*}[ht]
\renewcommand{\arraystretch}{1.2}
    \centering
    \captionsetup{font=small,skip=5pt}
    \small
\caption{
Excerpt from the Performance Optimization Pattern Catalog, showing representative patterns along with detection methods, examples, and targeted performance metrics. 
% Examples do not contain code snippets to conserve space. 
Detection methods are summarized and partially reflected in the examples.
This excerpt highlights the diversity of optimization strategies and shows how each is linked to specific code examples and corresponding performance metrics in our catalog.
The full catalog is available in the artifact (\cref{sec:DataAvailability}).
}
\begin{tabularx}{\textwidth}{
>{\RaggedRight\arraybackslash}p{1.8cm}
>{\RaggedRight\arraybackslash}p{2.2cm}
>{\RaggedRight\arraybackslash}p{3.5cm}
>{\RaggedRight\arraybackslash}X
>{\RaggedRight\arraybackslash}p{1.5cm}}
    \toprule
    \textbf{Optimization} & \textbf{Pattern} & \textbf{Detection Method} & \textbf{Example} & \textbf{Metrics} \\
    \toprule
    \rowcolor{gray!20}
    Algorithmic & Instruction-Level Parallelism (ILP) & Inspect loops for loop-carried dependencies $\rightarrow$ identify bottlenecks where CPI $ > 1.0 $  $\rightarrow$ apply loop unrolling and independent accumulators & \textbf{Profiling Result:} Loop-carried dependency blocked ILP. \textbf{Analysis:} Each iteration depended on previous result. \textbf{Fix:} Unroll loop and use multiple accumulators to increase instruction throughput. & $ \uparrow $ ILP, $ \uparrow $ throughput, $ \downarrow $ CPI \\
    \rowcolor{white}
    Control-Flow and Branching & Remove Branches by Doing Extra Work & Examine code for unpredictable conditionals in tight loops $\rightarrow$ consider optimization when branch misprediction rate $ > 0.05 $ & \textbf{Profiling Result:} Branch misprediction rate $ > 0.05 $ due to unpredictable alpha values. \textbf{Analysis:} Conditional blending caused frequent mispredictions. \textbf{Fix:} Apply blending unconditionally, removing branch and reducing misprediction. & $ \downarrow $ branch mispredictions, $ \downarrow $ latency \\
    \rowcolor{gray!20}
    Loop Transformations & Remove Conditional by Loop Unrolling & Look for conditionals dependent on loop index $\rightarrow$ eliminate by restructuring loops $\rightarrow$ check applicability when loop trip count is small and predictable & \textbf{Profiling Result:} Loop contained predictable conditional on index. \textbf{Analysis:} Branch mispredictions introduced latency. \textbf{Fix:} Unroll loop by two iterations, removing conditional checks and improving ILP. & $ \downarrow $ branch mispredictions, $ \uparrow $ ILP, $ \uparrow $ throughput \\
    \rowcolor{white}
    Memory and Data Locality & Avoid Cache Capacity Issues & Use ``1st level cache misses retired event counter'' $ \rightarrow$ Identify the cache miss sites & \textbf{Profiling Result:} multiplyMatrix is the site leading to a lot of cache misses. \textbf{Analysis:} Modify algorithm to fit in block. \textbf{Fix:} Use tile based flow, to make the mem tile block = L1 cache size of target system & $ \downarrow $ L1 cache misses, $ \uparrow $ throughput \\
    \bottomrule
\end{tabularx}

\label{tab:pattern-merged}
\end{table*}

Our catalog is designed to be language-agnostic and to 
encompass patterns at the algorithmic, memory, control-flow, and I/O levels. 
These patterns are expressed independently of specific syntax or language constructs.
However, certain optimizations—such as those involving low-level memory access (\eg store forwarding avoidance) or hardware-aware transformations (\eg loop peeling)—may not be directly applicable in all languages due to differences in memory models, type systems, or runtime abstractions.
Additionally, some patterns synthesized from papers are specific to a single language like Java, reflecting its object lifecycle, memory management, or standard libraries (\eg RTSJ memory or collection framework choices).
While these patterns may not directly apply to other languages, they convey general principles that can inform analogous optimizations in different programming environments.

\section{SysLLMatic: Design and Implementation}
\label{sec:methodology}
Building on the optimization pattern catalog, this section describes how SysLLMatic operationalizes these patterns in an automated optimization pipeline.
We first present the system architecture in~\cref {sec:Design-overview} and key components in~\cref{sec:perf_hotspot,sec:formulation,sec:Instrumentation,sec:data-collection-refinement}, followed by implementation details for prompts, profiling, and correctness validation.

\subsection{Design Overview}
\label{sec:Design-overview}
\begin{figure*}[t]
\centering

\begin{minipage}{\textwidth}
    \centering
    \includegraphics[width=\textwidth, trim={1.2cm 18.5cm 1.2cm 4.5cm}, clip]{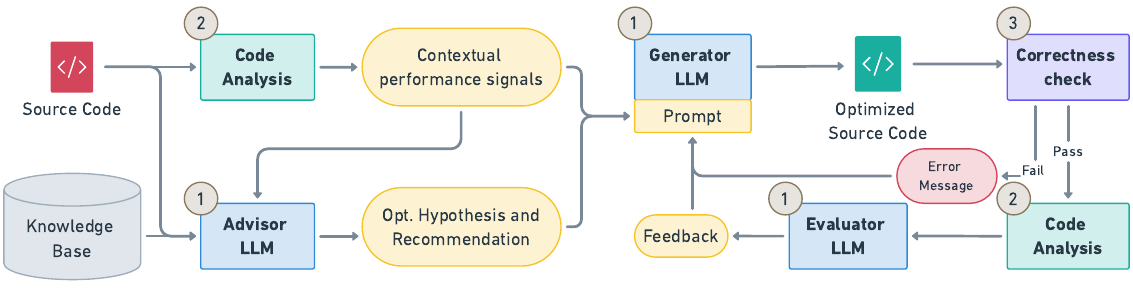}
    \caption{
    Overview of SysLLMatic, our automated LLM-based optimization framework (formalized in \cref{alg:sysllmatic}).
    SysLLMatic integrates domain-specific performance knowledge and follows an iterative Diagnosis Cycle (\cref{fig:cycle})to optimize performance. 
    Numbered circles denote major components. 
    Component \ding{172} encompasses three distinct LLM roles (Advisor, Generator, Evaluator). Component \ding{173} is Code Analysis, and Component \ding{174} is Correctness Check.
    }
    \label{fig:design-merged}
\end{minipage}

\vspace{0.5em}

\begin{minipage}{\textwidth}
% \captionsetup{type=algorithm}
\begin{algorithm}[H]
\caption{
The algorithm outlines the end-to-end process by which SysLLMatic transforms an input application $\mathcal{A}$ into an optimized version $\mathcal{A}^\star$, as illustrated in the system overview diagram (\cref{fig:design-merged}).
This algorithm approximates performance hotspot identification by profiling once and then applying iterative LLM-guided optimizations. While re-profiling after each optimization step could reveal newly emerging hotspots, the dominant hotspots remained largely stable in our benchmarks. We discuss this design choice and its limitations in \cref{sec:multi_objective_discussion}.
% \JD{I left a handwritten comment on this algorithm that said `If we profile to collect hotspots only once (line 2, line 4), isn't this potentially incorrect because the hotspots may change as we resolve the first several?' Seems like our algorithm should iteratively update the profile. Do we actually do that (and Figure 4 is simplified to avoid excessive loop, or it's contained within Code Analysis in the bottom right?), or we don't at all?}
% \JD{Why do we use the same variable name H in step 2 and also step 3? It's confusing to me, seems like we should have another name maybe? H-priority?}
}
\label{alg:sysllmatic}

\begin{algorithmic}[1]
\Require Application/source $\mathcal{A}$, pattern catalog $\mathcal{C}$,
         target metrics $\mathcal{M}$, hotspot budget $K$, iteration budget $T$,
         \textsc{Advisor}, \textsc{Generator}, \textsc{Evaluator}
\Ensure Optimized Application $\mathcal{A}^\star$
\State $Status \gets \Call{BuildAndRun}{\mathcal{A}}$ \Comment{Verify original application builds and runs}

\State $\mathcal{H} \gets \Call{Profile}{\mathcal{A}}$ \Comment{Collect performance hotspot}
\State $\mathcal{H}_{\text{priority}} \gets \Call{RankAndSelect}{\mathcal{H}, K}$

\For{\textbf{each} hotspot $h \in \mathcal{H}_{\text{priority}}$}
    \State $\textit{context} \gets \Call{CodeAnalysis}{h}$ \Comment{Static + dynamic context}
    % \State $\textit{sig} \gets \Call{PerfSignals}{h}$
    \State $(\mathsf{hyp}, \mathsf{pat}) \gets \Call{FormulateHypothesis}{\textsc{Advisor}, \textit{context}, \mathcal{C}}$ \Comment{Recommend patterns from catalog}
    \State $t \gets 1$

    \While{$t \le T$} \Comment{Up to $T$ attempts to optimize this hotspot}
        % \State $(\mathsf{hyp}, \mathsf{pat}) \gets \Call{FormulateHypothesis}{\textsc{Advisor}, \textit{context}, \mathcal{C}}$
        \State $\textit{prompt} \gets \Call{BuildPrompt}{\mathcal{A}, h, \textit{context}, \mathsf{hyp}, \mathsf{pat}}$
        \State $P \gets \Call{GenerateCandidate}{\textsc{Generator}, \textit{prompt}}$ \Comment{Candidate code}
        \State $(\mathcal{A}', \textit{Compile}) \gets \Call{ApplyPatch}{\mathcal{A}, P}$

        \If{\textbf{not} $\textit{Compile}$ \textbf{or not} \Call{CorrectnessCheck}{$\mathcal{A}'$}}
            \State $e \gets e + 1$
            \State \textbf{continue} \Comment{Discard failed candidate and fix error}
        \EndIf

        \State $\mathbf{m} \gets \Call{Measure}{\mathcal{A}', \mathcal{M}}$ \Comment{Latency, throughput, energy...}
        % \State $(\textit{context}) \gets \Call{UpdateContext}{P}$
        \State $\textit{fb} \gets \Call{EvaluateOptimization}{\textsc{Evaluator}, \mathbf{m}, P}$
        \State $\mathcal{A} \gets \mathcal{A}'$
        \State $t \gets t + 1$
    \EndWhile
\EndFor

\State $\mathcal{A}^\star \gets \mathcal{A}$
\State $Status \gets \Call{BuildAndRun}{\mathcal{A}^\star}$ \Comment{Verify final optimized application builds and runs}

\State $\mathbf{m}^\star \gets \Call{Measure}{\mathcal{A}^\star, \mathcal{M}}$ \Comment{Final updated performance data}

\end{algorithmic}
\end{algorithm}
\end{minipage}

\end{figure*}

% \JD{Is the relation in the next sentence correct? It sounds backwards to me.}
% Our system applies the Diagnosis Cycle (\cref{fig:cycle}) to guide LLM-based software optimization (\cref{fig:design-merged}).
Figure~\ref{fig:design-merged} presents the design of SysLLMatic. The system operationalizes the Diagnosis Cycle (Figure~\ref{fig:cycle}) to guide LLM-driven software optimization.
To ensure scalability to real-world software systems, we begin by identifying performance-critical hotspots and extracting the corresponding source code for optimization (\cref{sec:perf_hotspot}). 
The optimization workflow proceeds through the three stages of the Diagnosis Cycle.
In the hypothesis formulation stage (\cref{sec:formulation}), the Advisor component proposes optimizations based on high-level transformation patterns in~\cref{fig:patterncatalog-merged}. 
During instrumentation (\cref{sec:Instrumentation}), SysLLMatic collects structural and dynamic context by extracting AST and runtime traces using performance profiling. These artifacts inform the Generator's prompt construction. In the data collection and refinement stages (\cref{sec:data-collection-refinement}), the new code variants generated by the Generator are verified for correctness and re-profiled. 
Finally, the Evaluator analyzes the updated profiling data to assess improvements and guide further iterations.
\Cref{alg:sysllmatic} formalizes the end-to-end optimization procedure.
% \GKT{If I do not comment on these short changes, please assume that I approve of them and find them to be responsive to review feedback. Anything of a more critical nature will have a commment from me or "I approve".}
% \JD{Consider declaring figure 4 and algorithm 1 as minipages so that they appear together.}

\subsection{Performance Hotspot Identification}
\label{sec:perf_hotspot}
% \JD{These opening remarks feel like you are skipping ahead to the *implementation* instead of the *design*. Flame Graphs are not mandatory --- the concept of finding hot spots is the design approach you're taking. Talk about that first. Two paragraphs later you are talking about the design stuff/conceptual approach, after we've talked about the implementation. Feels mixed up.}

Large software systems contain thousands of methods and classes, but only a small fraction of them meaningfully impact runtime and energy consumption.
Optimizing arbitrary code is unlikely to yield measurable improvements. 
What is needed is a principled way to identify performance hotspots---the specific locations where execution disproportionately accumulates, and where optimization is likely to deliver the greatest end-to-end benefit.
% In practice, a performance hotspot is a function that receives a significantly larger share of CPU samples compared to others. 
In this work, we define a performance hotspot as a function that receives a significantly larger share of CPU samples compared to others.\footnote{Other definitions include functions dominated by I/O blocking, synchronization waits, or memory stalls.}
Identifying these hotspots requires two things: (1) a way to observe where execution time is spent, and (2) a strategy to aggregate these observations into a ranked list of candidate optimization targets. 
% The key design principle is to locate bottlenecks at the level of program execution.
Guided by the principle of identifying bottlenecks at the level of program execution, we assume that the target program is executed under a representative workload that covers performance-relevant code paths, either via associated test suites or benchmark-defined commands that execute the application with predefined inputs.
% \GKT{Good}

To operationalize this idea, we leverage sampling-based profiling to collect stack traces during program execution. 
Each trace pinpoints where the CPU was active at a particular moment.
We start by getting the collapsed text format of the Flame Graph (\cref{sec:rel-work-systems-performance}), where each line represents a stack trace. 
For example, \texttt{funcA;funcB;funcC\ N} represents a stack trace where \texttt{funcA} calls \texttt{funcB}, which in turn calls \texttt{funcC}, and this trace occurred \texttt{N} times.
The rightmost function (\texttt{funcC}) is the location where the CPU was active when the sample was taken, and aggregating these across the run yields a distribution of hotspots. 
Ranking functions by their aggregated counts gives us the Top-$K$ targets for optimization.

Formally, let $\mathcal{L}$ denote the set of profiling entries in the Flame Graph, where each element $(s, c)$ comprises a stack trace $s$ and a corresponding sample count $c \in \mathbb{N}$. 
Each stack trace $s = (f_1, f_2, \dots, f_k)$ is defined as an ordered sequence of function calls $f_i$.
% {\large
% \begin{center}
%     $ \displaystyle
%     \mathcal{L} = \{ (s,c) \mid s = \{ f_1, f_2, \dots, f_k \}, c\in\mathbb{N} \}
%     $
% \end{center}
% }
To focus the analysis on application code, we exclude system-level invocations from the stack trace. We designate $s_{\text{rightmost}} \in s$ as the rightmost element of the sequence. This function represents the application-level method that initiated the system-level call, corresponding to the location where the CPU sample was attributed.
% \GKT{Good. This is responsive to the reviewer comment.}
% Let $t$ represent a designated application-level method used to filter out system-level invocations.
% For any stack trace $s$ containing $t$, we define $f_{\text{rightmost}} \in s$ as the rightmost function call in the trace—this corresponds to the location where the CPU sample was attributed.
The function 
% $C(f_{\text{rightmost}})$ 
$C(f)$
denotes the cumulative sample count associated with
% $f_{\text{rightmost}}$
a specific function $f$
across all relevant stack traces:
% {\large
% \begin{center}
%     $ \displaystyle
%     \forall (s,c) \in \mathcal{L}:\exists f \in s \mid (t \in f) \implies C(f_{\text{rightmost}}) = \sum_{f_{\text{rightmost}} \in s_i} c_i
%     $
% \end{center}
% }

% {\large
% \begin{center}
% $C(f) = \sum_{(s,c) \in \mathcal{L} \atop s_{rightmost=f}} c$
% \end{center}
% }

    \[
    C(f) = \sum_{ (s,c) \in \mathcal{L} \,:\, s_{\text{rightmost}} = f } c
    \]

\begin{figure}[htbp]
  \centering
  % Left side: code listing
  \begin{minipage}[t]{0.48\textwidth}
    \vspace{0pt}
    \begin{lstlisting}[language=Java, basicstyle=\ttfamily\scriptsize, breaklines=true, columns=fullflexible]
[('org/biojava/nbio/core/sequence/template/SequenceMixin.toStringBuilder', 238),
  ('org/biojava/nbio/core/sequence/compound/AminoAcidCompoundSet.getCompoundForString', 128),
  ('org/biojava/nbio/aaproperties/PeptidePropertiesImpl.getMolecularWeight', 75),
  ('org/biojava/nbio/core/sequence/storage/ArrayListSequenceReader.setContents', 65),
  ('org/biojava/nbio/aaproperties/PeptidePropertiesImpl.getInstabilityIndex', 51),
...]
    \end{lstlisting}
  \end{minipage}\hfill
  % Right side: figure
  \begin{minipage}[t]{0.48\textwidth}
    \vspace{0pt}
    \includegraphics[width=\textwidth, trim={1cm 17cm 1cm 2.5cm}, clip]{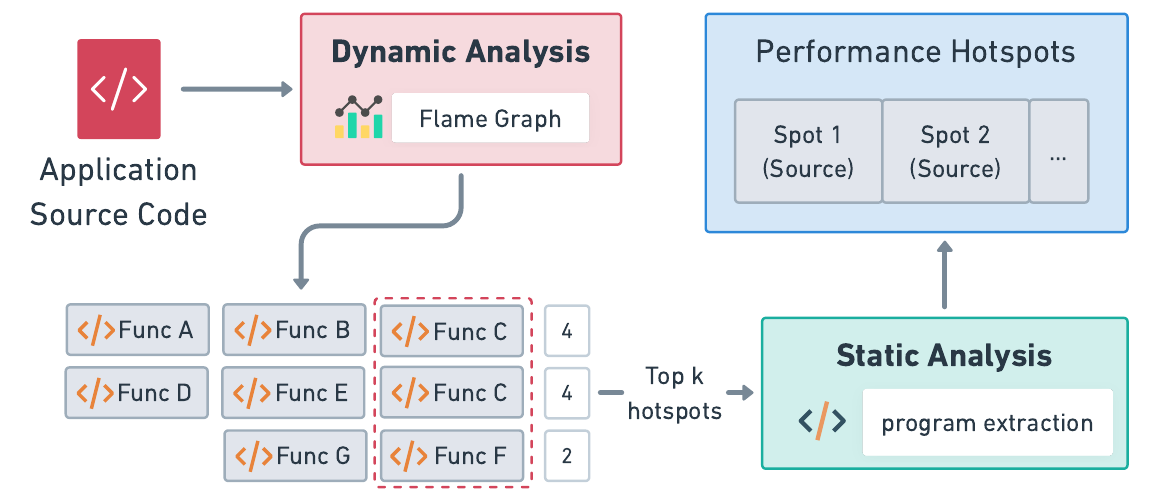}
  \end{minipage}
  \caption{Hotspot identification pipeline and example profiling output. 
  Dynamic analysis collects stack traces with Flame Graphs, which are aggregated to rank functions by execution frequency (left shows an excerpt from BioJava with counts). 
  The Top-$K$ hotspots are then extracted to form a candidate set of performance bottlenecks (\cref{sec:optimization_granularity}). 
  This process ensures that optimization focuses on code where performance cost is most concentrated.}
  \label{fig:hotspot}
\end{figure}

Finally, we get the set of $K$ functions 
% $f_{\text{rightmost}}$ 
$f$
with the highest aggregated sample counts 
% $C(f_{\text{rightmost}})$.
$C(f)$.
After identifying relevant components, we extract relevant code snippets as self-contained units for optimization.
An example performance profile generated after applying our performance hotspot identification function is shown in~\cref{fig:hotspot}. 
Each tuple contains the target function name and its corresponding sample count from the profiling output.

\subsubsection{\textbf{Optimization Unit Selection and Granularity}}
\label{sec:optimization_granularity}
The performance hotspot identification module is designed to operate across multiple levels of optimization granularity, enabling it to adapt to programs ranging from single-function tasks to full-scale software systems.
For \textit{single-function programs}, the entire function serves as the optimization unit; thus, we do not perform code extraction. The full function, along with its profiling trace, is provided directly to the LLMs.
In our benchmarks, these functions are small (17.52 lines of code on average, see~\cref{tab:line_diff_stats}), making it practical to treat the function as the optimization unit.\footnote{In principle, a single function may still contain complex constructs such as nested functions, lambdas, or closures, which could introduce additional optimization scope. However, such patterns were rarely observed in the programs studied.}
For \textit{class-based programs}, the framework supports two granularities. 
In the function-level configuration, the identified hotspot method is extracted and optimized independently as a self-contained unit. 
In the class-level configuration, the entire class enclosing the hotspot method(s) is treated as the optimization unit (\ie no code extraction), and the profiling trace highlighting the hotspot is included to guide optimization decisions.

For \textit{full-scale applications}, the process is the same as class-based programs, with hotspot extraction becoming mandatory due to repository size. 
Hotspot methods or classes are identified through full-application profiling and then isolated as optimization units before reintegration into the repository.
To ensure reliable validation and semantic preservation at scale, we additionally filter out hotspot classes that lack test coverage, restricting optimization to components that can be verified through automated test execution.

\subsection{Diagnosis Cycle --- Hypothesis Formulation}
\label{sec:formulation}

As illustrated in Figure~\ref{fig:design-merged}, the Diagnosis Cycle begins with the Advisor LLM generating concrete optimization hypotheses.
Its inputs are: (1) the target source code, (2) contextual performance signals from code analysis, including performance hotspots and AST representations, and (3) the full catalog of performance optimization patterns, provided as a structured textual knowledge base (\cref{sec:perf_optim_pattern}). 
Using this information, the Advisor identifies inefficiencies in the source code and, drawing on examples from the catalog, selects the patterns most applicable to those inefficiencies. 
From these, the Advisor selects the top strategies most relevant to the observed bottlenecks, prioritizing those likely to yield measurable performance gains. 
It formulates each chosen strategy as an actionable suggestion, consisting of a hypothesis and a recommendation that describes what to change and why the change may improve performance. 
These hypotheses connect observed inefficiencies to corresponding patterns in the catalog. 
The resulting ranked list of suggestions is then passed to the Generator to guide software optimization in the next stage of the cycle.

\subsection{Diagnosis Cycle --- Instrumentation}
\label{sec:Instrumentation}

In the Instrumentation stage, SysLLMatic collects \textit{structural context}, including the source code and its language-specific Abstract Syntax Tree (AST), as well as \textit{dynamic context} from runtime performance data captured as Flame Graphs (\cref{fig:hotspot}).
Both are incorporated into the Generator’s prompt, along with the relevant optimization patterns, to guide targeted code transformations.
The Generator uses Zero-shot Chain-of-Thought (CoT) prompting, instructing the LLM to reason about why a code region may be inefficient and how it could be optimized before producing modified code. 
The generated code will serve as candidate optimizations, which are then passed forward for correctness check and evaluation in the next stage of the cycle.

We adopt Zero-shot CoT over few-shot prompting to avoid reliance on handcrafted examples or expensive fine-tuning, in line with prior work such as RAPGen~\cite{garg2025rapgenapproachfixingcode}.
This approach enables generalization across domains and languages by grounding optimization decisions in broadly applicable performance patterns.
All prompts and LLM outputs are stored in the Generator's memory to support iterative refinement.

\subsection{Diagnosis Cycle --- Data Collection \& Hypothesis Refinement}
\label{sec:data-collection-refinement}
In the final stage of the Diagnosis Cycle, the Evaluator LLM analyzes the optimized code and generates natural language feedback to inform further optimization.
SysLLMatic first validates the code's correctness via benchmark tests.
On compilation or runtime failure, the error is returned to the Generator.
The process terminates if no correct version is produced within the predefined attempts.
If validation succeeds, SysLLMatic profiles the code to collect performance metrics (\eg latency) to further identify performance bottlenecks.
The Evaluator LLM then compares the original and optimized code to assess whether the applied transformations reduce the identified inefficiencies, guided by before-and-after performance data.
It identifies both improvements and any remaining or newly introduced bottlenecks. 
Based on this analysis, the Evaluator provides targeted feedback indicating which optimizations were effective, which had limited or adverse effects, and where further modifications may be necessary.
This feedback closes the loop by informing the Generator, enabling the formulation of refined hypotheses for the next optimization round.

\subsection{Prototype Implementation}
Our prototype implementation consists of approximately 3,000 lines of Python code and three prompts.
% \JD{Here, you should Expand this size breakdown by component type, and also indicate the total length of the prompts.}
We first specify the performance optimization objective implemented in the system.
We then describe the implementation details of each component, where the numbered circles 
(\eg \ding{172}, \ding{173}) correspond to the components shown in~\cref{fig:design-merged}.
Finally, we provide an end-to-end example to illustrate the complete optimization workflow in practice.

\subsubsection{Optimization Objective}
In our evaluation, SysLLMatic prompts the model to improve overall performance, rather than explicitly constraining optimization to a single metric. 
Hotspot identification is guided by CPU-based profiling (\cref{sec:perf_hotspot}), which focuses on computational hotspots. Although this focus on computation is consistent with prior work~\cite{gong2025}, the design biases improvements toward compute-bound regions of the program; we further discuss this effect and its limitations in \cref{sec:multi_objective_discussion}.
In addition, we provide the Advisor with the full optimization pattern catalog, rather than restricting it to patterns aligned with a specific objective. 
The framework itself is agnostic to performance metrics and optimization objectives. 
Alternative optimization goals (\eg memory consumption) can be supported by providing the corresponding performance signals (\eg heap profiling), selecting objective-aligned patterns from the catalog, and adapting the prompt to reflect the desired optimization target.

\subsubsection{Component \ding{172}: Prompt Details}
\label{sec:prompt}
To illustrate the design of our system, we include a shortened version of prompts for the \textit{Advisor}, \textit{Generator}, and \textit{Evaluator}. 
These excerpts preserve the key instructions and style of the full prompts while omitting implementation-specific details (\eg inputs for ASTs or flame graphs) for readability in the manuscript.

\myprompt{Advisor Prompt}{
Your goal is to identify optimization patterns from a catalog that can lead to significant performance improvements in the provided source code. 
Do not suggest patterns that provide only marginal benefits or are irrelevant to the code’s primary bottlenecks\dots

\textbf{Task Instructions:}
\begin{enumerate}
  \item Analyze the provided code: review the code to find any inefficiencies or areas that can be improved\dots
  \item Filter for high-impact patterns: using the optimization pattern catalog (descriptions, examples, detection methods), identify only the patterns directly applicable and likely to yield substantial gains\dots
  \item Rank patterns by expected impact: assign \texttt{rank=1} to the highest-impact pattern and justify the ranking\dots
\end{enumerate}

\textit{Output Requirements}: [Output requirements...]

\textit{Input}: [Input items...]
}

\myprompt{Generator Prompt}{
You are tasked with optimizing the following code to improve its performance \dots

\textbf{Task Instructions:}
\begin{enumerate}
  \item Analyze the code: examine the provided code in detail.
  \item Analyze the optimization patterns and choose the most effective optimization strategy.
  \item Implement the chosen optimization strategy: rewrite the code with the chosen optimization strategies\dots
\end{enumerate}

\textit{Output Requirements}: [Output requirements...]

\textit{Input}: [Input items...]
}

% \JD{So these prompts don't actually constrain the optimization metric of interest, it just says `performance' and `time and space complexity'. We do have an optimization catalogue --- does our catalogue indicate the expected kind of improvement? Do we have any ablations on the effect when we change the prompt to request `latency' or `memory' or `energy' or `cpu utilization'? I guess the method in Figure 4 is general and we are constrained by the specific data we are providing via code analysis (which is only CPU? unless we also get end-to-end data from the benchmark applications themselves that we are supplying?). I think we need to be (much) clearer about this somewhere.}
\myprompt{Evaluator Prompt}{
You are a code optimization expert. Analyze the code from a performance engineering perspective\dots

\textbf{Task Instructions:}
\begin{enumerate}
  \item Performance analysis: assess time and space complexity; identify bottlenecks such as expensive operations, nested loops, memory-heavy structures, or control-flow inefficiencies\dots
  \item Comparison to original code: highlight changes that contribute to performance gains or regressions\dots
  \item Improvement suggestions: provide concrete, actionable ideas (\eg data structures, algorithmic replacements, parallelism, memoization, I/O/memory/synchronization tuning), with examples or pseudocode where useful\dots
\end{enumerate}

\textit{Output Requirements}: [Output requirements...]

\textit{Input}: [Input items...]
}

\subsubsection{Component \ding{173}: Program Analysis}
\label{sec:impl-analysis}
% \JD{Hah! This is the place we need to be much more emphatic about `SysLLMatic is agnostic w.r.t. the specific program analysis. We use CPU flamegraphs in our evaluation, which will bias the result toward compute-specific metrics (latency, CPU utilization). But this is a choice of convenience rather than a fundamental limit of our approach; in principle, any analysis can be placed here and used as feedback for the LLM (eg prior work used X and Y citing papers from \$2). However, the LLM's general knowledge, our catalogue in particular, and our prompt (`time and space costs', cf. 5.6.1) are capable of improving other metrics optimizations, eg catalogue entry \#XX is BLAH BLAH which we expect would improve memory utilization rather than time elements. But we should also probably have some forward and backward references to this subsubsection to avoid any confusion.}
For the Java benchmarks, we generate ASTs using ANTLR4 with the \texttt{JavaLexer} and \texttt{JavaParser} modules and collect runtime performance data using \texttt{async-profiler} to produce flame graphs. 
We use the top-$K$ hotspot threshold with $K=50$ in our DaCapo evaluation, as it covers all or most meaningful hotspots after excluding very large files or trivial functions that appear high in the profile simply due to repeated calls in a loop.
For the HumanEval\_CPP benchmark, implemented in \Cpp, we extract ASTs using Clang’s AST dump and use Linux \texttt{perf} to generate corresponding flame graphs.

\subsubsection{Component \ding{174}: Code Correctness}
\label{sec:impl-correctness}
We consider the generated code to be correct if it successfully compiles and passes all test cases associated with a particular benchmark.
The number and adequacy of these tests differ across benchmarks. 
For instance, the HumanEval\_CPP benchmark includes an average of 9.6 tests per program, while SciMark2 evaluates accuracy by comparing output quality against a predefined threshold.

In the DaCapo benchmark suite, we rely on JUnit tests to verify program correctness, consistent with prior real-world software optimization studies that use unit tests as the correctness oracle~\cite{Garg2022}.
This means that SysLLMatic can modify the code as long as the optimized program passes the tests. 
To assess the adequacy of the test suite as a correctness oracle, we measure its coverage using the \texttt{JaCoCo} framework to compute coverage statistics for each application.
% To measure the extent of test coverage, we use the \texttt{JaCoCo} framework to compute coverage statistics for each application. 
As shown in \cref{tab:test-coverage-merged-fullwidth}, the coverage varies substantially across benchmarks, from less than 11\% in GraphChi to nearly 94\% in ZXing. 
To avoid modifying untested code, we conservatively skip classes without dedicated tests in the test suites.\footnote{A class is considered to have a dedicated test if it can be reached by any test, either directly or transitively, during test execution.
\cref{tab:test-coverage-touched} shows the improved test coverage of the code regions modified by SysLLMatic.}
Finally, to further ensure correctness, we also manually inspect all outputs to verify that changes are reasonable.

\begin{table*}[h!]
    \centering
    \captionsetup{type=table}
    \captionof{table}{
    JUnit test coverage of DaCapo applications, measured across instructions, branches, lines, and methods. Coverage varies widely across benchmarks, with GraphChi exhibiting very low coverage ($\sim$10\%) while ZXing achieves high coverage ($\sim$90\%). 
    On average across all applications, instruction coverage is 59.4\%, branch coverage is 49.7\%, line coverage is 57.6\%, and method coverage is 57.2\%.
    The variation shows the limitations of test-based validation in real-world software optimization and the importance of future work in this domain to consider more rigorous correctness validation, which we discuss in~\cref{sec:discussion_correctness}.
    }
    \label{tab:test-coverage-merged-fullwidth}
    \renewcommand{\arraystretch}{1}
    \scriptsize 
    \begin{tabularx}{\textwidth}{lRRRR}
        \toprule
        \textbf{Application} & \textbf{Instruction} & \textbf{Branch} & \textbf{Line} & \textbf{Method} \\
        \midrule
        BioJava   & 67.78\% & 58.42\% & 64.98\% & 59.98\% \\
        Fop       & 74.71\% & 56.62\% & 71.53\% & 72.97\% \\
        PMD       & 50.05\% & 43.24\% & 52.16\% & 55.55\% \\
        GraphChi  & 10.56\% &  9.25\% & 10.60\% &  8.50\% \\
        ZXing     & 94.02\% & 80.80\% & 88.74\% & 88.88\% \\
        \bottomrule
    \end{tabularx}
\end{table*}

\subsubsection{End-to-End Optimization Example}
We illustrate an example interaction in our optimization pipeline in~\cref{fig:advisor-generator-evaluator}. 
The pipeline begins with the \textbf{Advisor}, which produces optimization hypotheses; the \textbf{Generator}, which applies these hypotheses to transform the code; the \textbf{Evaluator}, which analyzes performance impact and provides feedback; and finally, the \textbf{Generator}, which refines the code based on evaluation feedback.

\begin{figure}[htbp]
\centering
\resizebox{\linewidth}{!}{%
\begin{tikzpicture}[
  font=\small,
  >={Stealth},
  stage/.style={
    rectangle split, rectangle split parts=2,
    rectangle split part fill={#1,gray!7},
    draw, rounded corners,
    align=left, inner sep=4pt,
    text width=0.28\linewidth % larger box to hold more details
  },
  line/.style={->, thick}
]

% Advisor
\node[stage=blue!8] (advisor)
{\textbf{Advisor Output}\nodepart{second}\ttfamily\scriptsize
Observation: Temporary objects cause code bloat and GC overhead.\newline
In \texttt{compute(...)}: new \texttt{ArrayList<Double>} and other objects created every call.\newline
Hypotheses:\newline
1) \emph{Optimize object use}: reuse lists/structures, reduce temp allocs.\newline
2) \emph{Caching}: memoize repeated property computations (\eg molecular weight).};

% Generator Iteration 1
\node[stage=green!8, right=8mm of advisor] (genone)
{\textbf{Generator}\nodepart{second}\ttfamily\scriptsize
Applied: reuse \texttt{ArrayList}; move \texttt{PeptidePropertiesImpl} creation out of loop;\newline
use \texttt{StringBuilder} for output rows; fix I/O bug (don’t close \texttt{System.out}).\newline
\newline
Excerpt:\newline
ArrayList<Double> dList = new ArrayList<>(40);\newline
StringBuilder rowBuilder = new StringBuilder(4096);\newline
IPeptideProperties pp = new PeptidePropertiesImpl();\newline
for (Entry<String, ProteinSequence> entry : a.entrySet()) \{ ... \}};

% Evaluator
\node[stage=yellow!20, right=8mm of genone] (eval)
{\textbf{Evaluator Feedback}\nodepart{second}\ttfamily\scriptsize
Findings: only micro-optimizations achieved.\newline
Loop still single-threaded, all sequences loaded into memory, output per sequence, no memoization.\newline
Recommendations: introduce parallelism (thread pool), stream input instead of bulk load, buffer writes, use primitive arrays, cache repeated sequences.};

% Generator Iteration 2
\node[stage=green!8, right=8mm of eval] (gentwo)
{\textbf{Generator}\nodepart{second}\ttfamily\scriptsize
Applied: streaming FASTA/GenBank I/O (no bulk storage), parallel per-sequence computation via thread pool, batched buffered writes, primitive \texttt{double[]} arrays to avoid boxing, preserved output order.\newline
\newline
Excerpt:\newline
ExecutorService pool = Executors.newFixedThreadPool(THREADS);\newline
double[] dList = new double[64];\newline
results.add(pool.submit(() -> \{ ... \}));};

% Arrows
\draw[line] (advisor) -- (genone);
\draw[line] (genone) -- (eval);
\draw[line] (eval) -- (gentwo);

\end{tikzpicture}%
}
\caption{End-to-end example of the iterative optimization workflow shown in~\cref{fig:design-merged}. 
The Advisor identifies inefficiencies (temporary object creation and redundant property computations). 
The Generator applies object reuse and I/O fixes, then the Evaluator highlights deeper bottlenecks in parallelism, memory use, and output handling. 
The Generator further implements improvements such as streaming input, thread-pool parallelism, buffered output, and primitive arrays. 
Each stage presents a summarized model output together with representative code excerpts, illustrating how the Advisor-Generator-Evaluator-Generator loop evolves optimizations.}

\label{fig:advisor-generator-evaluator}
\end{figure}

\section{Experimental Setup}
\label{sec:experiment}
This section outlines the experimental methodology used to evaluate SysLLMatic, including the evaluation questions, benchmark suites, evaluation metrics, system configuration and baselines, ablation studies, and the experimental environment.

\subsection{Evaluation Questions}
We evaluate the effectiveness of our system in optimizing software performance through both quantitative and qualitative analyses.

\subsubsection{Qualitative Characterization of SysLLMatic’s Behavior}
% \JD{Characterization}

We begin with a manual, qualitative analysis to build confidence in our technique, examining how SysLLMatic behaves in practice before assessing performance improvements. 
To this end, we analyze the optimizations made by SysLLMatic along several dimensions:

\begin{enumerate}
    \item \textit{Types of optimizations.} We show example optimizations made by the system on SciMark2 and DaCapo benchmarks. In addition, we categorize the transformations according to our optimization catalog (\eg algorithmic optimizations, loop transformations) and summarize the frequency with which each category is applied in HumanEval.
    
    \item \textit{Line- and Function-Level Change Analysis.} We quantify the scope of modifications by measuring line-of-code (LOC) differences between original and optimized versions.
    For DaCapo, we further examine the distribution of changes across functions, demonstrating whether optimizations are concentrated in a single function or spread across multiple functions.
    We also compare results with and without catalog guidance.

    \item \textit{Hypothesis-recommendation alignment of Advisor.} To evaluate whether the \textit{Advisor}'s optimization recommendations are consistent with its own hypotheses, we randomly select 30 Advisors' outputs from the SciMark2 and DaCapo benchmarks. 
    For each case, we compare the hypothesized performance issue (\eg redundant computations) with the corresponding optimization recommendation.
    We also describe the selected optimization strategies with and without catalog guidance to illustrate the impact of the catalog.

    \item \textit{Correctness validation.} We perform a manual inspection of all optimized outputs, verifying semantic equivalence to the original implementations beyond the provided testcases.
\end{enumerate}

\subsubsection{Quantitative Analysis}
We measure the optimization effectiveness of our system using four evaluation questions:
{\begin{RQList}
\item[EQ1:] How effective is our overall LLM-based optimization approach in improving software performance compared to existing baselines?

\textit{Method:} We evaluate our system on HumanEval-CPP, SciMark2, and DaCapo benchmarks, measuring latency, throughput, CPU cycles, memory, and energy, comparing against compiler optimizations and prior LLM-based baselines. 
In addition, we evaluate four open-source models to broaden coverage and reproducibility.

\item[EQ2:] How well does SysLLMatic balance multiple, potentially competing performance objectives, and which performance metrics are most improved?

\textit{Method:} We analyze per-metric improvements to assess how SysLLMatic balances multiple objectives across benchmarks. 
In addition, we assess whether the optimizations make the code easier or harder to maintain using static analysis metrics (\eg cyclomatic complexity). 
This complements the performance view by revealing potential tradeoffs between performance gains and code maintainability.

\item[EQ3:] What is the contribution of each core component to SysLLMatic’s optimization effectiveness?

\textit{Method:} As detailed in~\cref{sec:ablation}, we conduct both component-wise and system-wide ablation studies across all benchmarks to evaluate each module’s contribution to optimization effectiveness.
We also test program context granularity (function vs.\ class level) on SciMark2 to inform design choices for large-scale DaCapo applications.

\item[EQ4:] Under what circumstances do the benefits of SysLLMatic outweigh its cost?

\textit{Method:} We conduct a break-even analysis, modeling time-to-payoff by comparing LLM inference costs against runtime and energy savings under different workload frequencies.

\end{RQList}}

\subsection{Benchmarks}
To evaluate our framework's effectiveness across languages, domains, and application scales, we use three benchmark suites:
  HumanEval\_CPP,
  SciMark2,
  and
  DaCapo.

\subsubsection{Micro-benchmarks}
These consist of isolated functions and small kernels, providing a controlled setting to test whether our framework can generate correct and effective optimizations before scaling to larger systems.

\begin{itemize}
    \item \textit{HumanEval\_CPP}: HumanEval-X~\cite{zheng2023codegeex} extends HumanEval~\cite{chen2021evaluatinglargelanguagemodels} to multiple languages and includes 164 function-level tasks. 
    We use the \Cpp version due to this language's relevance in performance-critical systems~\cite{Edwards2013}. 
    To assess runtime efficiency, we augment it with translated stress tests from the COFFE benchmark~\cite{peng2025coffecodeefficiencybenchmark}, originally designed for the Python version of HumanEval.
    \item \textit{SciMark2}~\cite{scimark2}: A benchmark suite of five Java scientific computing kernels—FFT, LU, Monte Carlo, SOR, and SparseCompRow—targeting computation-intensive numerical workloads. Each kernel is implemented as a standalone Java file (26-303 lines).
\end{itemize}

\subsubsection{Full-scale Benchmark}
This involves large, real-world applications that test the scalability and practicality of our framework in complex software systems with diverse workloads and challenging code structures.
We use \textit{DaCapo-23.11-Chopin (2024 edition)}~\cite{Blackburn2025}, a collection of 22 multithreaded, real-world Java applications designed to benchmark full-scale systems (8k-400k lines of code). 
For this study, we selected a subset of 5 applications: BioJava, Fop, PMD, Graphchi, ZXing, as summarized in~\cref{tab:dacapo-apps-merged}. 
Out of the 22 DaCapo applications, we were able to successfully build and run 20, of which 8 provide both source code and test suites. 
We focus on the 5 that use Maven as the build system, which facilitated automation, while the remaining applications posed challenges such as custom build processes and complex dependencies~\cite{garg2025rapgenapproachfixingcode, Gong2024}. 
In addition, some files in DaCapo exceed 2,000 lines of code, where LLM optimizations often degrade due to lost context, leading to reduced performance and correctness.
Recent long-context evaluations show that despite recent LLMs such as GPT-4o and GPT-4.1 claiming 128K and 1M context lengths, their effective usable lengths are only about 8K and 16K tokens, respectively, beyond which accuracy drops sharply~\cite{modarressi2025nolima}.   
Thus, we conservatively skip files over 1,000 lines in this work.\footnote{'Skip' means that we exclude these files from the set of candidate optimization targets.} This filtering only affected \texttt{Fop}, where 11\% of files are excluded.

\begin{table}[h]
\centering
\renewcommand{\arraystretch}{1.1} 
\caption{Selected DaCapo applications used in our evaluation, covering diverse domains and workloads. The applications vary widely in size, from smaller systems like GraphChi ($\sim$8k LOC) to large codebases like Fop ($\sim$400k LOC).}
\label{tab:dacapo-apps-merged}
\begin{tabular*}{\textwidth}{@{\extracolsep{\fill}}lll} 
\toprule
\textbf{Application} & \textbf{Domain / Workload} & \textbf{Size (LOC)} \\
\midrule
BioJava   & Bioinformatics / Scientific Computing & $\sim$300k \\
Fop       & Document Processing (XSL-FO $\to$ PDF) & $\sim$400k \\
PMD       & Static Analysis (Java Source)         & $\sim$120k \\
Graphchi  & Graph Mining / Analytics              & $\sim$8k \\
ZXing     & Image Processing / Barcode Scanning   & $\sim$48k \\
\bottomrule
\end{tabular*}
\end{table}

\subsection{Metrics}
\label{sec:perf measurement}
% \JD{These metrics seem disconnected from the actual prompts we give the LLM, no?}
We use the performance metrics listed in \cref{tab:metrics-merged}. 
Our evaluation is structured around three categories of metrics: correctness, performance metrics, and comparative metrics.
Correctness refers to whether the optimized program preserves functional behavior by passing all provided test cases.
We capture multiple \textit{performance metrics}---including latency, memory usage, CPU cycles, throughput, and energy consumption---to provide a comprehensive view of resource utilization and system efficiency. 
This multi-dimensional perspective follows standard practice in performance engineering, where improvements in one dimension (\eg latency) can trade off against another (\eg memory usage)~\cite{Gregg_2021}.
In addition, we report \textit{comparative metrics}, following prior work~\cite{shypula2024learning}, to normalize and summarize performance gains across benchmarks.
Relative Improvement quantifies the ratio of original to optimized performance (\textit{before/after}), while Percentage Optimized indicates the proportion of cases where SysLLMatic achieves a meaningful ($\geq 10\%$) relative improvement.

We measure performance using the following implementations:
\begin{itemize}
\item \textit{Latency:} We measure latency as the time elapsed from the start of program execution to its completion for a given input size or workload, reported in milliseconds.
\item \textit{Memory Usage:} We measure memory usage using the Linux \texttt{/usr/bin/time} utility, which reports the peak Resident Set Size (RSS) in kilobytes.
\item \textit{CPU Cycles:} We measure the total number of CPU cycles consumed during execution using hardware performance counters via \texttt{perf stat -e cycles}.
\item \textit{Throughput:} We measure throughput as the amount of system work completed per unit time. In SciMark2, this metric is reported as MFLOPS (million-FLOPS). DaCapo reports application-level operations per unit time as defined by the benchmark harness (\eg number of files processed).
HumanEval does not report throughput, as it evaluates standalone program executions without a meaningful notion of operations per unit time.
% \GKT{Good to here. Breaking for dinner soon.}
\item \textit{Energy Consumption:} We measure energy consumption in Joules using the Intel Running Average Power Limit (RAPL), which accesses energy usage data from Model-Specific Registers (MSRs)~\cite{thorat2017energy, hahnel2012measuring, hackenberg2015energy}.
\end{itemize}

\begin{table}[H]
\captionsetup{font=small,skip=5pt}
\centering
\caption{We categorize the evaluation metrics into three groups: (1) Correctness, to ensure functional equivalence with the original code; (2) Performance metrics, covering latency, memory usage, CPU cycles, throughput, and energy consumption (\cref{sec:perf measurement}); and (3) Comparative metrics, including Relative Improvement (before/after ratio) and Percentage Optimized (percentage of correct programs with $\geq 10\%$ performance improvement), following prior work~\cite{shypula2024learning}.}
\label{tab:metrics-merged}
\renewcommand{\arraystretch}{1.2} 
\begin{tabularx}{\textwidth}{l c X}
\toprule
\textbf{Metric} & \textbf{Units} & \textbf{Description} \\
\midrule
\multicolumn{3}{l}{\underline{\textit{Correctness}}} \\
Code Correctness & Pass/Fail & Compiles and passes all associated tests \\
\midrule
\multicolumn{3}{l}{\underline{\textit{Performance Metrics}}} \\
Latency & milliseconds & Total execution time of the program \\
Memory Usage & kilobytes & Peak resident memory usage during execution \\
CPU Cycles & -- & Total CPU cycles consumed \\
Throughput & -- & Units of work completed per second \\
Energy Consumption & Joules & Total energy consumed, reported by RAPL interface \\
\midrule
\multicolumn{3}{l}{\underline{\textit{Comparative Metrics}}} \\
Relative Improvement & -- & Ratio of performance before vs. after optimization \\
Percentage Optimized & -- & Percentage of correct programs with $\geq 10\%$ performance improvement \\
\bottomrule
\end{tabularx}
\end{table}

\subsection{System Configuration and Baseline Selection}
We first describe the configuration of SysLLMatic and then present the baseline methods used for comparison.
\subsubsection{SysLLMatic Configuration}
We configure SysLLMatic with fixed parameters to ensure consistency across all benchmarks. 
The hotspot budget is set to $K=50$, meaning the top 50 profiled hotspots are selected for optimization. 
The iteration budget is set to $T=2$, so that the evaluator will provide one round of feedback to help the Generator refine the code.
For the LLMs, we fix the temperature at $0.7$ following prior work~\cite{garg2025rapgenapproachfixingcode, shypula2024learning, Gao2025}. 
We report results using \emph{pass@1}, meaning that for each evaluation the LLM produces one solution.
To reduce hardware-level measurement variance, we perform 2 warm-up runs followed by 5 measured runs for microbenchmarks, reporting the average of the measured runs~\cite{rahman2025marcomultiagentoptimizinghpc}. 
For DaCapo applications, prior studies show that the number of iterations required for JVM warm-up varies substantially across benchmarks and platforms, often requiring multiple warm-up iterations to reach steady-state performance~\cite{Georges2007, Kalibera2013}. 
Therefore, we use a more conservative configuration of 10 warm-up runs followed by 20 measured runs, reporting the average of the measured runs.

\subsubsection{Baseline Selection and Configuration}
We evaluate SysLLMatic against state-of-the-art representatives from the two categories of baselines discussed in~\cref{sec:background}.

\paragraph{Local Optimization via Compiler.}
While traditional techniques include a range of approaches (\eg compiler, program analysis, genetic programming), we focus on compiler-based baselines because they are actively maintained and readily applicable to our benchmarks.
For the \Cpp benchmark, we compile the code with \texttt{gcc -O3} optimization.
To extend beyond fixed compiler heuristics, we additionally include OpenTuner~\cite{Ansel2014} on HumanEval\_CPP, which explores a larger configuration space of compiler flags.
% \JD{Why are we citing the flags instead of the final word `optimizations'?}
For the Java benchmarks, we rely on the \texttt{JIT} compiler, configured with the following flags to enable common performance-oriented optimizations~\cite{Paleczny2001, OracleVMOptions}:
\begin{tcolorbox}[colback=gray!10, colframe=gray!40, boxrule=0.5pt, arc=2pt, left=2pt, right=2pt, top=2pt, bottom=2pt]
\texttt{-server -XX:+UseSuperWord -XX:+TieredCompilation -XX:TieredStopAtLevel=4}\\
\texttt{-XX:MaxInlineSize=100 -XX:FreqInlineSize=100 -Xms2g -Xmx2g}
\end{tcolorbox}

\paragraph{Source Code Optimization via LLM.}
For LLM-based optimizations, we evaluate representative prompt-based and agent-based LLM baselines that operate at the source-code level.

\textbf{Function-Level Code Optimization: PerfCodeGen~\cite{peng2024perfcodegenimprovingperformancellm}.}
For function-level optimizations, we compare against the SOTA training-free framework PerfCodeGen~\cite{peng2024perfcodegenimprovingperformancellm}, which also incorporates execution feedback into the optimization loop.
At the time of our experiments, PerfCodeGen represented the state of the art in training-free prompt-based code optimization, having outperformed all existing strategies on competitive programming benchmarks, so we exclude others in our study.
We apply PerfCodeGen to HumanEval\_CPP and SciMark2, which align with its focus on small code units. 
We reimplemented PerfCodeGen to support our benchmarks, which target programming languages different from those evaluated in the original paper, while preserving its core prompting strategy based on the publicly available prompts.
To ensure a fair comparison, we used an identical setup across approaches: the same input code, LLM (with the same temperature), execution environment and hardware, and evaluation metrics.
% Following their setup, we set $T=2$, which allows one refinement iteration guided by evaluator feedback.
% Unlike PerfCodeGen, our methodology does not involve code generation, \ie producing candidate implementations from natural language descriptions of the task. 
% Instead, we directly use the ground truth implementation, bypassing PerfCodeGen's code generation phase and feeding the same ground truth as the initial correctness solution input.
PerfCodeGen generates candidate implementations from natural language task descriptions prior to optimization.
In our evaluation, we adapt PerfCodeGen to our benchmarks by using the provided reference implementation as the initial program, bypassing its code generation stage to ensure consistent input programs across methods.

\textbf{Large-Scale Code Optimization: Codex~\cite{chen2021codex}.}
For DaCapo, existing function-level LLM-based techniques do not scale to the size and complexity of full applications, making them inapplicable in our setting. For example, approaches such as PrefCodeGen rely on selecting the fastest-running test cases to guide optimization, which becomes non-trivial to control in large, real-world systems with complex build and execution pipelines. We also exclude prior large-scale optimization approaches from~\cref{tab:related-works-llm-opt-merged} as baselines because they either target static metrics such as maintainability~\cite{Choi2024}, or are tailored to languages and frameworks not used in our benchmarks~\cite{garg2025rapgenapproachfixingcode}.

Although compiler optimizations are widely used and can be applied automatically to large-scale applications, they operate at the intermediate representation or binary level rather than at the source-code level. 
To enable comparison with a source-level optimization technique, we introduce a coding-agent-based baseline for large-scale applications: Codex~\cite{chen2021codex}.
While SysLLMatic was originally evaluated using \texttt{gpt-4.1}, that model lacks robust native tool-use capabilities (\eg autonomous patch application and iterative build/test execution), requiring manual orchestration. 
To enable a fully automated repository-level optimization setting, we instead adopt \texttt{gpt-5.2-codex} as the underlying LLM for our agent baseline. \texttt{gpt-5.2-codex} is a state-of-the-art coding LLM capable of repository exploration, in-place patch application, build and test execution, and iterative refinement via tool use.
Specifically, we configure the agent to operate autonomously over the entire repository following a structured multi-phase prompt inspired by SysLLMatic’s design, including repository orientation, bottleneck discovery, pattern-guided optimization, implementation, correctness validation, and refinement.
The agent is instructed to preserve functional equivalence, apply edits directly to the workspace, and validate changes via builds and test execution without user intervention. 
For each application, we provide an additional app-specific prompt describing the exact commands for building, running workloads, and executing tests to ensure consistent evaluation. 
We set the temperature to 0.7, following our main evaluation configuration, and configure the agent under a similar monetary budget (\cref{sec:time_and_monetary_cost}) to ensure a fair comparison.

\subsection{Ablation Study}
\label{sec:ablation}
We conduct ablation studies to understand the contribution of individual components and system-level factors.

\subsubsection{Component Ablation}
To assess each component’s contribution, we perform an ablation study, incrementally enabling components to measure their individual impact. Each subsequent configuration includes all components from the previous setting.
\begin{itemize}
    \item \textit{Base:} The Generator is prompted with only the raw source code and a natural language instruction
    \item \textit{+Evaluator (Base + Evaluator)}: Add the Evaluator, which introduces iterative refinement by providing performance improvement suggestions to the Generator. The Generator uses this information to revise its output in subsequent iterations (\textit{Iterative Refinement}).
    \item \textit{+Context (Base + Evaluator + Context)}: Add static/dynamic code analysis to provide the Generator with contextual performance signals (\eg hotspot methods, ASTs). This augments the feedback loop by grounding the Generator’s edits in actual program behavior (\textit{Instrumentation}).
    \item \textit{+Advisor (Base + Evaluator + Context + Advisor)}: Add the Advisor, which incorporates our catalog of 43 patterns to formulate optimization hypotheses from source code and profiling results, and then guides the Generator accordingly (\textit{hypothesis-driven optimization}).
\end{itemize}
We adopt an incremental strategy to measure the marginal benefit of each component as it is introduced into the optimization pipeline.

\subsubsection{System Ablation}
We evaluate two additional factors affecting system-wide performance. 
First, we examine the impact of varying the number of feedback iterations ($T$=1--4) on SciMark2.
We exclude HumanEval from this analysis because its programs are relatively simple: LLM with a single feedback loop already achieves high performance, and results stabilize once \textit{Context} is added, as shown in our ablation study.
Additional rounds, therefore, provide little benefit. For full DaCapo applications, where repeated optimization incurs high computational and inference costs on large-scale software, we instead ablate system performance by varying the number of selected performance hotspots.
Specifically, we evaluate sensitivity by adjusting the Top-$K$ threshold across three values: $K = 50, 100, 150$, where $K=50$ matches our main evaluation setting and larger $K$ values test whether expanding hotspot coverage yields additional gains.

\subsection{LLM Selection}
We use GPT-4o for micro-benchmark optimization, as it was the state-of-the-art proprietary model available during our experimentation period. 
For the full-scale DaCapo benchmark, we use GPT-4.1 due to its superior performance on the SWEBench and larger context window~\cite{openai2024gpt4.1}.
For all benchmarks, we additionally evaluate open-source LLMs of varying sizes, including state-of-the-art models that achieve competitive performance on SWEBench~\cite{jimenez2024swebench}. 
Specifically, we include qwen3-coder:480b, gemma3:27b, deepseek-r1:70b, and llama4:latest.
Following prior work~\cite{garg2025rapgenapproachfixingcode, shypula2024learning, Gao2025}, we set the LLM inference temperature to 0.7 to balance diversity and determinism in generation.

\subsection{Experiment Environment}
All LLM-optimized programs are executed on a dedicated bare-metal server (Intel Xeon W-2295, 36 CPUs, 188 GB RAM) to ensure a consistent evaluation environment.
Inference with open-source LLMs via \texttt{ollama} is performed on a GPU cluster, 
where each node has eight NVIDIA H100 GPUs (80~GB each), 112 CPU cores, and 1032~GB of memory.
After inference, results are transferred back to the bare-metal server for measurement.

\section{Results}
\label{sec:result}
We begin by characterizing SysLLMatic’s behavior through qualitative examples and quantitative analyses of the optimized code. 
We then address each of the 4 evaluation questions in turn, presenting results on performance improvements, tradeoffs, ablation studies, and break-even analysis.
Table~\ref{tab:results_roadmap} provides a structured roadmap of this section.

\begin{table}[t]
\centering
\caption{Roadmap of the Results section. 
The table summarizes the organization of the results, outlining each subsection, the analyses performed, and the key findings.}
\label{tab:results_roadmap}
\small
\begin{tabular}{p{0.08\linewidth} p{0.42\linewidth} p{0.42\linewidth}}
\toprule
\textbf{Section} & \textbf{Description and Subsections} & \textbf{Main Results Summary} \\
\midrule

\textbf{Qualitative Study} (\S\ref{sec:qualitative_result}) &
Characterize SysLLMatic’s optimization behavior:
\newline --- Optimization types and before/after examples
\newline --- Line- and function-level change analysis
\newline --- Optimizations with/without catalog
\newline --- Manual correctness inspection
&
Characterizing SysLLMatic’s optimization behavior, illustrating transformations, granularity, and the role of catalog-guided recommendations (\cref{fig:jacobi-dual,fig:biojava-dual,fig:optimization_pattern_merged,tab:line_diff_stats,fig:frac_all,tab:qualitative-alignment-merged,tab:test-coverage-touched}).
\\

\textbf{EQ1} (\S\ref{sec:EQ1}) &
Evaluate overall performance gains:
\newline --- Comparison with baselines
\newline --- Evaluation on open-source LLMs
&
SysLLMatic improves performance across benchmarks and surpasses LLM baselines on most metrics, while compiler optimizations remain competitive (\cref{tab:rq1,tab:humaneval-scimark-open,fig:baseline_dacapo_original}). \\

\textbf{EQ2} (\S\ref{sec:EQ2}) &
Analyze balance across competing objectives:
\newline --- Performance metrics
\newline --- Code maintainability analysis
&
Latency and throughput improve consistently, while CPU efficiency varies and exhibits trade-offs. Maintainability varies across applications, with CCN changes of -0.95\% to +21.05\% and function count increases of +1.47\% to +10.81\% (\cref{fig:radar_chart,tab:maintainability-metrics-merged}).\\

\textbf{EQ3} (\S\ref{sec:EQ3}) &
Assess importance of system components:
\newline --- Component ablation
\newline --- System ablation
\newline --- Sensitivity to program context
&
Ablation studies show that each module contributes to SysLLMatic’s performance, while varying Top-$K$ yields only minor gains (\cref{tab:ablation_study_metrics,fig:top_k variation,fig:feedback_iteration,fig:class_method}). \\

\textbf{EQ4} (\S\ref{sec:eq4_cost}) &
Determine practical viability:
\newline --- Time and monetary cost
\newline --- Resource consumption
&
Time and energy break-even occurs within days–weeks for high-throughput workloads but may extend to months or years for low-frequency applications; monetary cost remains negligible (\cref{tab:optimization-time,fig:granularity_and_breakeven}). \\

\bottomrule
\end{tabular}
\end{table}

\subsection{Characterizing SysLLMatic's Behavior}
\label{sec:qualitative_result}
% \JD{This subsection now includes quantitative data, so perhaps you should change this subsection title (and the signpost before it) to say something like `*characterizing* the system's behavior}
% \JD{This subsection is now long enough to merit signposting here.}

This section characterizes how SysLLMatic optimizes programs, combining illustrative qualitative examples with quantitative evidence. 
We begin by showing concrete before-and-after
 code snippets to demonstrate the optimizations SysLLMatic performs in practice. 
Next, we analyze the distribution of optimization patterns applied from the catalog, followed by a systematic examination of line- and function-level changes to quantify the scope and granularity of edits. 
Finally, we assess the alignment between Advisor-generated hypotheses and chosen optimizations, and compare optimization strategies generated by SysLLMatic with and without catalog guidance. 
Together, these analyses provide a holistic view of both the style and extent of changes introduced by SysLLMatic.

\subsubsection{Types of Optimizations Made by SysLLMatic}
\label{sec:optimizations-made}
To understand how SysLLMatic improves code, we examined the optimizations it applied across benchmarks. 
We illustrate these transformations with before-and-after code examples from SciMark2 and DaCapo applications, and analyze the optimization patterns from our performance catalog that are most frequently applied in HumanEval.

\paragraph{Before-and-After Examples}
First, we show examples (before-and-after) of code optimization in \cref{fig:jacobi-dual} and \cref{fig:biojava-dual}.
Successive Over-Relaxation (SOR), one of the kernels in the SciMark2 benchmark, is an iterative technique used to solve large systems of linear equations, often arising from discretized partial differential equations.
It improves upon the Jacobi method by introducing a relaxation factor to accelerate convergence.
\cref{fig:Jacobi_before} shows the original inner loop, where each value in the grid is updated based on its neighbors from the previous iteration.
\cref{fig:Jacobi_after} applies a loop unrolling strategy from our pattern catalog, which updates two elements per iteration to reduce loop overhead and increase instruction-level parallelism. This optimization can lead to better performance on modern CPUs~\cite{wiki_loop_unrolling}.

BioJava, part of the DaCapo benchmark suite, is an open-source framework for processing biological sequence data, widely used in bioinformatics applications.
The original version in~\cref{fig:biojava_before} performs sequential computation over a collection of protein sequences, invoking a method that likely involves significant processing time.
In~\cref{fig:biojava_after}, the code is optimized by adding a thread pool to parallelize the computation across available CPU cores~\cite{wiki_thread_pool}. 
This change exploits the parallel structure of the workload, where each task is independent.
This optimization improves performance by distributing tasks across threads rather than transforming the loop structure or any internal details of the underlying computation, although other strategies from the pattern catalog could lead to optimizations elsewhere in BioJava.

\begin{figure}[ht]
\centering

% --- First listing (a) ---
\begin{subfigure}{\linewidth}
\begin{lstlisting}[label={lst:jacobi-before}, numbers=none]
for (int j = 1; j < Nm1; j++) {
    Gi[j] = omega_over_four * (Gim1[j] + Gip1[j] + Gi[j - 1] + Gi[j + 1]) + one_minus_omega * Gi[j];
}
\end{lstlisting}
\caption{Jacobi SOR from SciMark --- Before}
\label{fig:Jacobi_before}
\end{subfigure}
\hfill
% --- Second listing (b) ---
\begin{subfigure}{\linewidth}
\begin{lstlisting}[label={lst:jacobi-after}, numbers=none]
int j;
for (j = 1; j < Nm1 - 1; j += 2) {
    double newValue1 = omega_over_four * (Gim1[j] + Gip1[j] + Gi[j - 1] + Gi[j + 1]) + one_minus_omega * Gi[j];
    double newValue2 = omega_over_four * (Gim1[j + 1] + Gip1[j + 1] + Gi[j] + Gi[j + 2]) + one_minus_omega * Gi[j + 1];
    Gi[j] = newValue1;
    Gi[j + 1] = newValue2;
}
if (j < Nm1) {
    Gi[j] = omega_over_four * (Gim1[j] + Gip1[j] + Gi[j - 1] + Gi[j + 1]) + one_minus_omega * Gi[j];
}
\end{lstlisting}
\caption{Jacobi SOR from SciMark --- After}
\label{fig:Jacobi_after}
\end{subfigure}

\caption{Comparison of Jacobi SOR code versions. The original implementation (a) applies the Successive Over-Relaxation (SOR) method in a straightforward loop. The optimized version generated by SysLLMatic (b) introduces loop unrolling, which exposes instruction-level parallelism and reduces loop overhead.}
\label{fig:jacobi-dual}
\end{figure}
\begin{figure}[ht]
\centering

\begin{subfigure}{\linewidth}
\begin{lstlisting}[label={lst:biojava-before}, numbers=none]
for (Entry<String, ProteinSequence> entry : a.entrySet()) {
    compute(output, entry.getValue().getOriginalHeader(), ...);
}
\end{lstlisting}
\caption{Protein  loop from BioJava --- Before}
\label{fig:biojava_before}
\end{subfigure}

\begin{subfigure}{\linewidth}
\begin{lstlisting}[label={lst:biojava-after}, numbers=none]
final int THREADS = Runtime.getRuntime().availableProcessors();
...
ExecutorService pool = Executors.newFixedThreadPool(THREADS);
List<Future<String>> results = new ArrayList<>(totalSequences);
for (Entry<String, ProteinSequence> entry : entries) {
    results.add(pool.submit(() -> {
        StringBuilder row = new StringBuilder();
        computeRow(row, ..., entry.getValue().getSequenceAsString().trim(), ...);
        return row.toString();
    }));
}
pool.shutdown();
\end{lstlisting}
\caption{Protein processing loop from BioJava --- After}
\label{fig:biojava_after}
\end{subfigure}

\caption{
Comparison of protein-processing loop implementations from BioJava.
The original version (a) executes sequentially, iterating over protein sequences in a single thread.
The optimized version generated by SysLLMatic (b) introduces thread-based concurrency using a fixed thread pool, enabling parallel computation of sequence rows.
Here, the LLM proposed setting \texttt{THREADS} to \texttt{Runtime.getRuntime().availableProcessors()}, which scales the degree of parallelism to the number of available CPU cores—an effective strategy for maximizing hardware utilization.  
% \JD{What value of THREADS did the LLM propose, and why? Please add this detail into the caption, possibly with commentary about whether that value was correctly chosen or is `reasonable'.}
% \GKT{Although this comment has been addressed, I would like 100\% confirmation that the LLM suggested this. What we had earlier suggests otherwise (i.e. the Java Runtime was not consulted).
% Please also note that more often than not, this number should be divided by 2 as many systems have hyperthreading, where the available processors may not exactly be the same as the number of real cores.
% At a minimum, can someone please confirm that availableProcessors() matches the number of cores for the hardware we are using for the benchmarks.}
% \GKT{For example, on my Mac M2, I seem to get the correct answer when firing up jshell:
% jshell> Runtime.getRuntime().availableProcessors()
% \$2 ==> 8}
}
\label{fig:biojava-dual}
\end{figure}

\begin{figure}[htbp]
    \centering
    \includegraphics[width=0.6\textwidth]{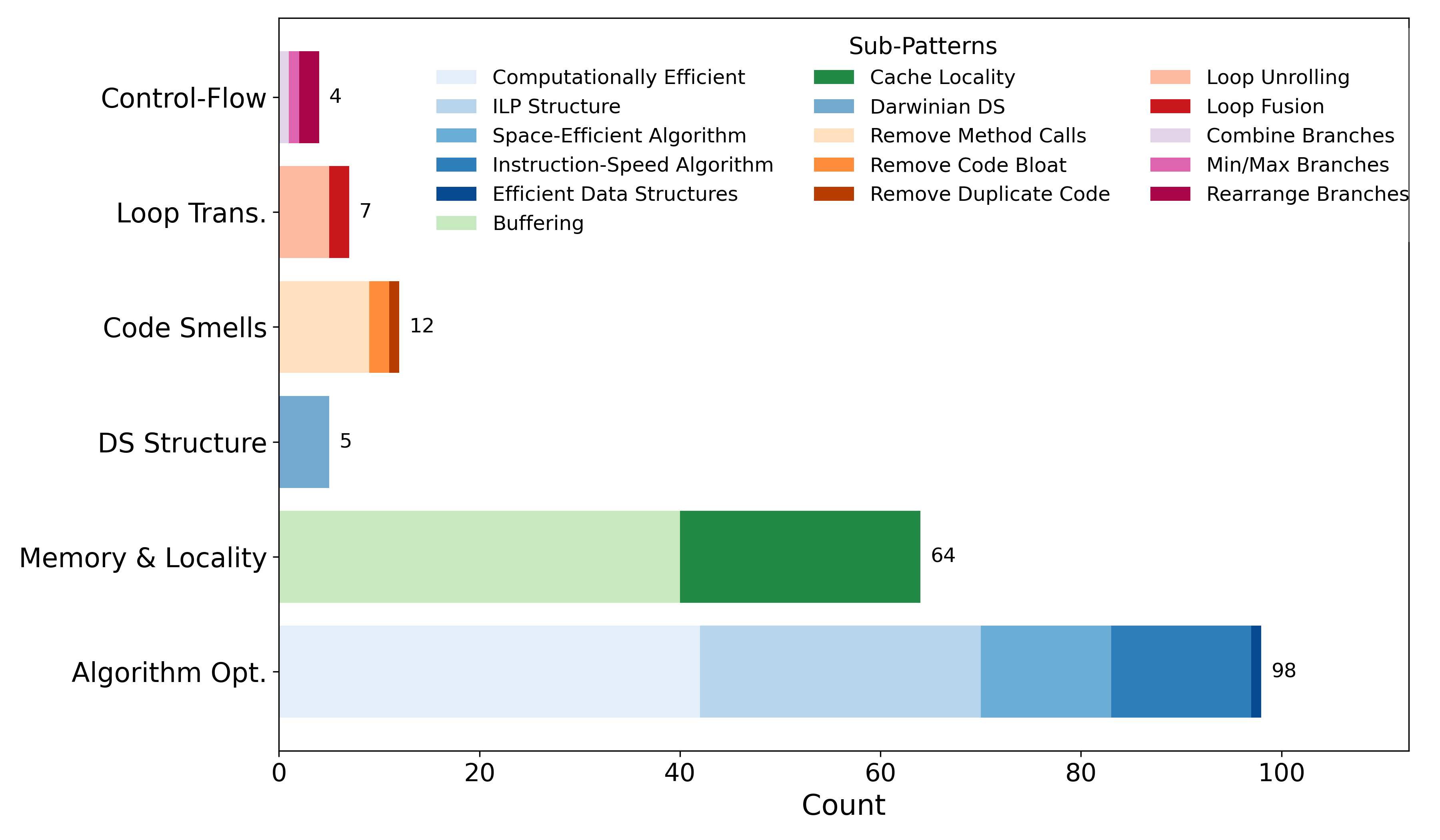}
    \caption{
    Distribution of optimizations applied to HumanEval\_CPP, categorized by our optimization catalog (\cref{sec:perf_optim_pattern}).
    Changes span 6 of the 7 themes.
    The I/O theme does not appear because HumanEval tasks contain no I/O operations, so SysLLMatic proposed no related optimizations.
    In total, 16/40 (40\%) patterns from these 6 themes were used. 
    }
    \label{fig:optimization_pattern_merged}
\end{figure}

\paragraph{Optimization Patterns}
Second, we analyzed SysLLMatic's optimizations on the HumanEval benchmark.
\cref{fig:optimization_pattern_merged} shows a stacked bar chart of the frequency of optimization patterns applied to each HumanEval program.
Algorithm-level optimizations dominated, appearing in 98 of 164 programs.
Common patterns including instruction speed (N=14), parallelism (N=28), and computational efficiency (N=42). 
Examples include using \texttt{OpenMP} for multithreading and applying techniques like two-pointers~\cite{algorithmic_technique_wikipedia} to reduce overhead.
Memory and locality patterns, such as buffering (N=40) and pre-allocating vectors (N=24), were also common.

\subsubsection{Line- and Function-Level Change Analysis}
\label{sec:line_and_func}

Understanding how SysLLMatic modifies programs requires analyzing both the extent of changes and their distribution across the codebase.
We examine line-level differences to capture the scale of edits, and function-level distributions to understand where and how optimizations are applied.
\textit{We do not observe a consistent quantitative trend across all studied applications, but we report the data here for completeness.}

% \JD{This section shows basically no trend, in either analysis, so I think we should be clear about that here. So add the following sentence. Keep the italics.``\textit{We do not observe a consistent quantitative trend across all studied applications, but we report the data here for completeness.}''}

\paragraph{Line of Code Difference After Optimization}
% \JD{This is with the catalog, right? Since we did an ablation in the previous section, and again in the next paragraph!, you should clarify here.}
At the line level, we measure how many lines were added, deleted, or modified across applications or programs, providing a direct view of the scope of SysLLMatic’s edits.
\cref{tab:line_diff_stats} reports line differences between the original and optimized programs for all benchmarks using the \texttt{cloc} tool.
In smaller benchmarks like HumanEval, modifications dominate.
In larger benchmarks, SysLLMatic makes substantial edits across the DaCapo applications, with the mix varying by application—some are addition-heavy, others deletion-heavy, and others primarily modification-heavy.
This suggests that SysLLMatic refines code in small programs but adopts more diverse editing strategies in large applications, ranging from targeted modifications to structural rewrites depending on the codebase characteristics.

\begin{table}[h]
\centering
\scriptsize
\caption{
Line-of-code (LOC) changes introduced by SysLLMatic across benchmarks, measured with \texttt{cloc}.
The columns indicate the size of the analyzed software and the number of lines that remained the same, or were modified, added, or deleted.
For HumanEval and SciMark2, values represent averages across programs, with “Avg. Total LOC’’ reporting the mean program size.
For DaCapo, ``Total LOC'' refers to the overall application size. 
The remaining columns capture the subset of lines that SysLLMatic selected for optimization. 
Within this processed subset, we report how many lines remained unchanged (Same), and how many were added, deleted, or modified.
In smaller benchmarks like HumanEval, modifications dominate, while in larger benchmarks the mix varies by application—some are heavy on additions, some on deletions, and others remain modification-heavy.
}
\label{tab:line_diff_stats}
\renewcommand{\arraystretch}{1.2}
\begin{tabular*}{\textwidth}{@{\extracolsep{\fill}}lrrrrr}
\toprule
\textbf{Benchmark} & \textbf{Avg. Total LOC} & \textbf{Avg. Same} & \textbf{Avg. Modified} & \textbf{Avg. Added} & \textbf{Avg. Deleted} \\
\midrule
HumanEval\_CPP & 17.52 & 3.30 & 11.66 (63.0\%) & 4.54 (24.5\%) & 2.32 (12.5\%) \\
SciMark2       & 132.20 & 75.50 & 9.00 (36.9\%) & 6.80 (27.9\%) & 8.60 (35.2\%) \\
\bottomrule
\end{tabular*}

\begin{tabular*}{\textwidth}{@{\extracolsep{\fill}}lrrrrrr}
\toprule
\textbf{DaCapo App} & \textbf{Total LOC} & \textbf{Total Same} & \textbf{Total Modified} & \textbf{Total Added} & \textbf{Total Deleted} \\
\midrule
BioJava   & 300k  & 133  & 1781 (89.1\%) & 210 (10.5\%) & 8 (0.4\%) \\
Fop       & 400k  & 3552 & 604 (33.7\%)  & 403 (22.5\%) & 784 (43.8\%) \\
PMD       & 120k  & 2164 & 198 (36.3\%)  & 127 (23.3\%) & 221 (40.5\%) \\
GraphChi  & 8k    & 224  & 47 (37.6\%)   & 61 (48.8\%)  & 17 (13.6\%) \\
ZXing     & 48k   & 1543 & 169 (33.8\%)  & 296 (59.2\%) & 35 (7.0\%) \\
\bottomrule
\end{tabular*}

\end{table}

\begin{figure*}[htbp]
    \centering
    \begin{subfigure}[t]{0.47\linewidth}
        \centering
        \includegraphics[width=\linewidth]{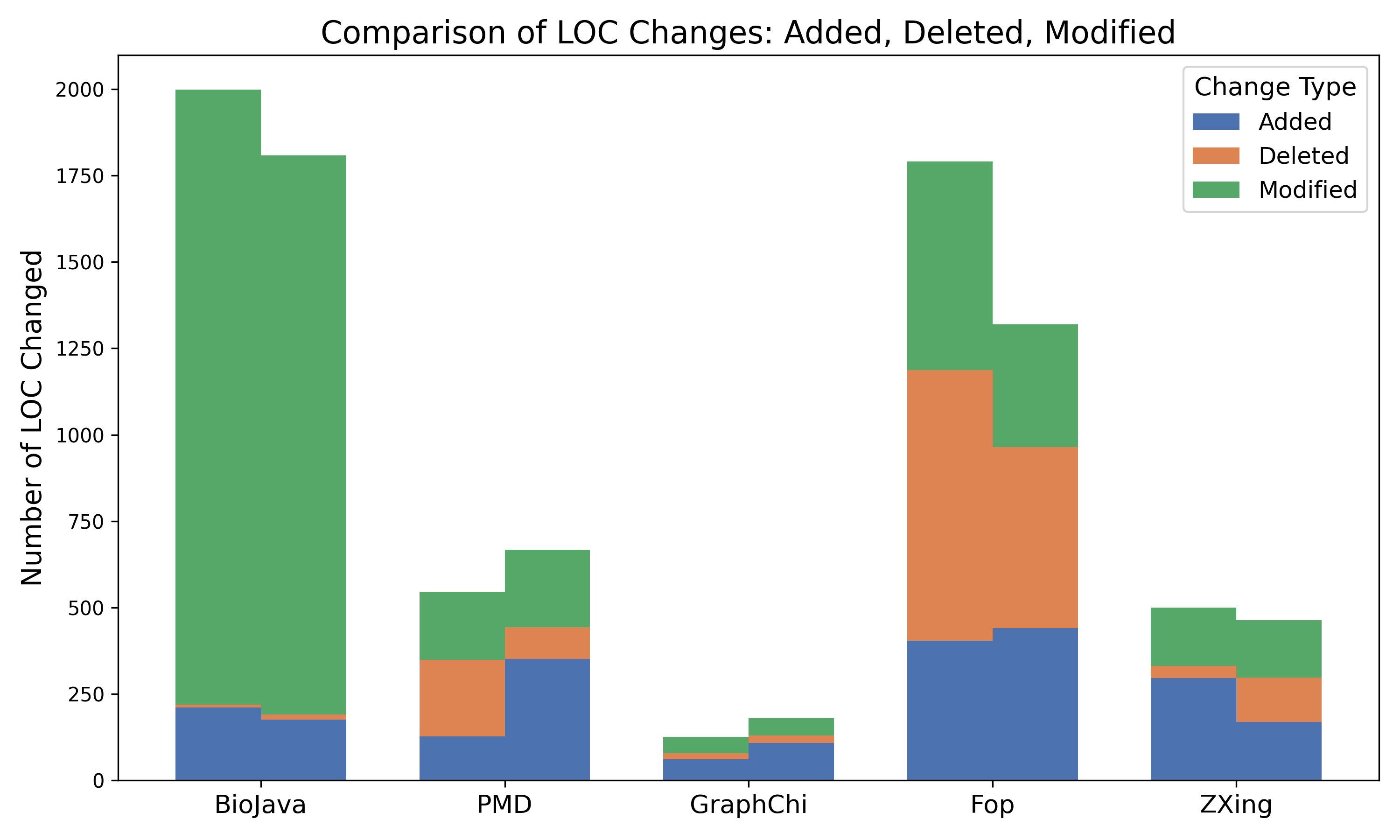}
        \caption{Line-level breakdown of added (blue), deleted (orange), and modified (green) code measured by \texttt{cloc}, with catalog (left bar) and without catalog (right bar). In general, using the catalog results in more overall changes, especially in larger applications such as BioJava and Fop. However, the composition of edits depends strongly on the characteristics of each application.}
        \label{fig:loc_change_compare}
    \end{subfigure}
    \hfill
    \begin{subfigure}[t]{0.47\linewidth}
        \centering
        \includegraphics[width=\linewidth]{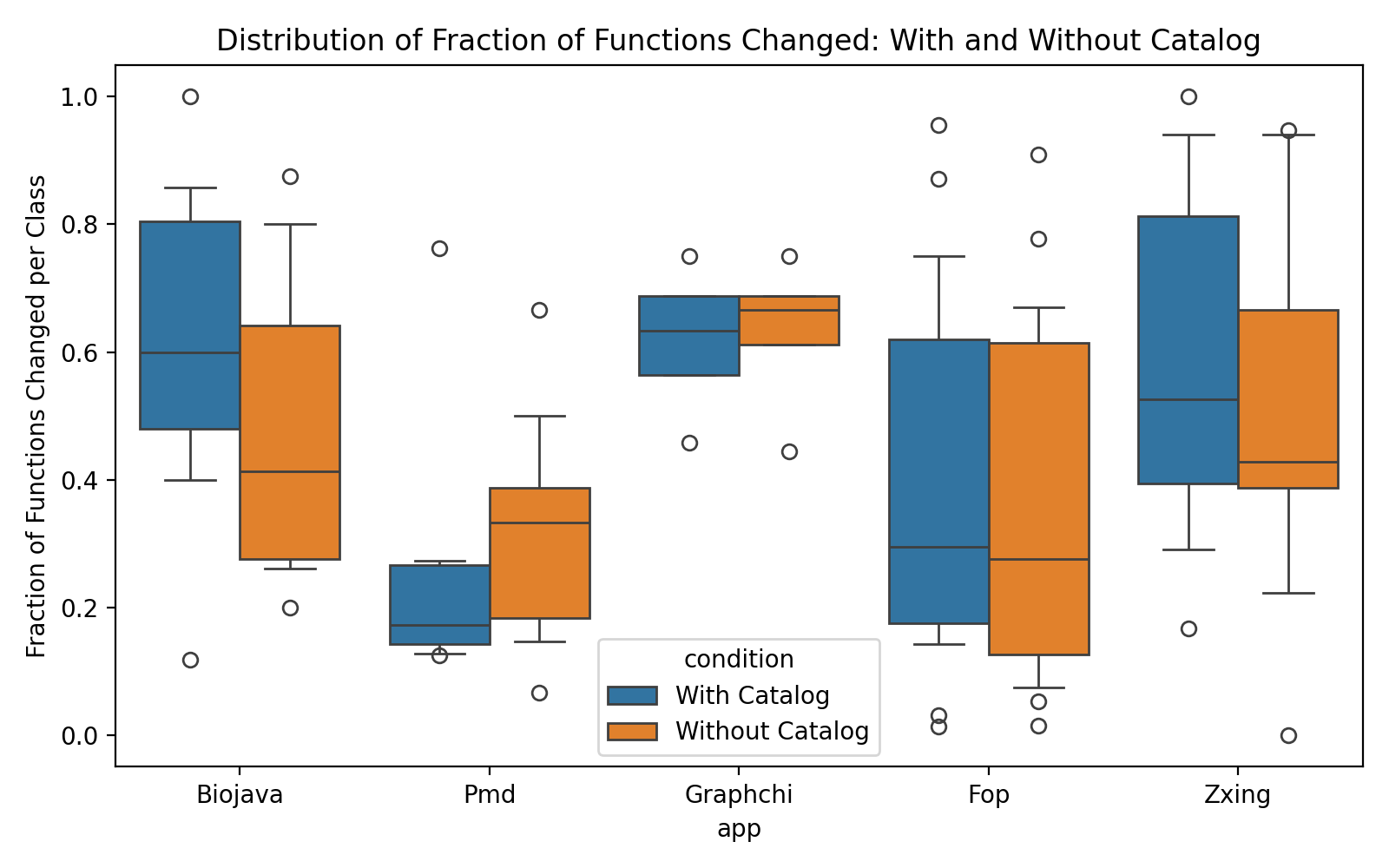}
        \caption{
        Fraction of functions changed per class (\(\text{FC}_{\text{union}}\)) across DaCapo applications, with catalog (left/blue) and without catalog (right/orange).
        Whiskers denote the 10th and 90th percentiles.
        Medians are higher with the catalog for BioJava, Fop, and ZXing, similar for GraphChi, and lower for PMD.}
        \label{fig:frac_1}
    \end{subfigure}
    \caption{Comparison of function- and line-level modifications with and without catalog guidance.
    Catalog use generally increases the total number of edits, particularly in larger applications such as BioJava and Fop.
    At the line level, the relative mix of modifications, additions, and deletions differs by application.
    At the function level, the breadth of changes also varies, showing that the effect of catalog guidance is application-dependent.}
    \label{fig:frac_all}
\end{figure*}

\paragraph{Line of Code Difference With and Without Catalog}
To examine the effect of the optimization pattern catalog on large real-world software systems, we analyze line-level changes in DaCapo applications, comparing results with and without catalog guidance in~\cref{fig:loc_change_compare}.
Adding the catalog produces a greater total volume of edits in the two larger systems, BioJava (300K LoC) and Fop (400K LoC), but we see greater variety in the smaller-size applications.
The balance of modifications, additions, and deletions is application-dependent, reflecting differences in code structure and optimization opportunities. 
% \JD{The line-level data does not tell us anything about targeted refinements vs. broader restructuring. Delete the next sentence or move it to the next part (but I do not think the next part supports the claim so I would delete it).}
% In practice, this means that catalog guidance does not enforce a uniform editing style: in some applications it amplifies targeted refinements, while in others it enables broader restructuring. 

\paragraph{Function Change Distribution With and Without Catalog}  
% \JD{Fill in the footnote with reference to a figure or subsection earlier in the paper. It's a big paper so the reader needs our help to avoid flipping around. Also remind the reader why we're talking about classes and functions, by telling them the *scope* that SysLLMatic is considering, \eg do we give the full class of a hotspot function to the LLM?}
Beyond raw line counts, we analyze how edits are distributed across functions within each class file on DaCapo, comparing optimizations performed with and without the pattern catalog.\footnote{SysLLMatic operates at the class level: when a hotspot function is selected, its full containing class is provided to the LLM, so edits may extend beyond the immediate hotspot function—a design decision we detail in \cref{sec:EQ3_context}.}
Normalizing changes at the function level reveals whether the catalog leads SysLLMatic to focus edits on a smaller subset of functions in each class or to spread modifications more broadly across the class.

\textit{Method:}
To quantify the extent of optimization-induced modifications, we first track the absolute total number of functions changed per application, comparing the averages with and without catalog guidance.
We then derive a normalized metric on a per-class basis: 
\begin{equation}
\text{FunctionChange}_{\text{union}} = \frac{\#\text{Functions}_{\text{added}} + \#\text{Functions}_{\text{deleted}} + \#\text{Functions}_{\text{modified}}}{\#|\text{Functions}_{\text{before}} \cup \text{Functions}_{\text{after}}|}
\end{equation}
where, for each class, the numerator counts the number of functions that were added, deleted, or modified by the optimization, and the denominator is the total number of distinct functions present across both the original and optimized versions of that class. 
We derive this metric using function signatures to classify functions as added, deleted, modified, or unchanged.
% \JD{How do you define this? Does cloc provide this metric directly for functions? Did you derive it? Did you use function *names* or function *contents*? (Tell us.}
This metric is bounded between 0 and 1, because every function must fall into one of four categories—added, deleted, modified, or unchanged—and the numerator only counts the first three, while the denominator counts all.
As a result, it provides a proportional view of how much of the program’s function set is touched by optimization.

\textit{Result:}  
At the function level, catalog guidance leads to more modifications overall, with an average of 93.4 functions changed per application compared to 73.6 without the catalog. When normalized by classes, however, the difference is much smaller.
\cref{fig:frac_1} plots distributions of \(\text{FunctionChange}_{\text{union}}\), the fraction of functions changed per class, across five DaCapo applications. BioJava and ZXing exhibit consistently higher values under catalog use, Fop and Graphchi remain similar across both conditions, and PMD shows the opposite trend with slightly higher values without the catalog, consistent with raw data where more lines of code were touched in~\cref{fig:loc_change_compare}. Overall, the catalog broadens the coverage of changes in 3 out of 5 applications, but the normalized averages are 48.9\% with the catalog and 45.8\% without, indicating no clear overall trend across the suite.

\subsubsection{Hypothesis-Recommendation by Advisor, and Chosen Optimization Strategies With/Without Catalog}
\label{sec:hypothesis-recoommendation-advisor}
We conducted a manual qualitative analysis to verify whether the Advisor’s hypotheses of performance bottlenecks align with the optimization recommendations.
From the evaluation set, we randomly sampled 30 cases across the SciMark2 and DaCapo benchmarks, covering approximately 50\% of the  SciMark2 and DaCapo evaluations.
In every case, the hypothesized performance issue and the corresponding recommendation were consistent, yielding a 100\% hypothesis-recommendation alignment rate. Table~\ref{tab:qualitative-alignment-merged} presents four examples across different optimization categories.

In addition, our previous analysis (\cref{sec:line_and_func}) showed that the impact of catalog guidance on program-level statistics—such as the nature and breadth of modifications—was mixed and application-dependent.
To better understand these differences, we examined example optimization strategies generated without catalog guidance and compared them to those produced with the catalog.
We observe that optimization suggestions generated without catalog guidance often diverge from catalog-driven recommendations, tending toward more ad-hoc and multi-dimensional changes. 
In contrast, catalog guidance narrows the focus to specific performance patterns, leading to more structured but narrower edits. 
This tradeoff is notable: with catalog, the optimizations are safer and more predictable but may leave out other costly operations (\eg parallelize the loop to distribute the workload), limiting overall gains. 
Without the catalog, recommendations can span multiple aspects and even include non-performance-related refactorings (\eg exception handling), which may yield broader improvements but at the cost of increased risk of errors. 
% Together, these findings highlight how catalog guidance provides structure and focus, while no-catalog runs explore a wider but less disciplined optimization space.
This qualitative distinction also mirrors the ablation study results in~\cref{tab:ablation_study_metrics}: for smaller applications, the broader no-catalog edits sometimes yield better improvements, whereas in larger applications, catalog guidance is essential to achieve consistent performance gains.
In addition, it reflects the line-of-code difference analysis above, where the result (\cref{fig:frac_all}) showed that with catalog guidance, SysLLMatic correctly optimized more of the codebase in larger benchmarks, aligning broader edits with meaningful performance benefits.

\subsubsection{Manual Inspection on Code Correctness}
\label{sec:manual-correctness}

\begin{table*}[h!]
    \centering
    \captionsetup{type=table}
    \captionof{table}{
    JUnit test coverage of \emph{touched (optimized)} code regions in DaCapo applications.
    Coverage is computed from JaCoCo reports by isolating coverage data for classes modified by SysLLMatic during optimization.
    This table quantifies the extent to which existing test suites exercise rewritten program regions.
    Across all benchmarks, coverage of touched code regions is substantially higher than overall application-level test coverage (\cref{tab:test-coverage-merged-fullwidth}), as the optimization process explicitly excludes classes that do not have dedicated test cases (\cref{sec:optimization_granularity}).
    For example, method coverage increases from 8.50\% to 67.65\% in GraphChi and from 88.88\% to 96.36\% in ZXing when considering only touched code regions.
    }
    \label{tab:test-coverage-touched}
    \renewcommand{\arraystretch}{1}
    \scriptsize
    \begin{tabularx}{\textwidth}{lRRRR}
        \toprule
        \textbf{Application} & \textbf{Instruction} & \textbf{Branch} & \textbf{Line} & \textbf{Method} \\
        \midrule
        BioJava   & 85.62\% & 76.20\% & 81.60\% & 85.12\% \\
        Fop       & 74.18\% & 68.29\% & 76.50\% & 83.56\% \\
        PMD       & 73.80\% & 66.48\% & 75.80\% & 75.30\% \\
        GraphChi  & 59.92\% & 26.00\% & 62.75\% & 67.65\% \\
        ZXing     & 95.85\% & 85.75\% & 94.72\% & 96.36\% \\
        \bottomrule
    \end{tabularx}
\end{table*}

A major threat to optimization is compromised correctness; for example, empty or hardcoded functions may execute quickly but fail to preserve program semantics.
We rely on existing test suites to validate functional correctness; however, such tests may not fully exercise rewritten code regions.
To quantify the adequacy of the test suite, we compute test coverage restricted to the classes modified by SysLLMatic using JaCoCo reports for DaCapo (\cref{tab:test-coverage-touched}), which measures the extent to which optimized code is exercised during testing. 
Despite relatively high coverage of touched code regions, automated tests are not sufficient to guarantee correctness.
To verify that performance gains were legitimate, we manually reviewed all outputs that passed automated tests.
We observed two cases where automated tests failed to catch incorrect optimizations.
In HumanEval, 1 out of 145 programs that passed the test cases had its core function removed.
No discrepancies were found in SciMark2.
In DaCapo, the optimized PMD application produced output that partially deviated from expected results: among 601 analyzed Java files, 2 files mismatched the ground truth.
While the overall functional behavior of the application was preserved, these mismatches suggest that SysLLMatic’s optimizations can occasionally alter subtle aspects of program logic or output formatting. 
Such deviations are rare but illustrate the importance of more rigorous correctness validation in large-scale benchmarks (\cref{sec:discussion_correctness}).

% This format is very careful not to put tables inline, which is annoying.
% clearpage helps.
% \clearpage

\begin{table}[!htbp]
\captionsetup{font=small,skip=5pt}
\centering
\newcolumntype{L}{>{\raggedright\arraybackslash}X}
\scriptsize
\caption{
Representative examples of optimization strategies with and without catalog guidance. 
Each row shows a benchmark, the hypothesis and recommendation generated by Advisor using the catalog, and the resulting optimization strategies. 
For comparison, we also include strategies generated without the catalog, which lack dedicated hypotheses and recommendations. 
The table illustrates how the catalog steers reasoning toward systematic optimization patterns, while the no-catalog setting often produces less targeted code modifications that increase the risk of introducing errors or unintended behavioral changes. 
This qualitative contrast also mirrors the performance results in Table~\ref{tab:ablation_study_metrics}: for smaller applications, the broader no-catalog edits sometimes yield larger improvements, whereas in larger applications, catalog guidance is essential for achieving consistent performance gains.
% \HP{this table has style issues and is hard-coded}
}
\label{tab:qualitative-alignment-merged}
\begin{tabularx}{\linewidth}{@{} p{1.5cm} | L p{2.5cm} L | L @{}}
\toprule
\textbf{Benchmark} & \textbf{Hypothesis from Catalog} & \textbf{Recommendation from Catalog} & \textbf{Selected Optimization Strategies (With Catalog)} & \textbf{Selected Optimization Strategies (Without Catalog)} \\
\midrule
MonteCarlo (SciMark) 
& The MonteCarloOptimized code performs a large number of iterations (over a billion). By restructuring the loop to reduce data dependencies, we could improve instruction-level parallelism and throughput, which is critical given the scale of iterations. 
& Algorithm-Level Optimizations: Structure Algorithm to Support Instruction-Level Parallelism (ILP). 
& A combination of algorithmic and loop-level optimizations.
1. Loop Unrolling with Multiple Accumulators: reduce the dependency between iterations and increase the throughput of the loop.
2. Optimization of RNG: Utilize a more efficient random number generator or alter the method to reduce the overhead of generating random numbers.
These changes will address the main performance bottlenecks identified in the flame graph and the analysis. 
& 1. Use ThreadLocalRandom for faster RNG: This will reduce the overhead.
2. Parallelize the loop with Java’s parallel streams to distribute the workload across multiple threads.
3. Avoid redundant calculations: Ensure that the calculation of x*x + y*y is performed only once per iteration.
4. Avoid creating a new Random instance: Use a ThreadLocalRandom instance to avoid the overhead of creating a new Random object every time integrate is called. \\
\midrule
LU (SciMark) 
& The \texttt{factor} method repeatedly applies row operations inside inner loops. These iterations are independent and suitable for loop unrolling, which reduces branch overhead and increases ILP. 
& Loop Transformations: Remove Conditional by Loop Unrolling. 
& A combination of loop unrolling and blocking for cache locality. Can address loop overhead and cache misses. By unrolling loops in the \texttt{factor} method, we reduce the number of iterations and improve instruction-level parallelism. Blocking the matrix operations will enhance cache usage, reducing latency due to cache misses. Additionally, reducing redundant copies by reusing data structures will help optimize memory usage. 
& The selected optimization strategy involves focusing on improving memory access patterns and reducing unnecessary operations in the matrix copying and LU factorization. Streamline and merge redundant methods, use \texttt{System.arraycopy} more effectively, and ensure loops are cache-optimized. Evaluate if parallel processing can be applied to matrix/vector initialization efficiently. \\
\midrule
ZXing - BitArray (DaCapo) 
& The \texttt{appendBitArray} method appends bits one by one via \texttt{appendBit}, leading to $O(n)$ overhead. Operating directly on word-level \texttt{int[]} blocks reduces complexity and memory traffic. 
& Algorithm-Level: Select Computationally Efficient Algorithms. 
& Replace per-bit appending in \texttt{appendBitArray} with word-level (\texttt{int[]}) copying, handling bits at word granularity where possible and falling back to per-bit appending only when necessary. 
& Optimize \texttt{appendBitArray} to copy full int words from the source, and only fall back to bitwise copy for remaining bits. Optimize \texttt{toBytes} to read bytes directly from the \texttt{int[]} array, avoiding per-bit \texttt{get} calls. Initialize \texttt{bits} to a minimal array in the no-arg constructor to avoid frequent resizing/copying. 
% These changes reduce latency, CPU cycles, and memory churn for large bit arrays.
\\
\midrule
FOP - PDFStream (DaCapo) 
& The code repeatedly creates temporary objects, notably in the \texttt{streamHashCode()} method (creates a \texttt{ByteArrayOutputStream} and a \texttt{byte[]} for every call), and in methods that may be called frequently (\eg \texttt{add(StringBuffer)} and dynamic resizing of \texttt{charBuffer}). Can reduce memory use and GC overhead, improving latency and throughput for large or frequent streams.
& Memory/Data Locality: Optimize Object Use 
& The most impactful optimization is to reduce temporary object creation in \texttt{streamHashCode()} by reusing the \texttt{MessageDigest} via \texttt{ThreadLocal}, and using a reusable buffer for encoding the MD5 result. Also, optimize the \texttt{add(StringBuffer)} method by using a reasonable default buffer size and minimizing resizes. 
& 1. Replace the manual \texttt{charBuffer} logic in \texttt{add(StringBuffer)} with a direct \texttt{sb.toString()} call for simplicity and efficiency.
2. Optimize \texttt{streamHashCode()} by using \texttt{DigestOutputStream} so we never materialize the full stream in memory.
3. Clean up exception handling in \texttt{add()} methods: propagate or wrap as unchecked, instead of swallowing.
4. Minor: Remove redundant buffer allocations and improve clarity of \texttt{setUp()}. \\

\bottomrule
\end{tabularx}
\end{table}

\subsection{EQ1: How effective is the SysLLMatic approach?}
\label{sec:EQ1}
We evaluate the effectiveness of SysLLMatic through two analyses: (1) comparison with baseline approaches and (2) evaluation using open-source LLMs to facilitate reproducibility.

\subsubsection{Comparison with Baselines}
\label{sec:comparision-with-baseline}
\cref{tab:rq1} presents the results of the baseline comparisons. 
On SciMark2, our system outperforms all baselines across all metrics, achieving higher relative improvements and a greater percentage of optimized programs (\%Opt), demonstrating effectiveness in optimizing numerical scientific kernels. 
On HumanEval, our system achieves notable gains over PerfCodeGen, with a 18.41\% improvement in correctness and 8.2\% more programs exhibiting latency improvements. Relative improvements in performance metrics are comparable between the two systems.
Interestingly, the compiler baseline, despite showing limited relative gains, achieves a higher \%Opt than our system in 3 out of 4 metrics on HumanEval. 
For instance, it yields improvements in CPU cycles for 31.12\% more programs than our method. 
This result indicates the continued strength of traditional compiler optimizations and emphasizes the importance of accounting for such techniques in LLM-driven software optimization workflows.

\begin{figure}[h]
    \centering
    \includegraphics[width=\textwidth]{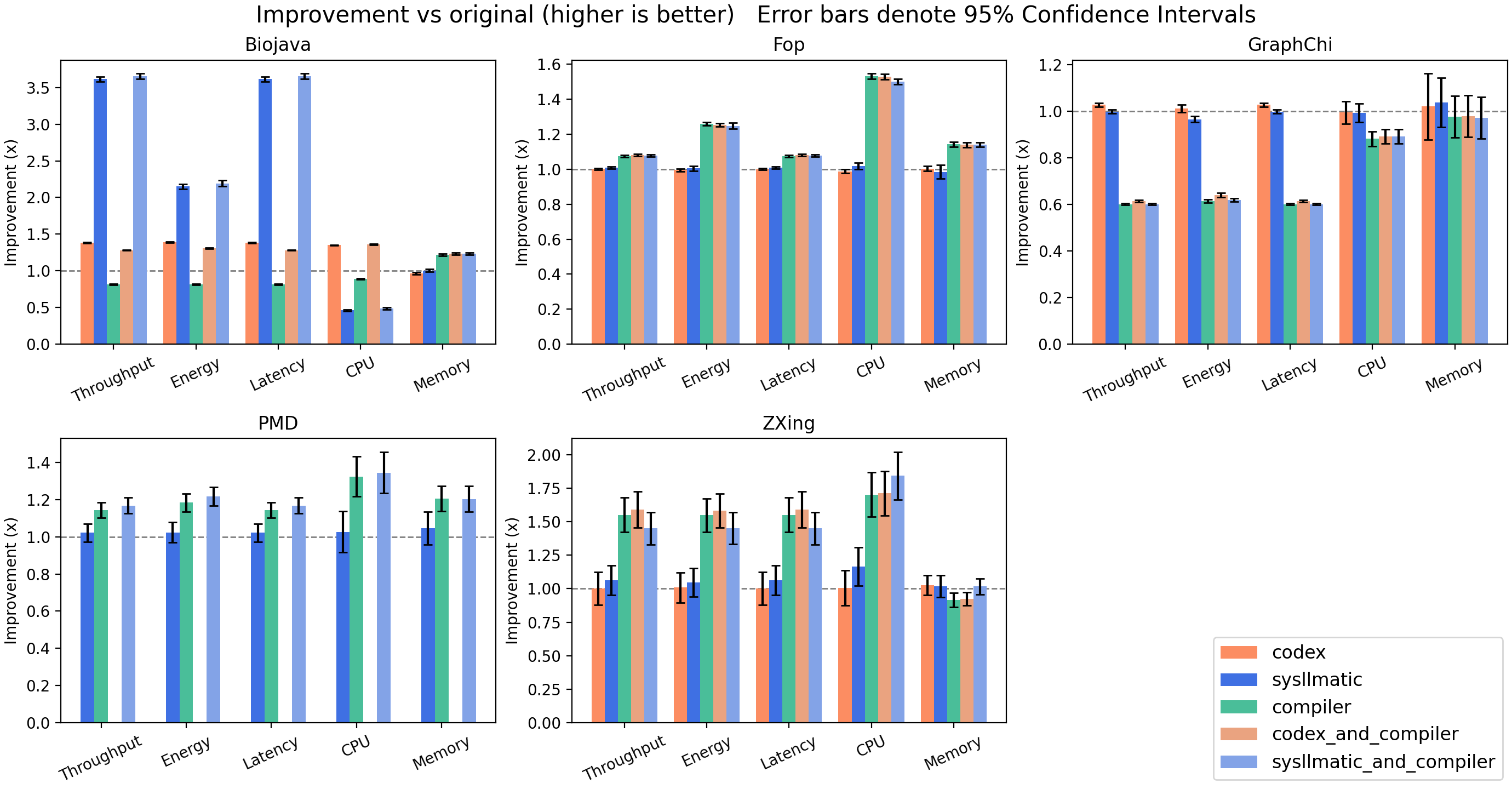}
    \caption{Relative performance improvements over the original baseline across five metrics for five DaCapo benchmarks. 
    Bars report mean improvement ratios (higher is better) computed after 10 warm-up iterations and 20 measured runs; error bars denote 95\% confidence intervals.
    Overall, Codex yields minor improvements across applications. SysLLMatic achieves substantial gains in BioJava and ZXing, with consistent improvements across most metrics. The compiler outperforms both approaches on Fop, PMD, and ZXing, delivering significant performance gains; however, it also incurs notable regressions (20-40\%) on BioJava and GraphChi, showing variability in compiler optimizations.
    Combining compiler optimizations with LLM-generated patches (Codex+compiler and SysLLMatic+compiler) yields mixed outcomes: in some cases, it further improves performance (Biojava, PMD, ZXing), while in others, it is unable to offset performance degradations introduced by the compiler (GraphChi).
    }
    \label{fig:baseline_dacapo_original}
\end{figure}

Next, we describe the effectiveness of SysLLMatic on five DaCapo applications: BioJava, Fop, PMD, ZXing, and Graphchi. 
We compare against both a Codex agent baseline (gpt-5.2-codex) and compiler optimizations.
We additionally combine compiler optimizations with LLM-generated patches.
\cref{tab:rq1} reports the average results across applications,
while \cref{fig:baseline_dacapo_original} presents the individual results for each of the five apps, showing relative improvements with 95\% confidence intervals computed from 10 warm-up runs followed by 20 measured runs to accommodate the high-variance nature of JVM execution.

\textbf{LLM-based approaches.}
The result shows that the Codex agent provides only modest performance gains over unoptimized code and has correctness issues on PMD. 
On average, SysLLMatic outperforms Codex, with lower latency (1.54$\times$ vs. 1.09$\times$) and energy (1.24$\times$ vs. 1.08$\times$).
BioJava exhibits significant improvements in latency (3.62$\times$) and energy (2.15$\times$), primarily due to the introduction of parallel processing. 
However, this also led to a significant degradation in CPU (0.46$\times$), reflecting increased concurrency overhead.
\textbf{Comparison with compiler optimizations.}
Compiler optimizations outperform both LLM-based approaches on several applications.
For ZXing, SysLLMatic achieves a 1.17$\times$ CPU improvement and consistent gains across other metrics; 
however, the compiler delivers stronger overall improvements, reaching up to 1.70$\times$ in CPU. 
Similarly, for Fop and PMD, compiler optimizations outperform both SysLLMatic and Codex across most metrics, while SysLLMatic provides marginal gains (up to 5\%) and Codex provides no improvement.
These results highlight that compiler optimization flags remain highly effective for real-world applications --- especially those with limited source-level optimization opportunities.
Moreover, compiler optimizations are deterministic and can apply effective transformations in a single optimization pass with negligible computational overhead.
However, they may also introduce severe performance regressions (\eg Graphchi, Biojava), whereas SysLLMatic consistently delivers performance improvements, except in cases where trade-offs occur among competing performance metrics (\eg Biojava).
We also evaluate configurations that combine compiler optimizations with LLM-generated patches. 
The interaction is mixed: in some cases, it provides additional improvements beyond the compiler, while in others, it fails to offset performance degradations introduced by compiler optimizations.

\begin{table}[ht]
\centering
\footnotesize
\caption{
Evaluation metrics for applications across three benchmarks after one feedback iteration, reported as performance ratios (higher is better).
Columns report correctness (Corr.), latency, memory usage, CPU cycles, throughput, and energy consumption. 
For each metric, \textit{gains} denotes the average improvement ratio relative to the original program, and \textit{\%opt} indicates the percentage of programs achieving $\geq$10\% improvement on that metric.
For DaCapo, correctness is reported as Correct\% / Output Match\%. Output Match is specific to DaCapo due to its high-workload nature.
Our approach consistently outperforms existing LLM-based baselines and surpasses compiler baselines on certain benchmarks and metrics, though the compiler remains highly competitive.
}
\label{tab:rq1}

%— First table: SciMark & HumanEval
\begin{tabular*}{\textwidth}{@{\extracolsep{\fill}} l l r
c@{}r
c@{}r
c@{}r
c@{}r
c@{}r @{}}
\toprule
\textbf{Benchmark} & \textbf{Opt.\ Approach} & \textbf{Corr. (\%)} &
\multicolumn{2}{c}{\textbf{Latency $ \uparrow $ }} & \multicolumn{2}{c}{\textbf{Memory $ \uparrow $}} &
\multicolumn{2}{c}{\textbf{CPU Cycles $ \uparrow $}} & \multicolumn{2}{c}{\textbf{Throughput $ \uparrow $}} &
\multicolumn{2}{c}{\textbf{Energy $ \uparrow $}} \\
\cmidrule(lr){4-5}\cmidrule(lr){6-7}\cmidrule(lr){8-9}\cmidrule(lr){10-11}\cmidrule(lr){12-13}
 & & & \textit{gains} & \textit{\%opt}
& \textit{gains} & \textit{\%opt}
& \textit{gains} & \textit{\%opt}
& \textit{gains} & \textit{\%opt}
& \textit{gains} & \textit{\%opt} \\
\midrule
\multirow{3}{*}{SciMark2}
 & PerfCodeGen & 100 & 1.00$\times$ & (0.0) & \textbf{1.00}$\times$ & (0.0) & 0.98$\times$ & (0.0) & 0.95$\times$ & (0.0) & 1.00$\times$ & (0.0) \\
 & Compiler & 100 & 0.89$\times$ & (0.0) & \textbf{1.00}$\times$ & (0.0) & 0.90$\times$ & (0.0) & 0.82$\times$ & (0.0) & 0.90$\times$ & (0.0) \\
 & SysLLMatic (Ours) & 100 & \textbf{1.55}$\times$ & (\textbf{40.0})& 0.98$\times$ & (0.0) & \textbf{1.52}$\times$ & (\textbf{40.0})& \textbf{1.39}$\times$ & (\textbf{40.0})& \textbf{1.51}$\times$ & (\textbf{40.0})\\
\cmidrule{2-13}
\multirow{4}{*}{HumanEval}
 & PerfCodeGen & 70 & \textbf{1.84}$\times$ & (15.0)& 1.00$\times$ & (0.0) & 9.31$\times$ & (21.0)& \multicolumn{2}{c}{--} & \textbf{2.07}$\times$ & (17.0)\\
 & Compiler & \textbf{100} & 1.36$\times$ & (\textbf{23.8}) & 1.00$\times$ & (0.0) & 11.60$\times$ & (\textbf{61.0})& \multicolumn{2}{c}{--} & 1.43$\times$ & (15.9)\\
 & Opentuner & \textbf{100} & 1.36$\times$ & (23.2)& 1.00$\times$ & (0.0) & \textbf{11.61}$\times$ & (57.3)& \multicolumn{2}{c}{--} & 1.47$\times$ & (\textbf{22.0})\\
 & SysLLMatic (Ours) & 88 & 1.69$\times$ & (23.2)& \textbf{1.01}$\times$ & (\textbf{1.2}) & 10.40$\times$ & (29.9)& \multicolumn{2}{c}{--} & 2.04$\times$ & (19.5)\\
 \cmidrule{2-13}
\multirow{3}{*}{DaCapo}
 & Codex Agent & 80/80 
 & 1.09$\times$ & (20.0) 
 & 1.00$\times$ & (0.0)
 & 1.06$\times$ & (20.0) 
 & 1.08$\times$ & (20.0) 
 & 1.08$\times$ & (20.0)  \\
 & Compiler & 100/100 
 & 1.03$\times$ & (40.0) 
 & \textbf{1.09$\times$} & (60.0) 
 & \textbf{1.26$\times$} & (60.0)  
 & 1.03$\times$ & (40.0) 
 & 1.08$\times$ & (60.0)\\
 & SysLLMatic (ours) & 100/80
 & \textbf{1.54$\times$} & (20.0)
 & 1.02$\times$ & (0.0)
 & 0.93$\times$ & (20.0) 
 & \textbf{1.54$\times$} & (20.0) 
 & \textbf{1.24$\times$} & (20.0) \\
\cmidrule{2-13}
\bottomrule
\end{tabular*}
\end{table}

% This format is very careful not to put tables inline, which is annoying.
% clearpage helps.
% \clearpage

\subsubsection{Evaluation on Open-source LLMs}
\label{sec:eq2-open-source}
We evaluate 4 open-source LLMs of varying sizes on three benchmarks: HumanEval\_CPP, SciMark2, and DaCapo 
(\cref{tab:humaneval-scimark-open}), revealing three key findings.
% We observe that open-source LLMs exhibit highly variable optimization performance.
% The results highlight three main findings. 

\begin{table}[ht]
\centering
\footnotesize
\begin{threeparttable}

\caption{
Results for open-source LLMs of different sizes on three benchmarks (HumanEval\_CPP, SciMark2, and DaCapo). 
Overall, larger models such as Qwen3-Coder:480B achieve stronger gains than smaller models across most metrics, achieving up to 1.38$\times$ on Biojava. 
While open-source models like Gemma and Qwen show strong performance on micro-benchmarks (HumanEval\_CPP, SciMark2), 
their advantages diminish on larger-scale benchmarks (DaCapo), where proprietary GPT models consistently perform better. 
We exclude Llama4 from DaCapo evaluation due to its poor performance on the micro-benchmarks.
}
\label{tab:humaneval-scimark-open}
\begin{tabular*}{\textwidth}{@{\extracolsep{\fill}} l @{\hskip 6pt} l @{\hskip 4pt} r @{\hskip 8pt}
c@{}r
c@{}r
c@{}r
c@{}r
c@{}r @{}}
\toprule
\textbf{Benchmark} & \textbf{Model} & \textbf{Correct (\%)}\tnote{1} &
\multicolumn{2}{c}{\textbf{Latency $ \uparrow $}} &
\multicolumn{2}{c}{\textbf{Memory $ \uparrow $}} &
\multicolumn{2}{c}{\textbf{CPU Cycles $ \uparrow $}} &
\multicolumn{2}{c}{\textbf{Throughput $ \uparrow $}} &
\multicolumn{2}{c}{\textbf{Energy $ \uparrow $}} \\
\cmidrule(lr){4-5}\cmidrule(lr){6-7}\cmidrule(lr){8-9}\cmidrule(lr){10-11}\cmidrule(lr){12-13}
 &  &  & \textit{gains} & \textit{\%opt}
  & \textit{gains} & \textit{\%opt}
  & \textit{gains} & \textit{\%opt}
  & \textit{gains} & \textit{\%opt}
  & \textit{gains} & \textit{\%opt} \\
\midrule
\multirow{4}{*}{HumanEval}
 & DeepSeek-r1:70b & 77 & 1.56$\times$ & (14.0) & 1.01$\times$ & (0.6)  & 6.64$\times$ & (26.2) & \multicolumn{2}{c}{--} & 1.76$\times$ & (15.9) \\
 & Gemma3:27b      & 88 & 1.61$\times$ & (15.2) & 1.00$\times$ & (0.0)  & 8.93$\times$ & (26.8) & \multicolumn{2}{c}{--} & 1.83$\times$ & (17.7) \\
 & Llama4:latest(17b) & 29 & 1.09$\times$ & (6.1)  & 1.00$\times$ & (0.0)  & 2.67$\times$ & (6.1)  & \multicolumn{2}{c}{--} & 1.10$\times$ & (7.3)  \\
 & Qwen3-Coder:480B & 94 & 1.93$\times$ & (14.6) & 1.01$\times$ & (1.8)  & 10.48$\times$ & (42.7) &   
 \multicolumn{2}{c}{--} & 2.03$\times$ & (15.2) \\
 & Gpt-4o & 88 & 1.69$\times$ & (23.2) & 1.01$\times$ & (1.2)  & 10.40$\times$ & (29.9) & \multicolumn{2}{c}{--} & 2.04$\times$ & (19.5) \\
\cmidrule(lr){2-13}
\multirow{4}{*}{SciMark2}
 & DeepSeek-r1:70b & 80 & 0.98$\times$ & (0.0)  & 1.00$\times$ & (0.0)  & 1.00$\times$ & (0.0)  & 1.00$\times$ & (0.0)  & 1.00$\times$ & (0.0)  \\
 & Gemma3:27b  & 100 & 0.99$\times$ & (0.0)  & 0.99$\times$ & (0.0)  & 1.00$\times$ & (0.0)  & 1.02$\times$ & (20.0) & 1.00$\times$ & (0.0)  \\
 & Llama4:latest(17b)   & 0  & 0.00$\times$ & (0.0)  & 0.00$\times$ & (0.0)  & 0.00$\times$ & (0.0)  & 0.00$\times$ & (0.0)  & 0.00$\times$ & (0.0)  \\
 & Qwen3-Coder:480B & 100 & 1.77$\times$ & (40.0) & 0.99$\times$ & (0.0)  & 1.84$\times$ & (40.0) & 1.43$\times$ & (20.0) & 1.66$\times$ & (60.0) \\
 & Gpt-4o          & 100   & 1.55$\times$ & (40.0) & 0.98$\times$ & (0.0)  & 1.52$\times$ & (40.0) & 1.39$\times$ & (40.0) & 1.51$\times$ & (40.0) \\
 \cmidrule(lr){2-13}
\multirow{4}{*}{DaCapo}
 & DeepSeek-r1:70b & 100   & 0.93$\times$ & (0.0)  & 1.01$\times$ & (0.0)  & 0.93$\times$ & (0.0)  & 0.93$\times$ & (0.0)  & 0.92$\times$ & (0.0)  \\
 & Gemma3:27b      & 60   & 0.98$\times$ & (0.0)  & 0.97$\times$ & (0.0)  & 0.97$\times$ & (0.0)  & 0.98$\times$ & (0.0)  & 0.98$\times$ & (0.0)  \\
 & Qwen3-Coder:480B & 100   & 1.05$\times$ & (20.0) & 1.01$\times$ & (0.0) & 1.03$\times$ & (20.0) & 1.05$\times$ & (20.0) & 1.02$\times$ & (20.0) \\
 & Gpt-4.1         & 100   & 1.54$\times$ & (20.0) & 1.02$\times$ & (0.0) & 0.93$\times$ & (20.0) & 1.54$\times$ & (20.0) & 1.24$\times$ & (20.0) \\
\bottomrule
\end{tabular*}
\begin{tablenotes}
\footnotesize
\item[1] For DaCapo, correctness is defined as the percentage of applications whose outputs are valid and included in the evaluation. 
All four models generated optimized code that compiled and passed application test cases. 
In terms of matching the ground truth outputs provided by the benchmark, DeepSeek and GPT-4.1 succeeded on 80\% of applications, Qwen on 60\%, and Gemma on 40\%. 
Two applications (Fop and Graphchi) failed at runtime under Gemma and were excluded.
\end{tablenotes}
\end{threeparttable}
\end{table}

First, larger models clearly outperform smaller ones across most metrics. 
Qwen3-Coder:480B stands out with 93.9\% correctness on HumanEval, along with the 
largest gains in latency (1.93$\times$) and CPU cycles (10.48$\times$). 
It also delivers strong improvements on SciMark2, where it achieves 100\% correctness 
and notable speedups (1.77$\times$ latency and 1.66$\times$ energy). 
By contrast, Llama4:17B performs poorly, reaching only 29.3\% correctness on HumanEval 
and yielding negligible gains, and 0\% in SciMark2. After manually examining the generated outputs, we found 
that Llama4 often fails to follow the required structured format: it mixes code with 
natural-language analysis, which results in numerous compiler errors even after automated 
cleanup and post-processing. 
Due to these issues, we exclude it from DaCapo evaluation.

Second, while mid-sized models perform well on micro-benchmarks, their performance does not carry over to larger software systems. 
For example, Gemma3:27B achieves 88.4\% correctness 
on HumanEval and 100\% on SciMark2, and DeepSeek-r1:70B reaches 76.8\% and 80\% respectively, both with noticeable performance gains on HumanEval. 
On DaCapo, improvements remain limited, with even Qwen3-Coder:480B achieving only small gains (\eg on average 1.05$\times$ latency 
and 1.02$\times$ energy). 
By contrast, proprietary GPT models are more effective on DaCapo, 
achieving an average of 1.54$\times$ latency and 1.24$\times$ energy improvements.
We also observe that open-source models, including Qwen, occasionally lead to performance degradations of around 8-30\% across metrics, whereas GPT models consistently deliver non-degrading or positive performance improvements.
This indicates that while open-source models are competitive on small-scale tasks, they remain less effective 
on complex, real-world applications.

Finally, in qualitative observations of the Advisor role, we find that proprietary GPT models and large open-source models (Qwen3-Coder:480B) demonstrate superior instruction following: they typically produce 2–3 concise, relevant optimization recommendations without requiring explicit prompting. 
Open-source models, by contrast, often deviate from the curated optimization pattern catalog. 
For example, some outputs included external references such as links to Wikipedia pages on generic software optimizations, while others produced more than 20 recommendations, the majority of which were irrelevant. 
This suggests that using the optimization catalog as a way to condition the optimization based on knowledge or details about the underlying system is more effective in larger models, which can reliably follow such guidance, whereas smaller models tend to generate less focused and more random suggestions. This observation is consistent with prior findings~\cite{wei2022emergent}.

\begin{tcolorbox}[colback=gray!5,colframe=gray!35!black,title={EQ1: How effective is the SysLLMatic approach?}]
\small
SysLLMatic is highly effective on micro-benchmarks, though compilers remain competitive in some cases. 
On DaCapo, SysLLMatic outperforms or matches LLM-based baselines across applications, although JIT compiler optimizations remain dominant in \texttt{Fop}, \texttt{PMD}, and \texttt{ZXing}. 
Open-source LLMs such as \texttt{Qwen3-Coder:480B} perform comparably to GPTs on micro-benchmarks but achieve only limited gains on DaCapo.
\end{tcolorbox}

% This format is very careful not to put tables inline, which is annoying.
% clearpage helps.
%\clearpage

\subsection{EQ2: How well does SysLLMatic balance multiple, potentially competing performance objectives, and which performance metrics are most improved?}
\label{sec:EQ2}
This section examines SysLLMatic’s ability to balance optimization across multiple performance dimensions and maintainability.
We explicitly articulate all target performance goals in the optimization prompts (see~\cref{sec:prompt}); however, SysLLMatic’s performance profiling relies on CPU-cycle sampling, and its feedback signal is derived from latency measurements.
No explicit objectives related to code quality or maintainability are specified in the prompts.

\subsubsection{Performance Metrics}
\label{sec:eq2-performance-metrics}
\begin{figure}[htbp]
    \centering
    \includegraphics[width=0.4\textwidth]{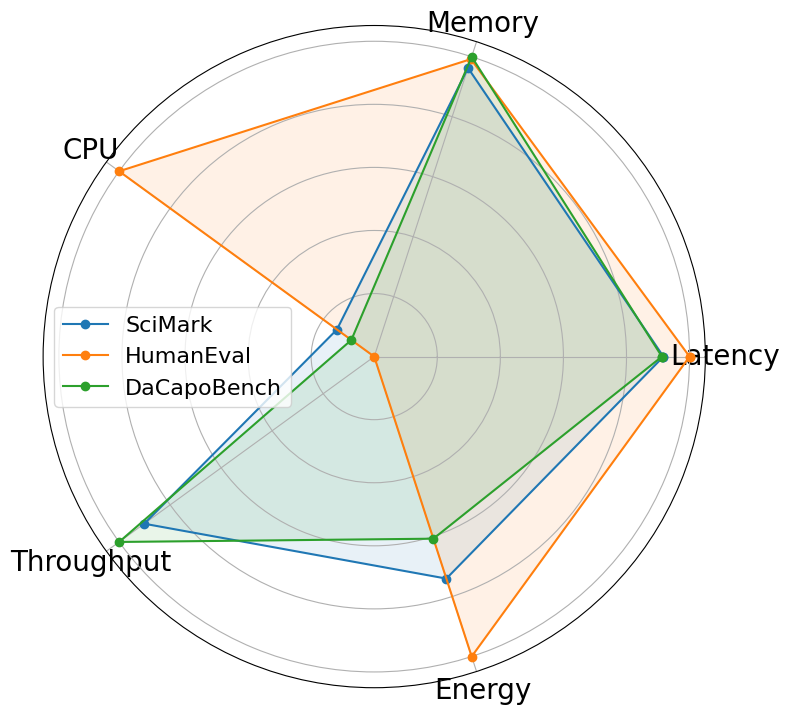}
    \caption{Normalized performance gains across five metrics on three benchmarks, based on values from~\cref{tab:rq1}. 
    Metrics are scaled by the maximum observed gain to enable comparison. The radar plot illustrates trade-offs in multi-objective optimization.
    SysLLMatic achieves complementary improvements across benchmarks rather than uniform scaling, with \Cpp workloads showing outsized CPU efficiency gains, while Java workloads (SciMark2, DaCapo) emphasize throughput and memory.
    }
    \label{fig:radar_chart}
\end{figure}

\cref{fig:radar_chart} illustrates SysLLMatic’s ability to optimize multiple performance metrics across the three benchmarks. 
SciMark2 and DaCapo, both Java-based, exhibit similar trends in metric-level trade-offs. 
In contrast, HumanEval\_CPP --- implemented in \Cpp --- shows substantially greater improvement in CPU efficiency (10.4$\times$), likely due to the lower-level expressiveness of \Cpp, which enables the LLM to apply more fine-grained instruction-level optimizations~\cite{qing2025effibenchx}.
Between the two Java benchmarks, DaCapo demonstrates stronger improvements in latency, memory, and throughput relative to SciMark2. 
However, this comes at the expense of a slight regression in CPU usage (0.93$\times$), suggesting a trade-off where SysLLMatic increases computational effort to enhance I/O and runtime efficiency. 
This behavior aligns with the structure of DaCapo applications, which benefit from parallel execution and data batching.

\subsubsection{Code Maintainability}
\label{sec:code-maintainability}
Balancing performance gains with code maintainability and understandability is an important concern, especially on real-world applications.
While our primary focus is on performance gains, we also examine maintainability on DaCapo to see if performance improvements come at the cost of overly complex or fragile code.
We assess the maintainability of the optimized code using Cyclomatic Complexity (CCN) and other structural metrics. 
We report five maintainability metrics averaged at the file level, measured with Lizard~\cite{lizard_terryyin_2025}. Avg Total NLOC captures the average file size in lines of code (excluding comments/whitespace). 
Avg.NLOC and Avg Tokens measure average function size, in lines and tokens, respectively. AvgCCN reflects the average cyclomatic complexity per function, indicating logical branching complexity. 
Avg Func Cnt reports the average number of functions per file, serving as a proxy for modularity.
As shown in Table~\ref{tab:maintainability-metrics-merged}, average maintainability decreased, with CCN generally increasing by up to 21.05\%. 
Function counts also increased (up to 10.81\%), reflecting additional decomposition, which may aid readability. 
These results indicate that while performance optimizations can streamline certain code structures, they often introduce additional control-flow complexity, with the impact varying across applications.

\begin{table}[h]
\centering
\scriptsize

\caption{
Per-class average maintainability metrics before and after optimization across five DaCapo applications, measured with Lizard~\cite{lizard_terryyin_2025}.
Metrics include Avg Total NLOC (file size), Avg.NLOC and Avg Tokens (function size in lines and tokens), AvgCCN (cyclomatic complexity per function), and Avg Func Cnt (functions per file, proxy for modularity). 
Orig and Opt denote original and optimized values; $\Delta$ is the absolute change (Opt – Orig) and $\Delta\%$ the relative change, with positive values indicating increases and negative values reductions.
Results vary across applications, with CCN changes ranging from –0.95\% to +21.05\% and function count changes from +1.47\% to +10.81\%, reflecting a trade-off between increased control-flow complexity and improved modular structure.
% \JD{ALL CAPTIONS SHOULD END WITH SOME INTERPRETATION. I HAVE SAID THIS MANY TIMES I WANT TO STOP SAYING IT. Please review all captions, starting with this one :-P.}
}
\label{tab:maintainability-metrics-merged}
\renewcommand{\arraystretch}{1.2} % Adds a bit of vertical space to rows
\begin{tabular*}{\textwidth}{@{\extracolsep{\fill}}llrrrr}
\toprule
\textbf{Application} & \textbf{Metric} & \textbf{Orig} & \textbf{Opt} & \textbf{$\Delta$} & \textbf{$\Delta$\%} \\
\midrule
\multirow{5}{*}{BioJava}
  & Avg Total NLOC & 192.200 & 212.300 & 20.100  & 10.46 \\
  & Avg.NLOC       & 15.340  & 14.660  & -0.680  & -4.43 \\
  & AvgCCN         & 4.380   & 4.470   & 0.090   & 2.05  \\
  & Avg Tokens     & 140.160 & 132.020 & -8.140  & -5.81 \\
  & Avg Func Cnt   & 16.900  & 18.400  & 1.500   & 8.88  \\
\midrule
\multirow{5}{*}{Fop}
  & Avg Total NLOC & 274.444 & 253.278 & -21.167 & -7.71 \\
  & Avg.NLOC       & 8.944   & 8.744   & -0.200  & -2.24 \\
  & AvgCCN         & 2.578   & 2.622   & 0.044   & 1.72  \\
  & Avg Tokens     & 56.400  & 57.550  & 1.150   & 2.04  \\
  & Avg Func Cnt   & 33.278  & 33.889  & 0.611   & 1.84  \\
\midrule
\multirow{5}{*}{PMD}
  & Avg Total NLOC & 287.000 & 276.556 & -10.444 & -3.64 \\
  & Avg.NLOC       & 8.222   & 7.911   & -0.311  & -3.78 \\
  & AvgCCN         & 2.344   & 2.322   & -0.022  & -0.95 \\
  & Avg Tokens     & 55.833  & 54.211  & -1.622  & -2.91 \\
  & Avg Func Cnt   & 30.222  & 30.667  & 0.444   & 1.47  \\
\midrule
\multirow{5}{*}{GraphChi}
  & Avg Total NLOC & 72.000  & 83.000  & 11.000  & 15.28 \\
  & Avg.NLOC       & 5.350   & 6.000   & 0.650   & 12.15 \\
  & AvgCCN         & 1.425   & 1.725   & 0.300   & 21.05 \\
  & Avg Tokens     & 41.350  & 43.200  & 1.850   & 4.47  \\
  & Avg Func Cnt   & 9.250   & 10.250  & 1.000   & 10.81 \\
\midrule
\multirow{5}{*}{ZXing}
  & Avg Total NLOC & 218.125 & 250.750 & 32.625  & 14.96 \\
  & Avg.NLOC       & 19.588  & 21.375  & 1.788   & 9.13  \\
  & AvgCCN         & 5.488   & 5.675   & 0.188   & 3.42  \\
  & Avg Tokens     & 134.262 & 146.175 & 11.913  & 8.87  \\
  & Avg Func Cnt   & 12.875  & 13.375  & 0.500   & 3.88  \\
\bottomrule
\end{tabular*}
\end{table}

\begin{tcolorbox}[colback=gray!5,colframe=gray!35!black,title={EQ2: How well does SysLLMatic balance multiple, potentially competing performance objectives?}]
\small
Latency and throughput improve most consistently, with \texttt{HumanEval\_CPP} showing large CPU efficiency gains (10.4$\times$). 
\texttt{DaCapo} improves other metrics but regresses slightly in CPU usage (0.93$\times$), while \texttt{SciMark2} achieves steadier, more balanced improvements without strong trade-offs. 
In terms of maintainability on DaCapo, performance optimizations generally increase code complexity (CCN up to 21\%) while also raising function counts (up to 10.8\%), suggesting a trade-off between added control-flow complexity and improved modularity.
\end{tcolorbox}

\subsection{EQ3: What is the contribution of each core component to SysLLMatic's optimization effectiveness?}
\label{sec:EQ3}

To better understand how SysLLMatic achieves its performance gains, we conduct an ablation study that isolates the contribution of each core component. Our analysis proceeds along three axes. 
First, we perform component-wise ablations, removing or disabling individual roles to quantify their impact.
Second, we examine system-level variants, comparing different configurations to assess overall effectiveness.
Third, we analyze program context granularity on SciMark2, evaluating whether SysLLMatic behaves differently when optimizing at the function level versus the class level.
Insights from this analysis guide our choice of optimization granularity for large-scale DaCapo applications.

\begin{table}[ht]
\centering
\footnotesize
\caption{Contributions of our core components --- Evaluator, Context, and Advisor (which incorporates the optimization pattern catalog) --- to optimization effectiveness across three benchmarks.
Components are incrementally enabled, such that each setting includes all components from the preceding one (\eg +Context = Base + Evaluator + Context).
For details on the ablation setup, refer to~\cref{sec:ablation}. 
For DaCapo, Context (Performance Hotspot Identification) is required for large-scale application optimization, as optimizing the whole application at once will exceed the LLM context window, thus cannot be ablated.
The results show the incremental value of adding Evaluator and Context in SciMark2 and HumanEval.
In contrast, adding the Advisor degrades performance on these microbenchmarks, but provides clear benefits for DaCapo.
}
\label{tab:ablation_study_metrics}
\begin{tabular*}{\textwidth}{@{\extracolsep{\fill}}l l r
             r r
             r r
             r r
             r r
             r r @{}}
\toprule
\textbf{Bench.} & \textbf{comp.}
 & \textbf{Corr.\ (\%)} 
 & \multicolumn{2}{c}{\textbf{Latency $ \uparrow$}} 
 & \multicolumn{2}{c}{\textbf{Memory $ \uparrow$}} 
 & \multicolumn{2}{c}{\textbf{CPU $ \uparrow$}} 
 & \multicolumn{2}{c}{\textbf{Throughput $ \uparrow$}} 
 & \multicolumn{2}{c}{\textbf{Energy $ \uparrow$}} \\
\cmidrule(lr){4-5}\cmidrule(lr){6-7}\cmidrule(lr){8-9}\cmidrule(lr){10-11}\cmidrule(lr){12-13}
& &  & \textit{gains} & \%\textit{opt}
     & \textit{gains} & \%\textit{opt}
     & \textit{gains} & \%\textit{opt}
     & \textit{gains} & \%\textit{opt}
     & \textit{gains} & \%\textit{opt} \\
\midrule
SciMark2 & Base & 60    & 0.98$\times$ & 0.0  & 0.99$\times$ & 0.0  & 0.99$\times$ & 0.0  & 1.00$\times$ & 0.0  & 1.05$\times$ & 0.0  \\
& +Evaluator & 100   & 1.18$\times$ & 20.0 & 0.81$\times$ & 0.0  & 0.68$\times$ & 0.0  & 0.89$\times$ & 20.0 & 0.89$\times$ & 0.0  \\
& +Context & 100   & 6.25$\times$ & 60.0 & 0.95$\times$ & 0.0  & 1.12$\times$ & 20.0 & 3.29$\times$ & 60.0 & 3.05$\times$ & 60.0 \\
& +Advisor & 100   & 1.55$\times$ & 40.0 & 0.98$\times$ & 0.0  & 1.52$\times$ & 40.0 & 1.39$\times$ & 40.0 & 1.50$\times$ & 40.0 \\
\cmidrule(lr){2-13}
HumanEval & Base & 65 & 1.69$\times$ & 12.8  & 1.01$\times$ & 1.2  & 10.39$\times$ & 18.3 & --     & --     & 1.92$\times$ & 14.0 \\
& +Evaluator & 87 & 1.72$\times$ & 17.0 & 1.01$\times$ & 1.2  & 10.71$\times$ & 27.4 & --     & --     & 2.07$\times$ & 20.1 \\
& +Context & 90 & 1.76$\times$ & 24.4 & 1.01$\times$ & 1.2  & 9.57$\times$  & 31.1 & --     & --     & 2.02$\times$ & 23.8 \\
& +Advisor & 88 & 1.69$\times$ & 23.2 & 1.01$\times$ & 1.2  & 10.39$\times$ & 29.8 & --     & --     & 2.04$\times$ & 19.5 \\
\cmidrule(lr){2-13}
DaCapo & +Context & 80    & 1.05$\times$ & 20.0 & 1.03$\times$ & 20.0  & 1.06$\times$ & 20.0 & 1.05$\times$ & 20.0 & 1.03$\times$ & 20.0 \\
& +Advisor        & 100   & 1.54$\times$ & 20.0 & 1.02$\times$ & 0.0 & 0.93$\times$ & 20.0 & 1.54$\times$ & 20.0 & 1.24$\times$ & 20.0 \\
\bottomrule
\end{tabular*}
\end{table}

\subsubsection{Component Ablation}
\label{sec:component-ablation}
\cref{tab:ablation_study_metrics} presents the results of our ablation study across three benchmarks.
For SciMark2 and HumanEval\_CPP, code correctness improves significantly with the inclusion of additional components --- rising from 60\% to 100\% in SciMark2 and from 65\% to 90\% in HumanEval\_CPP. 
In SciMark2, adding the Evaluator alone yields modest improvements, but performance gains across all metrics become substantial once context information is incorporated, with latency increasing by 6.25$\times$ and throughput by 3.29$\times$. 
However, the subsequent addition of the Advisor component primarily improves CPU efficiency while reducing gains in other metrics. 
HumanEval\_CPP exhibits notable speedups even in the Base configuration, aligning with the findings in the previous work~\cite{peng2024perfcodegenimprovingperformancellm}.
For DaCapo, we only report the +Context and +Advisor configurations because the Context module is required to identify the relevant code regions for optimization. 
Since the full application exceeds the LLM’s context window, we cannot evaluate SysLLMatic without this component. 
Result shows that adding the Advisor improves correctness from 80\% to 100\% and yields substantial performance gains, including lower latency (1.54$\times$), and improved energy efficiency (1.24$\times$), despite a slight increase in CPU usage (0.93$\times$).
We further observe that without the Advisor, LLM-generated transformations are unstable and inconsistent across applications: memory usage degrades by 23\% on BioJava, while GraphChi shows a 35\% memory improvement but suffers 10-18\% degradation in other metrics. 
In contrast, with the Advisor enabled, no metric degrades by more than 4\% across applications, except for a 50\% increase in CPU usage on BioJava, which is attributable to aggressive parallelization that delivers a 3.6$\times$ latency and throughput improvement.

These results reveal two key insights. First, structured inputs like ASTs and Flame Graphs provide crucial semantic context, enabling more effective LLM-driven optimizations. 
Second, the Advisor component, based on a general pattern catalog, causes regressions in SciMark2 and HumanEval\_CPP but improves performance in DaCapo. 
This is likely due to the catalog’s language-agnostic nature, which overlooks fine-grained heuristics needed for simpler benchmarks but proves beneficial in complex systems by promoting consistent, correctness-preserving transformations.

\textbf{Cost of the Catalog.} In our implementation, the full optimization pattern catalog is provided to the Advisor as part of the input (\cref{sec:formulation}). For the largest application (Fop), the catalog accounts for approximately 10\% of total input tokens; this proportion increases for smaller programs where the code context is shorter. Nevertheless, the absolute monetary cost remains modest: across DaCapo, end-to-end optimization costs remain under \$6 per application (\cref{sec:time_and_monetary_cost}). Future work could reduce this overhead through retrieval-augmented generation to dynamically select relevant patterns, or by filtering the catalog based on program characteristics and optimization objectives.

\subsubsection{System Ablation}
\label{sec:eq3-system-ablation}
We additionally analyze system-wide performance across 1–4 Evaluator feedback iterations on SciMark2. 
As shown in \cref{fig:feedback_iteration}, the Evaluator-driven feedback loop yields substantial improvements in latency, MFLOPS, and energy over successive iterations. CPU and memory performance decline slightly, due to parallelism-induced trade-offs. Improvements saturate after the third feedback iteration, suggesting diminishing returns.

\begin{figure}[htbp]
    \centering
    \includegraphics[width=\textwidth]{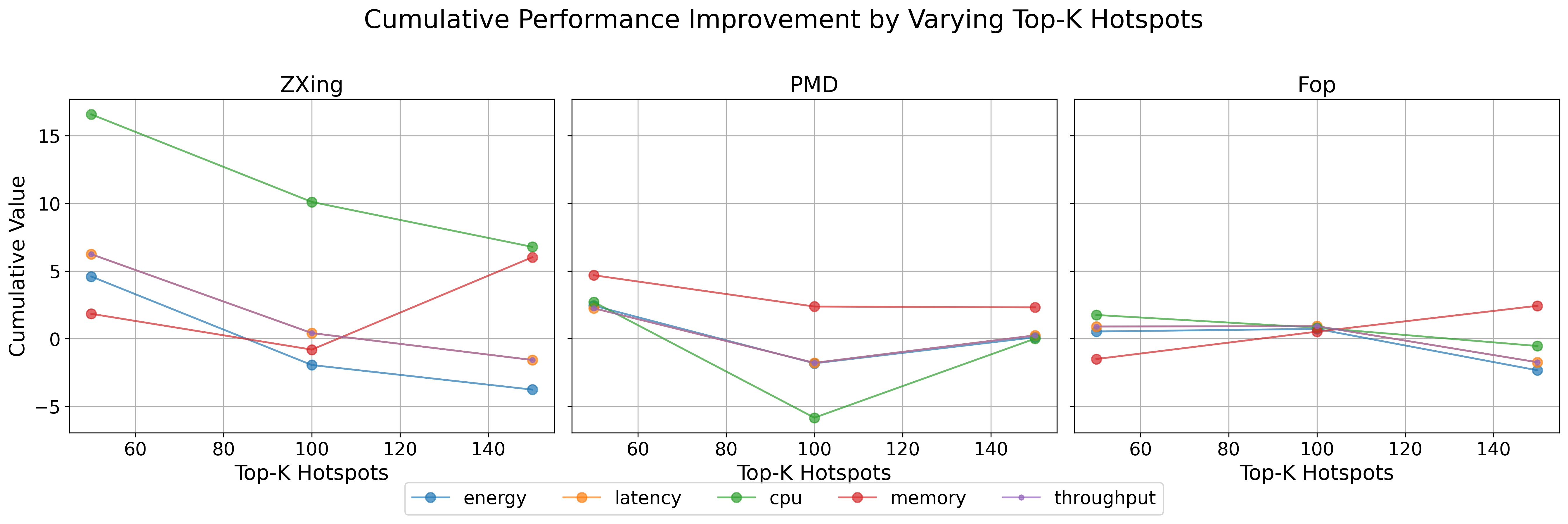}
    \caption{Cumulative performance improvements across applications when varying the number of profiled hotspots ($K$). 
    BioJava and Graphchi are excluded because $K$ = 50 already covers all their hotspots captured in execution. 
    Across ZXing, PMD, and Fop, increasing $K$ yields only small performance gains, and in some cases degrades performance, indicating limited benefit from expanding optimization beyond the top-ranked hotspots.}
    \label{fig:top_k variation}
\end{figure}

This analysis is limited to SciMark2 due to the high cost of repeated profiling and optimization on other benchmarks. 
We exclude HumanEval from this analysis because its programs are relatively simple: LLM with a single feedback loop already achieves high performance, and results stabilize once context is added, as shown in our ablation study. Additional rounds therefore provide little benefit. 
For full DaCapo applications, we instead ablate system performance by varying the number of selected performance hotspots. Specifically, we evaluate sensitivity by adjusting the Top-$K$ threshold across three values: $K = 50, 100, 150$.
As shown in Figure~\ref{fig:top_k variation}, expanding the hotspot set beyond the top 50 has limited impact on overall performance. 
For BioJava and Graphchi, $K=50$ already covers all profiled hotspots, so these applications are excluded. 
For ZXing, PMD, and Fop, increasing $K$ produces only marginal additional gains and, in several cases, degrades performance. 
For example, ZXing and Fop show improved memory performance at 
$K=150$, but decreased CPU efficiency.
% For example, FOP shows higher latency improvements at $K=100$, but energy and CPU performance degrade as more hotspots are included. Similarly, in PMD, memory efficiency decreases as $K$ grows, despite improvements in CPU. 
These results suggest that optimization benefits are concentrated in a small number of critical hotspots, and expanding the scope to less frequently executed functions adds noise, increasing inference and computational cost without commensurate benefits.

\begin{minipage}[t]{0.48\textwidth}
    \centering
    \includegraphics[width=\linewidth]{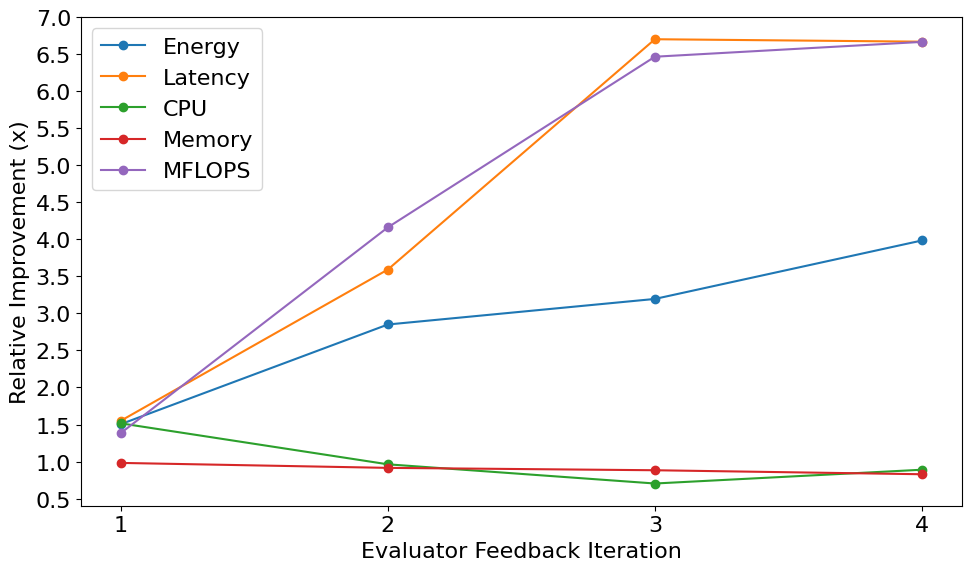}
    \captionof{figure}{Performance improvements across 1–4 Evaluator feedback iterations on SciMark2. 
    Latency and MFLOPS show the largest gains, with improvements stabilizing after the third iteration.}
    \label{fig:feedback_iteration}
\end{minipage}
\hfill
\begin{minipage}[t]{0.48\textwidth}
    \centering
    \includegraphics[width=\linewidth]{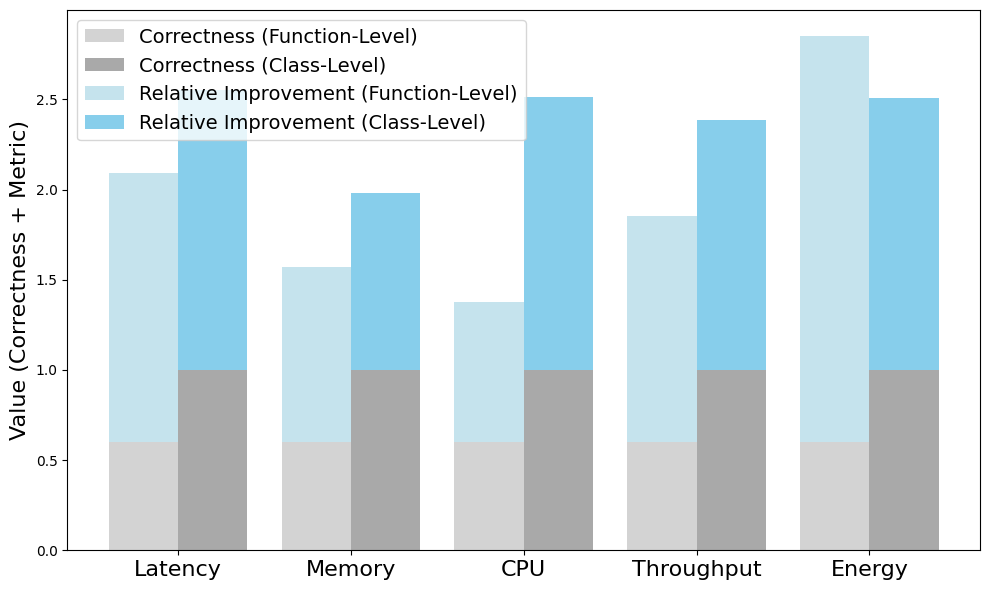}
    \captionof{figure}{Comparison of function-level and class-level optimization on SciMark2 across five performance metrics. 
    Bars show combined values of correctness and relative metric improvement. Class-level optimization consistently achieves higher correctness and greater gains across most metrics.}
    \label{fig:class_method}
\end{minipage}

\subsubsection{How does SysLLMatic performance vary based on the program context it receives?}
\label{sec:EQ3_context}
When optimizing large-scale systems, code modularity enables applying transformations at varying granularities—such as functions, classes, or namespaces. 
To evaluate in real-world code, we apply our framework to SciMark2 at both function and class levels. 
Flame Graph reveals that each kernel has a dominant function consuming over 50\% of CPU time, indicating that optimizing this function could yield substantial performance benefits for the whole program. As shown in~\cref{fig:class_method}, function-level optimization achieves greater performance improvements in some cases, but at the cost of correctness—dropping from 100\% to 60\%. 
For example, the LU kernel sees significant latency and energy improvements only at the function level via parallelization ( \texttt{IntStream.parallel()}).
However, manual inspection reveals that function-level optimization often breaks interface contracts or implicit dependencies due to a lack of broader context.
In contrast, class-level optimization preserves structural integrity and maintains full correctness while still delivering consistent performance gains. This illustrates a trade-off between optimization effectiveness and semantic correctness. 
Consequently, we adopt class-level granularity for DaCapo to ensure correctness while achieving meaningful improvements in complex applications.

\begin{tcolorbox}[colback=gray!5,colframe=gray!35!black,title={EQ3: What is the contribution of each core component to SysLLMatic’s optimization
effectiveness?}]
\small
Ablation results show that each module contributes to SysLLMatic’s performance. 
Increasing the number of evaluator feedback iterations yields significant improvements, whereas varying the Top-$K$ threshold provides minor gains. 
Finally, function-level optimization reduces overall effectiveness, so we adopt class-level granularity for large-scale applications.
\end{tcolorbox}

\subsection{EQ4: Under what circumstances do the benefits of SysLLMatic outweigh its cost?}
\label{sec:eq4_cost}
We evaluate the cost-effectiveness of SysLLMatic from two perspectives: time taken to optimize entire applications, and resource consumption under different deployment settings.

\subsubsection{Time and Monetary Cost}
\label{sec:time_and_monetary_cost}
Table~\ref{tab:optimization-time} reports the time taken for SysLLMatic to optimize four real-world applications of varying sizes. 
These measurements reflect end-to-end runtime, including profiling, performance hotspot selection, iterative optimization, as well as the time required to build each application and execute its test suite.
While larger applications generally take longer, the scaling is not strictly linear with code size. 
For instance, ZXing (48K LOC) required 90 minutes, nearly twice as long as BioJava (300K LOC), where the entire process was completed in 47 minutes.
This contrast demonstrates how build and test overhead, together with structural complexity, can outweigh raw code size in determining total optimization time.

\begin{center}
\small
\begin{tabular}{lcccc}
\toprule
\textbf{Approach} & \textbf{Model} & \textbf{Token Pricing (per 1M)} & \textbf{Cost / App} & \textbf{Time (min)} \\
\midrule
SysLLMatic & \texttt{gpt-4.1} 
& In:\$2.00 / Cached:\$0.50 / Out:\$8.00 
& $\leq$\$5.64 & 11--206 \\
Codex Agent & \texttt{gpt-5.2-codex} 
& In:\$1.75 / Cached:\$0.175 / Out:\$14.00 
& $\leq$\$5.51 & 3.5--8.6 \\
\bottomrule
\end{tabular}
\end{center}

In addition to time cost, we analyze the monetary cost of LLM usage by reporting an upper bound across our benchmarks.
For Fop, the largest evaluated application (400K LOC), the complete end-to-end optimization required 116 LLM queries and incurred a total API cost of \$5.64.
SysLLMatic uses \texttt{gpt-4.1} (priced at \$2.00 / 1M input tokens, \$0.50 / 1M cached input tokens, and \$8.00 / 1M output tokens).
Although model pricing varies by provider, this absolute cost is small relative to typical software development and deployment expenses.

For comparison, the baseline Codex agent incurs a monetary cost ranging from \$1.43 to \$5.51 per application, which is comparable to SysLLMatic.
Codex agent uses \texttt{gpt-5.2-codex}, which is priced at \$1.75 / 1M input tokens, \$0.175 / 1M cached input tokens, and \$14.00 / 1M output tokens.
However, its time cost is significantly lower, ranging from 3.5 to 8.6 minutes, as the scale of modifications is substantially smaller (only 1-4 files changed per application).
In contrast, SysLLMatic performs larger-scale, multi-file optimizations, resulting in end-to-end runtimes ranging from 11 to 206 minutes.

\begin{table}[htbp]
\centering
\caption{End-to-End Optimization Time and Resource Consumption for Real-World Applications. 
Energy consumption is shown with a low bound (0.24 Wh/query) reflecting hyperscale-optimized serving efficiency, 
and a high bound (3--6 Wh/query) reflecting less efficient or on-premises deployments.
The results show that optimization time and energy cost vary not only with application size, and that time cost is not necessarily proportional to energy cost. 
In practice, usage should weigh whether the long-term efficiency benefits justify the initial resource costs, taking into account application characteristics, deployment context, and workload intensity.}
\label{tab:optimization-time}
\scriptsize
\begin{tabular*}{\textwidth}{@{\extracolsep{\fill}}lrrrrr}
\toprule
\textbf{Application} & \textbf{Total LOC} & \textbf{Time Taken} & \textbf{\# Queries} & \textbf{Low (Wh)} & \textbf{High (Wh)} \\
\midrule
GraphChi & 8K   & 11 mins  & 18  & 4.3  & 54--108  \\
ZXing    & 48K  & 90 mins  & 60  & 14.4 & 180--360 \\
PMD      & 120K & 45 mins  & 65  & 15.6 & 195--390 \\
BioJava  & 300K & 47 mins  & 54  & 13.0 & 162--324 \\
Fop      & 400K & 206 mins & 116 & 27.8 & 348--696 \\
\bottomrule
\end{tabular*}
\end{table}

\subsubsection{Resource Consumption}
\label{sec:eq4_resource_consumption}
In addition to time, we also assess the energy consumption of using SysLLMatic, as resource efficiency is a critical consideration --- especially for large-scale software.

\textit{Method:}
The total energy cost comprises three main components: (1) LLM inference, (2) build and execution of the target application (\eg DaCapo), and (3) internal operations within SysLLMatic. Among these, LLM inference dominates the cost and is the most persistent source of computational overhead. Thus, we approximate total resource consumption based on the energy usage of LLM inference. To estimate whether SysLLMatic is a worthwhile investment, we model the net performance benefit as:

\begin{equation}
\text{NetGain} = (\Delta P \times N) - C_{\text{SysLLMatic}}
\end{equation}
where:
\begin{itemize}
    \item $\Delta P$ is the performance improvement per execution (\eg a $2\times$ speedup),
    \item $N$ is the number of times the optimized code is executed,
    \item $C_{\text{SysLLMatic}}$ is the one-time resource cost (\eg energy) of applying SysLLMatic.
\end{itemize}

Because the optimization cost is incurred once while the performance benefit compounds over repeated executions, this model allows us to estimate a break-even point. The amortized gain outweighs the initial cost when
\begin{equation}
N > \frac{C_{\text{SysLLMatic}}}{\Delta P}
\end{equation}
% \AG{I think we should make it clear where those numbers are coming from}

\textit{Result:}
For BioJava, we observed that the energy break-even occurs after 226 repeated executions, while latency improvements break even after 416 executions. 
For ZXing, energy improvements require 2442 executions and latency improvements require 10789 executions to break even.
These thresholds are computed directly from our evaluation results using the LLM inference energy cost of 0.3 Wh/query~\cite{You2025}.

\begin{figure}[htbp]
\centering
\begin{subfigure}[b]{0.48\textwidth}
\centering
\includegraphics[width=\textwidth]{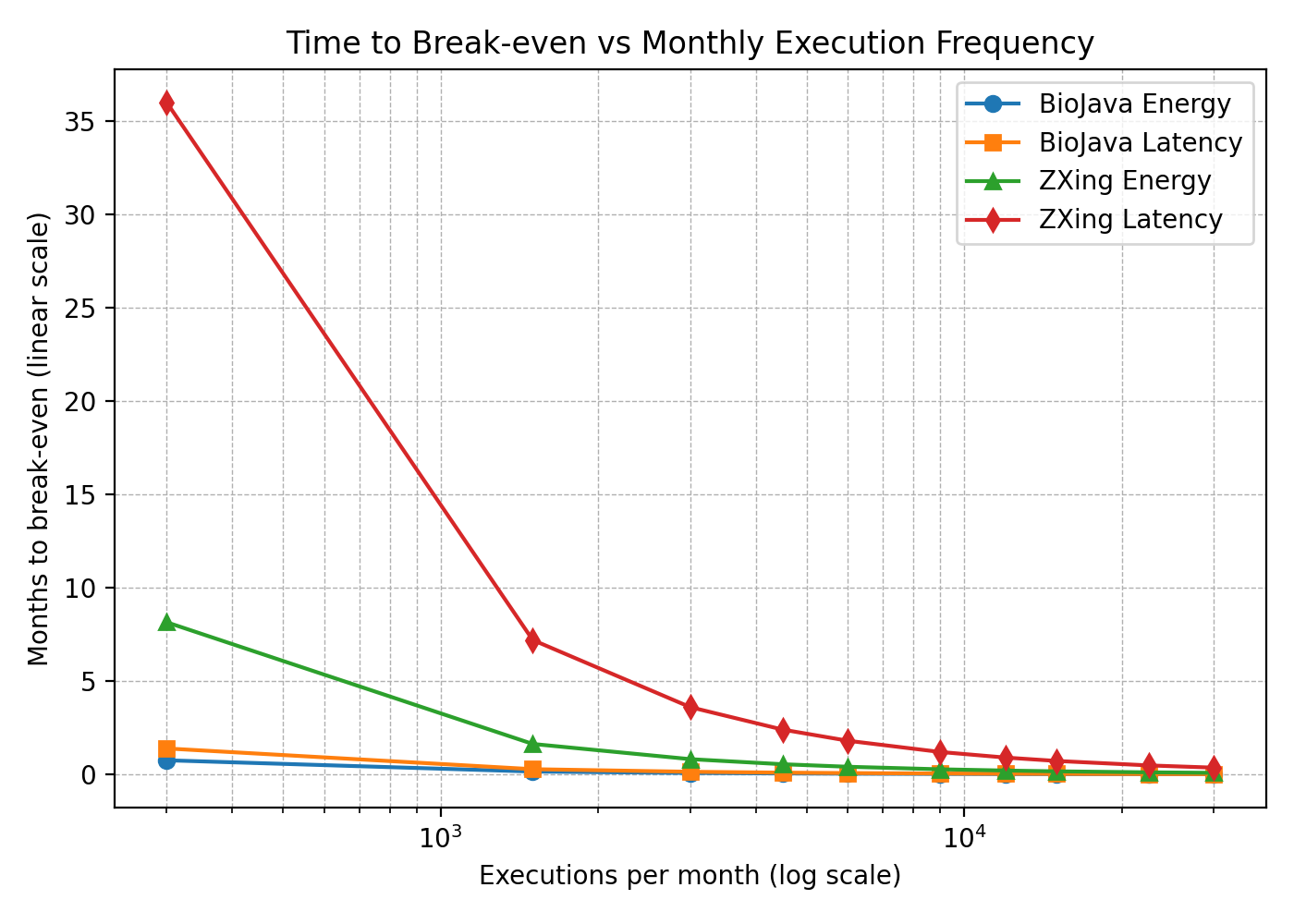}
\caption{Time-to-break-even for BioJava and ZXing as monthly execution frequency increases, assuming a ChatGPT inference energy of 0.3Wh/query~\cite{You2025}. 
High-throughput workloads amortize the one-time optimization cost within days to weeks, whereas low-frequency workloads may require months to years.}
\label{fig:break_even}
\end{subfigure}
\hfill
\begin{subfigure}[b]{0.5\textwidth}
\centering
\includegraphics[width=\textwidth]{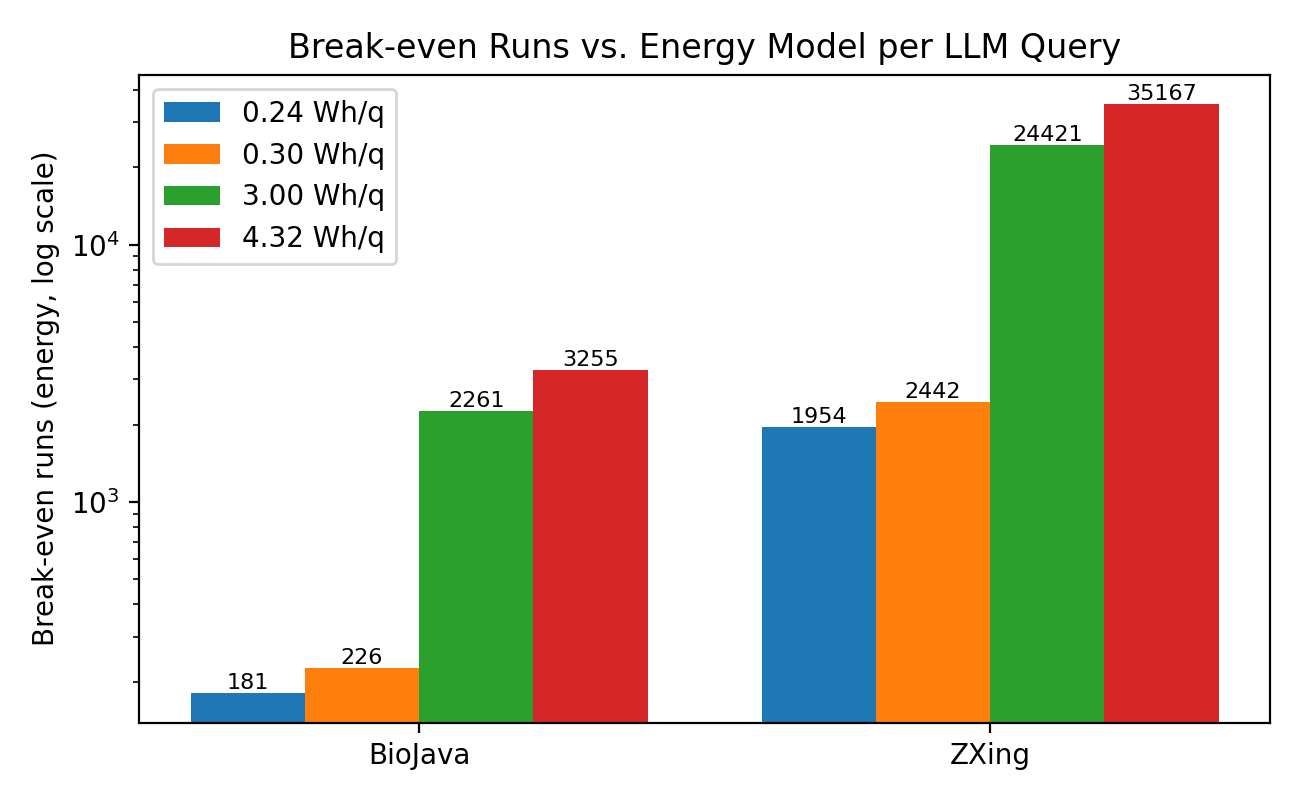}
\caption{Break-even analysis under three alternative LLM energy models. 
We show three representative scenarios to match different deployment settings: a cloud-optimized environment (0.24\ Wh/query~\cite{elsworth2025measuringenvironmentalimpactdelivering}, a baseline inference energy estimate for GPT models (0.30\ Wh/query~\cite{You2025}), representative of public providers such as GCP), a moderate-cost setting (3.00\ Wh/query~\cite{DEVRIES20232191}), and a pessimistic on-premises deployment (4.32\ Wh/query~\cite{oviedo2025energyuseaiinference}).}
\label{fig:break_even_energy_models_sub} 
\end{subfigure}
\caption{
Analysis of SysLLMatic’s cost-effectiveness under different usage and energy models. \cref{fig:break_even} shows time-to-break-even as a function of monthly execution frequency (log scale). 
\cref{fig:break_even_energy_models_sub} shows break-even runs under three additional LLM energy models representing different deployment settings.
These findings suggest that the viability of deploying SysLLMatic depends on workload intensity and the efficiency of the underlying inference infrastructure.
}
\label{fig:granularity_and_breakeven}
\end{figure}

To translate these break-even points into time-to-payoff, we assume different execution frequencies.
To illustrate the results, we conjecture a conservative scenario of 50 executions/day, under which BioJava amortizes the inference cost within 5 days (energy) and 8 days (latency), while ZXing requires 1.6 months (energy) and 7.2 months (latency).
In contrast, under a heavy-usage scenario (500 executions/day), all energy improvements break even within a week, and latency improvements for both applications within 1 month.
\cref{fig:break_even} visualizes time-to-break-even versus monthly execution frequency.
Reported energy usage for LLM inference, however, varies widely across deployment environments, from optimized public clouds to much less efficient on-premises clusters. 
To capture this uncertainty,~\cref{fig:break_even_energy_models_sub} extends the analysis with three additional models:
  a cloud-optimized scenario (0.24Wh/query) from Elsworth~\etal in 2025~\cite{elsworth2025measuringenvironmentalimpactdelivering},
  a moderate-cost setting (3.0Wh/query) measured by De Vries, A. in 2023
  ~\cite{DEVRIES20232191},
  and
  an on-premises case (4.32Wh/query) reported by Oviedo~\etal in 2025~\cite{oviedo2025energyuseaiinference}.

Overall, high-throughput workloads recover SysLLMatic’s one-time optimization cost quickly, whereas low-frequency workloads may require months or even years to break even. 
While upfront costs remain a consideration in low-usage scenarios, the cumulative benefits in performance-critical or frequently executed code paths make SysLLMatic a net-positive investment in high-throughput environments for both energy and performance efficiency.

\begin{tcolorbox}[colback=gray!5,colframe=gray!35!black,title={EQ4: Under what circumstances do the benefits of SysLLMatic outweigh its cost?}]
\small
SysLLMatic pays off quickly for high-throughput or frequently executed applications, where one-time optimization costs are amortized within days to weeks. 
For lower-frequency workloads, break-even can extend to months or years.
Our modeling permits engineers to assess SysLLMatic for their circumstance.
\end{tcolorbox}

\section{Discussion}
\label{sec:Discussion}
% \JD{Do we need to add `and Limitations' here? It is odd because \$9 is Threats and the two titles seem redundant. This section is mostly Discussion IMHO}
This section discusses the broader implications and limitations of our work. 
Section~\cref{sec:Discussion_A} reflects on lessons learned, limitations, and future directions for automated software optimization, while Section~\cref{sec:discussion_2} examines the practical costs of applying SysLLMatic, including engineering effort, maintainability tradeoffs, and computational overhead.

% \JD{Far too little connection to our results in here. Open every paragraph with a statement of our relevant results. Only then do you build credibility to Discuss them and connect to other ideas.}

\subsection{How Far Along Are We Toward Push-Button Software Optimization?}
\label{sec:Discussion_A}
This section reflects on the extent to which current techniques move us toward push-button software optimization. 
We first summarize key lessons learned from applying SysLLMatic to real-world softwares, then discuss the limitations of the current method, and finally outline future directions for advancing automated software optimization.

\subsubsection{Advancing LLM-Based Software Optimization: Lessons Learned}

\paragraph{Progress Beyond Small-Scale Examples.}

Ideally, software optimization would require only a source input, \eg a GitHub repository URL, to obtain an optimized version.
% \JD{The next sentence does not flow from the previous one. Try again.}
However, the controlled environments typically used in research differ substantially from production software, and most prior work evaluates optimization techniques on synthetic or small-scale benchmarks rather than real-world software systems.
% \JD{The crefs in the next sentence feel a bit lazy. Let's state succinctly the claim that is substantiated in those crefs. The other crefs in this sentence do a good job at this.}
Our work takes a step toward bridging this gap by moving from small-scale programs to DaCapo applications, demonstrating the potential of LLMs to improve performance in complex software systems (\cref{tab:rq1}).

Nonetheless, several challenges remain.
First, our approach assumes access to comprehensive test suites for correctness validation, which may not hold in practice (\cref{sec:impl-correctness}).
Second, real-world applications often have complex build and configuration dependencies. 
In our experiments, DaCapo applications were pre-built, but automating this for arbitrary software remains difficult~\cite{garg2025rapgenapproachfixingcode, Gong2024}. 
These difficulties in automating compilation and test execution can limit the generalizability and practical utility of the approach. 
However, in practice, many engineering teams already rely on CI/CD pipelines that automate these workflows. 
We believe our approach can be effectively integrated into such existing infrastructure with reasonable effort.
Third, our method relies on profiling data to guide optimization;
while informative, such data can be time-consuming to collect at scale (see~\cref{sec:perf_hotspot} and~\cref{sec:impl-analysis}).
Real-world adoption must weigh the optimization gains against these computational and operational costs.

\paragraph{Toward Scalable and Generalizable Software Optimization.}
% \JD{Add a topic sentence please. Or re-organize this paragraph-level material (next 2 paragraphs) because I see a topic sentence in the *second* paragraph here.}
Our results reveal a \textit{scalability and generalizability challenge}: the models that deliver the highest optimization quality (GPTs) are proprietary, while the largest open-source alternatives (\eg Qwen-480B) demand extreme computational resources that limit practical deployment.
Our evaluation shows that proprietary GPT models achieve an average of 1.54$\times$ latency and 1.24$\times$ energy improvements on DaCapo, while the largest open-source model we tested (Qwen-Coder:480B) achieves only 1.05$\times$ latency and 1.02$\times$ energy (\cref{tab:humaneval-scimark-open}).
This gap indicates that, although open-source models are highly competitive on micro-benchmarks (matching or even exceeding GPTs on SciMark2 and HumanEval), they fall short on complex, large-scale applications.
% This calls into question conclusions from prior work that report strong performance of open models on synthetic benchmarks, since those results may not generalize to complex, real-world applications.
As a result, strong performance reported on synthetic benchmarks from prior work may not generalize to real-world systems.
Future research should evaluate optimization methods on full-scale software systems and explore hybrid strategies that combine small specialized models with large general-purpose ones. 
Improving the efficiency of open-source models so they can run on modest resources is also critical for enabling practical, push-button software optimization.

\paragraph{Context Matters: Rethinking Function-level Code Generation with LLMs.}

Most LLM4Code efforts have targeted isolated, self-contained functions, overlooking the common case where functions are embedded within larger codebases with dependency constraints~\cite{huang2024,peng2024perfcodegenimprovingperformancellm,rahman2025marcomultiagentoptimizinghpc}.
Prior work also shows LLMs perform better at function-level code optimization than at broader scopes~\cite{Du2024}.
However, our results show that ignoring such context leads to poor code optimization outcomes (\cref{sec:EQ3_context,fig:class_method}).
We therefore advocate for repository-level code generation and optimization as a target for future work.

\iffalse
\paragraph{Toward Next Generation Optimization with LLMs.}
\JD{I read this and it's good --- but since our work does nothing architecture-specific, we cannot say this material here without the reader feeling we are pontificating instead of explicating our results. Save it for a workshop paper or another work.}
\JD{No connection to our results in here.}
The trajectory of algorithm development offers a useful lens for situating our work. 
Early contributions to computer science emphasized algorithms that were human-readable, even if not always optimal for the hardware of the time. 
As architectures diversified—from multicore CPUs to GPUs, FPGAs, and today’s tensor-oriented accelerators—these algorithms often required performance-driven transformations that rendered them unrecognizable and difficult to maintain. 
Traditional attempts to refactor through abstractions or design patterns have frequently produced code that is not comprehensible. 
Our approach suggests that LLMs may offer a new path forward: enabling the generation and maintenance of architecture-specialized implementations without demanding constant expert intervention, while still relying on specialists to validate correctness and robust testing.
\fi

\subsubsection{Limitations of the Current Approach}
\label{sec:multi_objective_discussion}
\begin{figure*}[h]
    \centering
    \begin{subfigure}{\textwidth}
        \centering
        \setlength{\tabcolsep}{3pt}
        \begin{tabular}{ccccc}
            \adjustbox{max width=0.19\textwidth}{\includegraphics{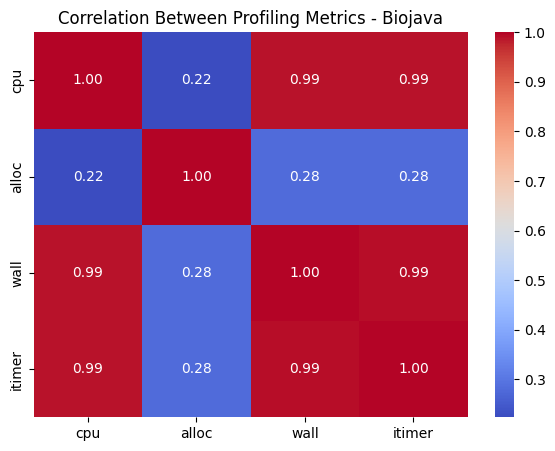}} &
            \adjustbox{max width=0.19\textwidth}{\includegraphics{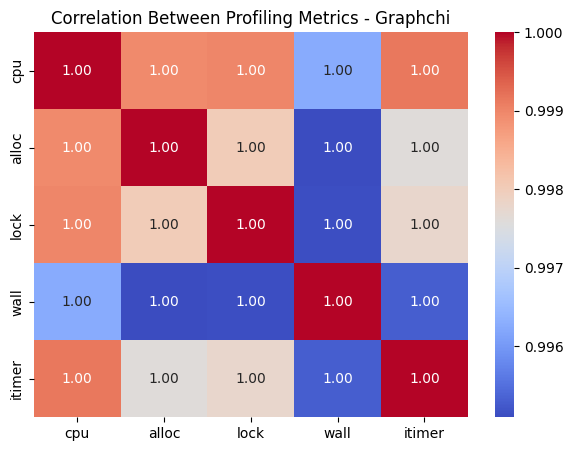}} &
            \adjustbox{max width=0.19\textwidth}{\includegraphics{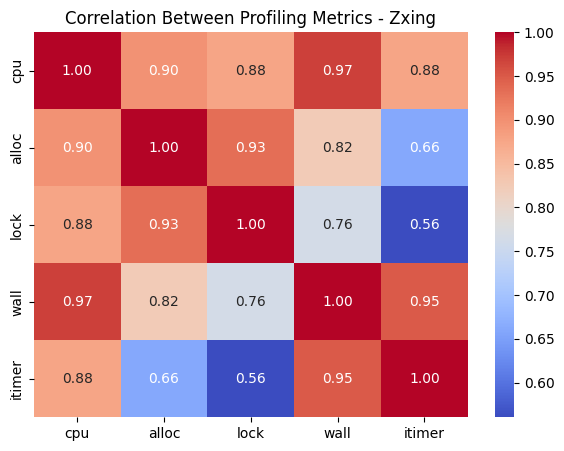}} &
            \adjustbox{max width=0.19\textwidth}{\includegraphics{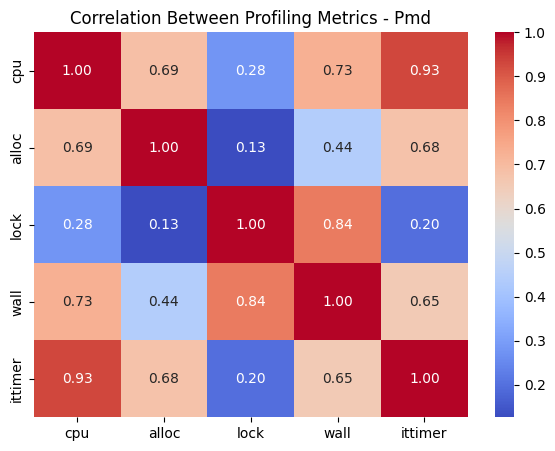}} &
            \adjustbox{max width=0.19\textwidth}{\includegraphics{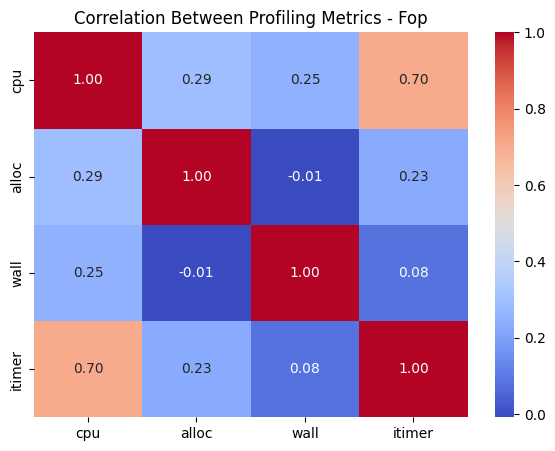}} \\
        \end{tabular}
        \caption{
            Hotspot distribution correlation (Pearson) across profiling metrics for each benchmark.
            High correlations indicate that wall-clock time is dominated by CPU activity, while lower
            correlations suggest additional effects from allocation or lock.
        }
        \label{fig:heatmaps}
    \end{subfigure}

    \vspace{1em}

    \begin{subfigure}{\textwidth}
        \centering
        \setlength{\tabcolsep}{3pt}
        \begin{tabular}{ccccc}
            \adjustbox{max width=0.19\textwidth}{\includegraphics{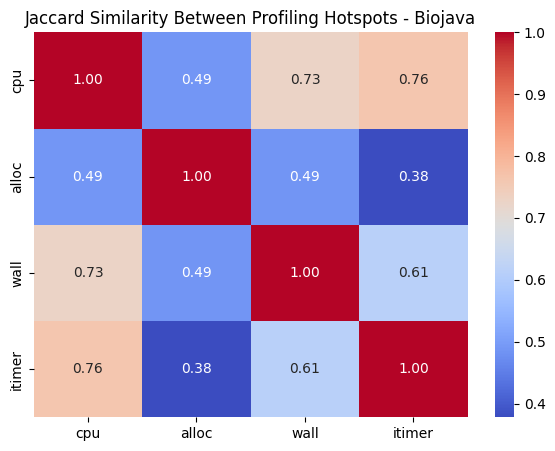}} &
            \adjustbox{max width=0.19\textwidth}{\includegraphics{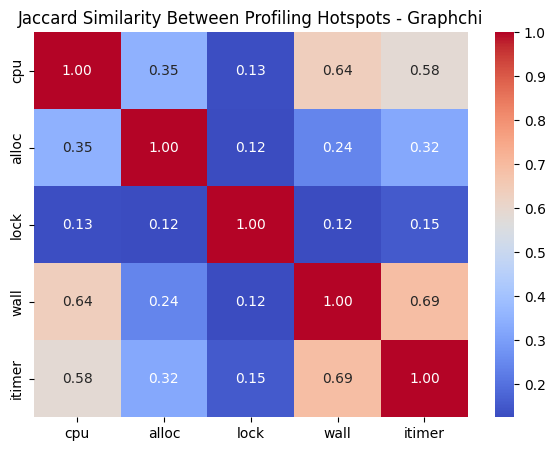}} &
            \adjustbox{max width=0.19\textwidth}{\includegraphics{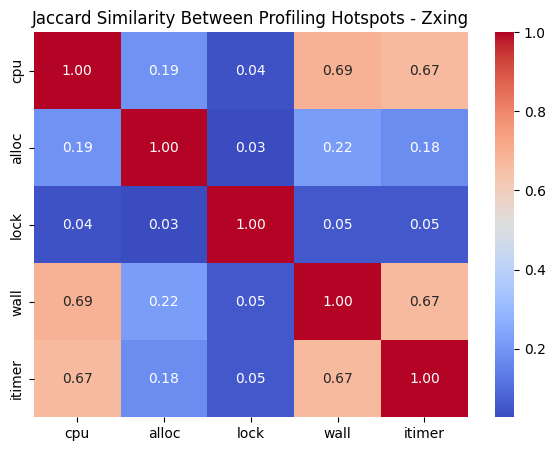}} &
            \adjustbox{max width=0.19\textwidth}{\includegraphics{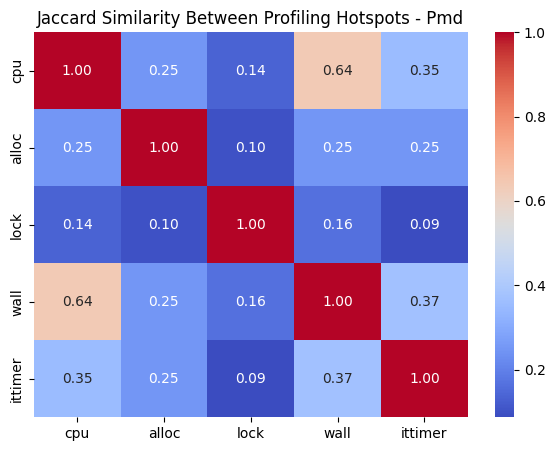}} &
            \adjustbox{max width=0.19\textwidth}{\includegraphics{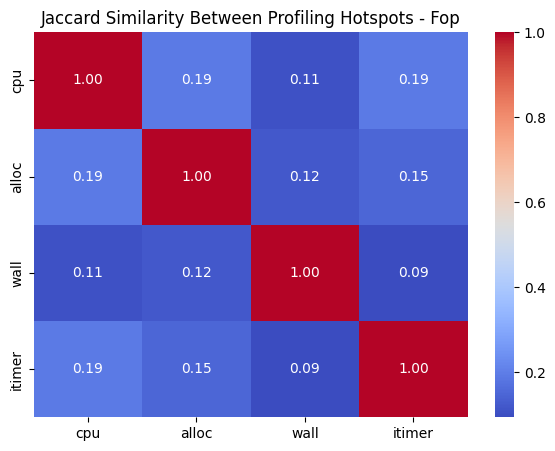}} \\
        \end{tabular}
        \caption{
            Hotspot set overlap (Jaccard similarity) between profiling metrics for each benchmark. 
            High overlap means the same methods dominate across metrics, while low overlap indicates different code paths are responsible for CPU, allocation, or lock costs.
        }
        \label{fig:heatmaps-jaccard}
    \end{subfigure}

    \caption{Comparison of hotspot analysis across benchmarks using Pearson correlation (\cref{fig:heatmaps}) and Jaccard similarity (\cref{fig:heatmaps-jaccard}). Results show that CPU cycles are strongly correlated with wall time and itimer in compute-bound workloads (\eg BioJava, Graphchi), while divergence emerges in memory- or I/O-intensive applications (\eg Fop, PMD), where allocation and synchronization events reveal additional hotspots not captured by CPU-centric profiling. This shows the importance of multi-objective optimization.}
    \label{fig:heatmaps-combined} % A new label for the entire figure
\end{figure*}

\paragraph{Multi-Objective Optimization.}
Our hotspot identification relies on CPU cycle sampling, yet real-world performance optimization often involves multiple dimensions such as memory allocation, synchronization overhead, and I/O latency.
% \JD{Add a sentence here that says `obvs these might interact (cite cite)' and then the next sentence should say `to illustrate that rule in our context, we compared...'}
These dimensions can interact in complex ways, where improving one may worsen another~\cite{Gregg_2021}.
To illustrate this in our context, we compared hotspot profiles obtained under different event types (\texttt{cpu}, \texttt{memory}, \texttt{lock}, \texttt{wall}, \texttt{itimer}), measuring both Pearson correlation of hotspot distributions and Jaccard overlap of hotspot sets across five DaCapo applications (\cref{fig:heatmaps-combined}).
The results show that CPU cycles are strongly correlated with wall time and \texttt{itimer} in compute-bound workloads (\eg BioJava, Graphchi), while divergence emerges in memory- or I/O-intensive applications such as Fop and PMD, where memory allocation and synchronization events reveal additional hotspots not captured by CPU-centric profiling.

These findings reveal a limitation of single-metric optimization: improvements guided solely by CPU usage may overlook bottlenecks in memory or concurrency.
The lower Jaccard similarity across event types further indicates that hotspot rankings can differ substantially, even when distribution-level correlations remain high.
Prior work has largely focused on optimizing with respect to a single metric and has evaluated success mainly on synthetic benchmarks~\cite{gong2025}.
Our results show that with sufficient scaffolding, the same insights can be generalized to real-world applications. 
However, our study also reveals that this generalization is not sufficient: real-world optimization is inherently multi-objective, since bottlenecks may shift between computation, memory allocation, synchronization, or I/O. 
Accordingly, we view multi-objective optimization as an essential next step.

Future work should explore adaptive profiling and optimization strategies that target different metrics depending on application characteristics—for instance, prioritizing memory events for memory-bound programs, lock events for synchronization-heavy workloads, and CPU cycles for compute-bound kernels.
While such considerations are less critical in micro-benchmark settings such as competitive programming or HumanEval tasks, they are indispensable for scaling LLM-based optimization to large, real-world applications.
% \JD{Right now I read this and say (1) this is obvious, and (2) if you're so smart, do it. To address that, we need to do two things. First, we need to cite prior work and draw out the fact that they didn't do the large-scale apps (we showed their local insight can generalize with enough scaffolding!) and having shown that, that we aren't done yet because real-world optimization *is* multi-objective. So we're leaving this for future work but you're welcome everyone for pointing it out.}

% Jaccard

\paragraph{Limitations of Workload-Driven Hotspot Detection.}

% \JD{Hmm, I think we are cref'ing to this when we talk about using CPU traces. This paragraph talks about *workload-driven* but that's not the same thing as *which data we give to optimize*. We can do both...but we need to make the second point here to make the earlier cref to this point useful. Or maybe that earlier cref should be to 8.1.3 instead? But maybe 8.1.3.1 and this part overlap a lot, so should we merge them? Please consider and use your judgment.}
SysLLMatic relies on profiling to identify performance hotspots, and therefore the optimization opportunities it discovers are inherently dependent on the executed workloads (\cref{sec:perf_hotspot}). 
For HumanEval and SciMark2, profiling is driven by test cases or microbenchmark workloads. 
Because these benchmarks consist of small, standalone programs, the tests typically exercise most relevant code paths, making hotspot identification relatively comprehensive. 
However, this assumption may not hold for larger real-world applications. 
For DaCapo, we rely on its benchmark harness with predefined inputs and execution phases, but hotspot identification still depends on the representativeness of these workloads. 
Methods not exercised under the selected workloads will not appear in profiling results and therefore cannot be optimized, potentially missing performance-critical behaviors that arise under alternative workloads. 
Future work could mitigate this limitation by incorporating multiple workload configurations, workload synthesis, or coverage-guided profiling to expose a broader set of performance behaviors.

\paragraph{Impact of Modularity on Optimization Effectiveness}
Our hotspot identification strategy prioritizes code regions whose optimization can yield measurable end-to-end performance improvements; consequently, software modularity can influence optimization effectiveness.
The DaCapo benchmarks used in our evaluation exhibit reasonably good modular structure: as shown in \cref{tab:maintainability-metrics-merged}, average per-file NLOC ranges from 5-20 lines and average CCN remains low (1.4-5.5), suggesting relatively small, well-contained units of functionality despite large overall codebases.
This level of granularity supports targeted optimization and helps constrain the search space presented to the LLM, thereby reducing iteration overhead. However, this assumption may not hold for all real-world applications, particularly those with large monolithic components or tightly coupled logic. In such settings, improved modularity could facilitate more precise and efficient transformations by isolating performance-critical behavior into well-defined components, narrowing the optimization search space.
This suggests an important direction for future work: explicitly incorporating modularity into the optimization process itself by assessing architectural structure prior to optimization and improving it when necessary (\eg through automated refactoring) to better support downstream transformations. This question is particularly important as software engineering becomes increasingly automated, where code generation, refactoring, and performance optimization are likely to operate as components of a unified pipeline rather than isolated stages.

\paragraph{Single-pass Hotspot Profiling as an Approximation.}
In its current embodiment, SysLLMatic identifies performance hotspots using a single profiling pass before the optimization loop (\cref{alg:sysllmatic}). 
In principle, hotspot profiles may change as optimizations remove existing bottlenecks and expose others, suggesting that an ideal implementation would re-profile the program after each optimization step. 
In this work, we approximate this process by profiling once, which simplifies the pipeline and avoids repeated profiling overhead. 
Empirically, we observed that the dominant hotspots in our benchmarks remained largely stable across optimization iterations: although the ordering of hotspot methods may vary between runs, the set of methods responsible for most execution time changes little. 
Re-profiling also introduces challenges when comparing methods across runs, since flame graph frequencies reflect relative sampling counts rather than absolute execution time. 
After an optimization, a method may appear more frequently simply because it executes more times within the same sampling window, making direct comparisons of profiles misleading. 
An extension could explore dynamic hotspot tracking strategies, \eg periodically re-profile the program to detect newly emerging hotspots and update the optimization queue accordingly.

\subsubsection{Toward Autonomous Optimization Systems}
\label{sec:discussion_agent}
\paragraph{Toward System-Level Reasoning for Performance Optimization.}
Although SysLLMatic targets performance optimization in complex software systems, our current design does not fully realize system-level reasoning. 
Optimization opportunities are identified by selecting CPU-intensive functions or classes based on call-stack hotspots, which biases the system toward localized, largely independent code regions. 
Consequently, SysLLMatic optimizes components in isolation and does not explicitly reason about cross-class dependencies, component interactions, or architectural decisions. 
In practice, many performance bottlenecks arise from interactions across components and end-to-end execution paths rather than isolated code inefficiencies~\cite{smith2000}. 
Addressing such system-level effects requires reasoning over broader program structure and inter-component relationships.
Future work could incorporate architectural information such as call graphs and dependency graphs, and explore agentic approaches to enable multi-component optimization across larger system-level search spaces.

\paragraph{Agentic LLMs for Optimization.}
While our system uses a structured, feedback-driven loop, the LLM is not autonomous—it depends on external control (\cref{sec:problem_Statement_our_approach}).
In particular, we anticipate that the operator is a human developer or a CI/CD system that integrates profiling and test execution into the workflow. 
We make this assumption because changing complex applications without strong guardrails is dangerous: they may have inadequate test coverage or domain-specific correctness constraints, and leaving the LLM fully unsupervised can introduce subtle errors (as we observed occasionally, see~\cref{sec:impl-correctness} and \cref{sec:manual-correctness}). 
However, there are many engineering contexts where a higher level of test coverage may be available or a lower degree of criticality may be assumed.
For these cases, future work might explore agentic approaches, where LLM-based optimizers are equipped with decision-making, long-term memory, and planning capabilities~\cite{luo2025largelanguagemodelagent, dong2025surveycodegenerationllmbased}. 
Such agents could operate in a self-directed manner, learning to prioritize optimization opportunities, balance performance trade-offs, and adapt strategies over time with minimal human intervention.
The potential advantages include scalability to larger and more complex systems, continuous adaptation as workloads evolve, and reduced reliance on human steering~\cite{ConcoLLMic}. 
However, agentic systems also introduce risks, such as reduced controllability, difficulty in attributing performance changes to specific decisions, and the possibility of unsafe or suboptimal optimizations if the agent’s objectives are misaligned~\cite{wang2025aiagenticprogrammingsurvey, sajadi2025aigeneratedfixessecureanalyzing, peng2025agentsperformcodeoptimization}. 

Our comparison with a fully automated Codex-based agent baseline further illustrates this trade-off.
While the agent achieves substantially lower end-to-end optimization time (\cref{sec:time_and_monetary_cost}) and requires minimal implementation efforts, it delivers only moderate performance improvements and exhibits correctness issues on one application (\cref{tab:rq1}). 
These results suggest that structured, feedback-driven optimization as employed by SysLLMatic enhances reliability and improvement magnitude, whereas greater autonomy reduces engineering overhead and turnaround time, potentially at the cost of stability and optimization quality.
Balancing these opportunities and risks will be crucial in determining how far LLM-based optimizers can evolve toward fully autonomous software improvement.

\paragraph{Compiler Optimization in the Era of LLMs.}
Our results illustrate a fundamental trade-off between classical compiler optimizations and LLM-driven optimization approaches such as SysLLMatic (\cref{tab:rq1}). 
Classical compiler optimizations are grounded in decades of systematic engineering, in which transformation patterns are manually designed and rigorously validated to ensure semantic preservation~\cite{Bacon1994}. 
This results in a high implementation cost for each optimization pass, but also strong correctness guarantees: compilers are deterministic and do not rely on external validation infrastructure. 
These properties make compilers highly reliable and efficient for production use, as identical inputs consistently yield identical optimized outputs. 
In several settings, we observe that classical compiler optimizations outperform SysLLMatic (\cref{tab:rq1}). 
In contrast, SysLLMatic leverages the general reasoning capabilities of LLMs to propose broader source-level transformations without requiring the manual construction and formal verification of new optimization passes, thereby lowering the barrier to introducing new optimization strategies. 
However, these transformations are probabilistic and lack intrinsic correctness guarantees, requiring a high-coverage test suite and iterative feedback to ensure semantic preservation, which incurs additional computational overhead (\cref{tab:optimization-time}).
Future research should further examine this trade-off between the stability and determinism of compiler optimizations and the adaptability and generalizability of LLM-driven methods.

\subsection{Costs of Applying SysLLMatic}
\label{sec:discussion_2}
This section examines the costs associated with applying SysLLMatic in practice. 
We discuss the engineering effort required for reliable correctness validation, the tradeoffs between performance improvements and code maintainability, the role of software modularity in shaping optimization effectiveness, and the computational overhead introduced by LLM-driven optimization workflows.

\subsubsection{Engineering Costs of More Rigorous Correctness Validation}
\label{sec:discussion_correctness}
In LLM-for-Code research, like previous approaches to refactoring, test case validation is the standard for assessing code correctness. 
This approach works well for competitive programming tasks and small-scale programs with controlled inputs. 
However, its sufficiency is questionable when scaling to large systems (see~\cref{sec:manual-correctness}). 
For example, in our experiments with DaCapo---a well-established benchmark suite with JUnit test coverage---we observed substantial variation across applications (Table~\ref{tab:test-coverage-merged-fullwidth}). 
GraphChi exhibits very low coverage ($\approx$10\% across instructions, branches, lines, and methods), while ZXing achieves high coverage ($\approx$90\%). 
On average across all applications, instruction coverage is 59.4\%, branch coverage is 49.7\%, line coverage is 57.6\%, and method coverage is 57.2\%. 
Despite this level of coverage, some optimized programs still passed all tests but failed output validation due to parallelism-induced nondeterminism and race conditions.

This highlights the limitations of test-based validation in complex systems. 
When combined with SysLLMatic’s capacity for large-scale rewrites, the risks are amplified: even subtle functional errors can propagate widely and become difficult to detect or localize.
Mitigating this risk requires engineering strategies that constrain the scope of errors and facilitate debugging. 
One practical approach is decomposing large transformations into smaller, incremental patches. 
Beyond such engineering safeguards, ensuring reliability in complex software requires stronger correctness guarantees, such as automated test case generation to broaden validation coverage and formal verification to provide rigorous assurances~\cite{shashidar2005}.
% \NJE{This is quite a lot of future work - it makes it seem as though our system is quite risky, and has uncertain tradeoffs.}
% \JD{I think it's OK, more directions are better than fewer ones.}
% \HP{I revised a little}

\subsubsection{Tradeoff of Maintainability and Performance}
Balancing performance gains with code maintainability and understandability is an important concern. Our maintainability analysis (\cref{tab:maintainability-metrics-merged}) shows mixed outcomes. 
Cyclomatic complexity (CCN) was largely stable, with four applications changing only slightly (\(-0.95\%\) to \(+3.42\%\)) and one application showing a larger increase of 21.05\%. 
In contrast, function counts consistently grew across all applications (from 1.47\% to 10.81\%), reflecting finer-grained decomposition that may enhance modularity. 
These findings suggest that SysLLMatic’s optimizations generally preserve code structure, but they tend to introduce more functions and, in some cases, add control-flow complexity.

Qualitative observations (\cref{sec:qualitative_result,fig:jacobi-dual}) further illustrate this tradeoff: 
optimizations like loop unrolling and parallelism improved performance but yielded denser control structures, which may reduce code understandability and maintainability. 
Overall, the maintainability impact is modest rather than uniformly negative, yet recurring costs can arise when developers need to modify, debug, or integrate optimized code. 
Unlike the one-time computational cost of running SysLLMatic, these human costs may resurface repeatedly during a system’s lifecycle. 
Thus, the tradeoff between performance and maintainability must be carefully weighed in real-world adoption, particularly for systems that evolve rapidly or require frequent manual intervention. 
Looking ahead, agentic systems may perform such optimizations independently, but for now, the maintainability of LLM-optimized code remains an important engineering concern.

\subsubsection{Computational Costs of Using SysLLMatic}
LLM-based optimization delivers measurable performance improvements, but it also introduces resource consumption costs that cannot be ignored (\cref{sec:eq4_cost}).
For example, optimizing BioJava required 54 LLM queries and 47 minutes of optimization time (\cref{tab:optimization-time}).
These costs are amortized differently across workloads. 
In high-frequency settings, the upfront overhead is repaid quickly, but for applications that run infrequently, the break-even point may require months or longer (\cref{fig:granularity_and_breakeven}).
These findings raise broader sustainability concerns. The trade-off between optimization quality, model size, and energy use remains an open challenge.
Lightweight alternatives—such as smaller models for refinement or more energy-efficient deployment platforms—could make LLM-driven optimization more accessible.
It is the engineering team's responsibility to assess whether the long-term efficiency benefits justify the initial resource costs in their own deployment context.

\section{Threats to Validity}
\label{sec:Threats}

We discuss construct, internal, and external threats to validity~\cite{Wohlin2012}. Following the guidance of Verdecchia \etal~\cite{VERDECCHIA2023107329}, we focus on substantive threats that might influence our findings.

\underline{Construct Threats} are potential limitations of how we operationalize concepts.
We define \textit{performance} in terms of runtime metrics such as latency and memory footprint.
We omit other factors like code security. 
We define ``\textit{correctness}'' based on existing test cases.
Although this is a reasonable choice and consistent with prior work~\cite{jakobs2022peqtest}, of course it may not fully capture semantic equivalence~\cite{shashidar2005}.
As a mitigation, for DaCapo we automatically exclude files without unit tests from the list of optimization targets, resulting in substantially higher test coverage (\cref{tab:test-coverage-touched}).
% \JD{cref misbehavior w.r.t. paragraph-level cref}
In addition, we rely primarily on CPU-based performance profiling to guide hotspot identification, rather than memory, lock, or I/O events, which we detail in~\cref{sec:multi_objective_discussion}.

\underline{Internal Threats} are those that affect cause-effect relationships.
We benchmark against compiler and state-of-the-art LLM baselines under controlled settings, using a fixed LLM temperature of 0.7, following prior work that adopts this value as a standard setting for code generation and optimization tasks~\cite{garg2025rapgenapproachfixingcode, shypula2024learning, Gao2025}. 
While this encourages diverse solution exploration, it also introduces nondeterminism: identical prompts may yield different outputs across runs. 
In addition, environment-level factors may introduce additional noise into performance measurements. 
To mitigate hardware-level variance, we perform two warm-up runs followed by five measured runs for microbenchmarks, reporting the average of the measured executions.
For DaCapo, since Java applications are known to exhibit substantial performance variability~\cite{Kalibera2013}, we increase the repetitions to 10 warm-up runs and 20 measured runs, and report the mean with 95\% confidence intervals (\cref{fig:baseline_dacapo_original}). 
While this configuration improves statistical stability, residual measurement noise may still affect the precision of the reported performance improvements.

\underline{External Threats} may impact generalizability.
We evaluated across three benchmarks, observing similar results across programming languages and software types.
However, there are several aspects of generalizability that are not assessed, most notably to other
  programming languages,
  application types,
  and hardware platforms.
% The SysLLMatic approach may not generalize in these cases because of incomplete training data for the underlying LLM.
In particular, our evaluation is limited to \Cpp and Java, which share a C-family syntax and similar imperative programming models; 
this may bias the results toward languages with similar execution semantics, memory management mechanisms, and performance-relevant coding patterns.
Although C-family languages are widely used in software engineering practice, the observed effectiveness of SysLLMatic may not directly generalize to languages with substantially different paradigms or runtime behaviors.

\section{Conclusion}
\label{sec:Conclusions}

We present SysLLMatic, a performance-guided software optimization system that integrates performance domain knowledge with LLMs to improve program efficiency across diverse benchmarks, including real-world applications.
Our system outperforms state-of-the-art LLM-based approaches on key metrics such as latency and throughput.
In addition, we contribute a catalog of 43 performance optimization patterns, which grounds SysLLMatic’s decision-making and provides a reusable resource for the community. 
Beyond performance gains, SysLLMatic highlights both the opportunities of automated software optimization at scale and the challenges of ensuring correctness, balancing performance with maintainability, and managing the cost of large-scale deployment. 
Future work includes integrating energy models to guide optimizations toward energy efficiency and enhancing correctness guarantees through formal verification, paving the way for more sustainable and trustworthy automated software optimization systems.

\section{Data Availability}
\label{sec:DataAvailability}
Our artifact is: \url{https://github.com/sysllmatic/sysllmatic}.

\section*{Acknowledgments}

Davis acknowledges support from NSF award \#2343596.
Thiruvathukal and Läufer acknowledge support from NSF award \#2343595.

\clearpage

%% Loading bibliography style file
\bibliographystyle{IEEEtran}

% Loading bibliography database
\bibliography{references,laufer-manual}

\end{document}